%
%
\input harvmac
\newcount\yearltd\yearltd=\year\advance\yearltd by 0

\noblackbox

\input epsf

\def\tilde{\widetilde}
\def\hat{\widehat}
\newcount\figno
\figno=0
\def\fig#1#2#3{
\par\begingroup\parindent=0pt\leftskip=1cm\rightskip=1cm\parindent=0pt
\baselineskip=11pt
\global\advance\figno by 1
\midinsert
\epsfxsize=#3
\centerline{\epsfbox{#2}}
\vskip 12pt
{\bf Figure\ \the\figno: } #1\par
\endinsert\endgroup\par
}
\def\figlabel#1{\xdef#1{\the\figno}}
\def\encadremath#1{\vbox{\hrule\hbox{\vrule\kern8pt\vbox{\kern8pt
\hbox{$\displaystyle #1$}\kern8pt}
\kern8pt\vrule}\hrule}}

\def\half{{\textstyle{1\over2}}}

\def\apm{{\alpha^{\prime}}}

\def\inbar{\vrule height1.5ex width.4pt depth0pt}

\def\half{{1\over 2}}

 \def\ep{{\epsilon}}
 \def\d{{\delta}}
 \def\rh{\rho}
 
 \def\a{{\alpha}}
 
 \def\frac#1#2{{#1\over #2}}

 \def\CO{{\cal O}}

 \def\CN{{\cal N}}
 \def\CD{{\cal D}}
 \def\p{\partial}
 \def\pslash{\p \llap{/}}

 \def\apm{{\a^{\prime}}}

\def\bd{{\bar{\delta}}}

\def\IR{\relax{\rm I\kern-.18em R}}
\def\IC{\relax\hbox{$\inbar\kern-.3em{\rm C}$}}
\def\IZ{\relax\ifmmode\hbox{Z\kern-.4em Z}\else{Z\kern-.4em Z}\fi}


\lref\SundborgUE{
B.~Sundborg,
``The Hagedorn transition, deconfinement and $N = 4$ SYM theory,''
Nucl.\ Phys.\ B {\bf 573}, 349 (2000)
[arXiv:hep-th/9908001].
}

\lref\GopakumarNS{
R.~Gopakumar,
``From free fields to AdS,''
arXiv:hep-th/0308184.
}

\lref\creutz{
M.~Creutz,
``On invariant integration over $SU(N)$,''
J.\ Math.\ Phys.\  {\bf 19}, 2043 (1978).
}

\lref\AharonyTT{
O.~Aharony and T.~Banks,
``Note on the quantum mechanics of M theory,''
JHEP {\bf 9903}, 016 (1999)
[arXiv:hep-th/9812237].
}

\lref\HawkingDH{
S.~W.~Hawking and D.~N.~Page,
``Thermodynamics of black holes in anti-de Sitter space,''
Commun.\ Math.\ Phys.\  {\bf 87}, 577 (1983).
}

\lref\WittenZW{
E.~Witten,
``Anti-de Sitter space, thermal phase transition, and confinement in
gauge theories,''
Adv.\ Theor.\ Math.\ Phys.\  {\bf 2}, 505 (1998)
[arXiv:hep-th/9803131].
}

\lref\PolyakovJU{
A.~M.~Polyakov,
``The wall of the cave,''
Int.\ J.\ Mod.\ Phys.\ A {\bf 14}, 645 (1999)
[arXiv:hep-th/9809057].
}

\lref\PolyakovAF{
A.~M.~Polyakov,
``Gauge fields and space-time,''
Int.\ J.\ Mod.\ Phys.\ A {\bf 17S1}, 119 (2002)
[arXiv:hep-th/0110196].
}

\lref\PolchinskiTW{
J.~Polchinski,
``High temperature limit of the confining phase,''
Phys.\ Rev.\ Lett.\  {\bf 68}, 1267 (1992)
[arXiv:hep-th/9109007].
}
\lref\KutasovGQ{
D.~Kutasov,
``Two-dimensional QCD coupled to adjoint matter and string theory,''
Nucl.\ Phys.\ B {\bf 414}, 33 (1994)
[arXiv:hep-th/9306013].
}

\lref\tdstring{
D.~J.~Gross,
``Two-dimensional QCD as a string theory,''
Nucl.\ Phys.\ B {\bf 400}, 161 (1993)
[arXiv:hep-th/9212149]; ~~~
D.~J.~Gross and W.~I.~Taylor,
``Two-dimensional QCD is a string theory,''
Nucl.\ Phys.\ B {\bf 400}, 181 (1993)
[arXiv:hep-th/9301068]; ~~~
D.~J.~Gross and W.~I.~Taylor,
``Twists and Wilson loops in the string theory of two-dimensional QCD,''
Nucl.\ Phys.\ B {\bf 403}, 395 (1993)
[arXiv:hep-th/9303046].
}

\lref\AtickSI{
J.~J.~Atick and E.~Witten,
``The Hagedorn transition and the number of degrees of freedom of
string theory,''
Nucl.\ Phys.\ B {\bf 310}, 291 (1988).
}

\lref\adscft{
J.~M.~Maldacena,
``The large $N$ limit of superconformal field theories and supergravity,''
Adv.\ Theor.\ Math.\ Phys.\  {\bf 2}, 231 (1998)
[Int.\ J.\ Theor.\ Phys.\  {\bf 38}, 1113 (1999)]
[arXiv:hep-th/9711200]; ~~~
S.~S.~Gubser, I.~R.~Klebanov and A.~M.~Polyakov,
``Gauge theory correlators from non-critical string theory,''
Phys.\ Lett.\ B {\bf 428}, 105 (1998)
[arXiv:hep-th/9802109]; ~~~
E.~Witten,
``Anti-de Sitter space and holography,''
Adv.\ Theor.\ Math.\ Phys.\  {\bf 2}, 253 (1998)
[arXiv:hep-th/9802150].
}

\lref\adsreview{
O.~Aharony, S.~S.~Gubser, J.~M.~Maldacena, H.~Ooguri and Y.~Oz,
``Large $N$ field theories, string theory and gravity,''
Phys.\ Rept.\  {\bf 323}, 183 (2000)
[arXiv:hep-th/9905111].
}

\lref\SvetitskyGS{
B.~Svetitsky and L.~G.~Yaffe,
``Critical behavior at finite temperature confinement transitions,''
Nucl.\ Phys.\ B {\bf 210}, 423 (1982).
}
\lref\ThornIV{
C.~B.~Thorn,
``Infinite $N_c$ QCD at finite temperature: is there an ultimate
temperature?,''
Phys.\ Lett.\ B {\bf 99}, 458 (1981).
}

\lref\YaffeQF{
L.~G.~Yaffe and B.~Svetitsky,
``First order phase transition in the $SU(3)$ gauge theory at finite
temperature,''
Phys.\ Rev.\ D {\bf 26}, 963 (1982).
}

\lref\HallinKM{
J.~Hallin and D.~Persson,
``Thermal phase transition in weakly interacting, large $N_c$ QCD,''
Phys.\ Lett.\ B {\bf 429}, 232 (1998)
[arXiv:hep-ph/9803234].
}
\lref\PisarskiDB{
R.~D.~Pisarski,
``Finite temperature QCD at large $N$,''
Phys.\ Rev.\ D {\bf 29}, 1222 (1984).
}

\lref\sundfree{
 P.~Haggi-Mani and B.~Sundborg,
``Free large $N$ supersymmetric Yang-Mills theory as a string
theory,'' JHEP {\bf 0004}, 031 (2000) [arXiv:hep-th/0002189]; ~~~
B.~Sundborg, ``Stringy gravity, interacting tensionless strings
and massless higher spins,'' Nucl.\ Phys.\ Proc.\ Suppl.\  {\bf
102}, 113 (2001) [arXiv:hep-th/0103247].
}
\lref\thorn{
K.~Bardakci and C.~B.~Thorn,
``A worldsheet description of large $N_c$ quantum field theory,''
Nucl.\ Phys.\ B {\bf 626}, 287 (2002)
[arXiv:hep-th/0110301]; ~~~
C.~B.~Thorn,
``A worldsheet description of planar Yang-Mills theory,''
Nucl.\ Phys.\ B {\bf 637}, 272 (2002)
[Erratum-ibid.\ B {\bf 648}, 457 (2003)]
[arXiv:hep-th/0203167]; ~~~
K.~Bardakci and C.~B.~Thorn,
``A mean field approximation to the world sheet model of planar $\phi^3$
field theory,''
Nucl.\ Phys.\ B {\bf 652}, 196 (2003)
[arXiv:hep-th/0206205]; ~~~
S.~Gudmundsson, C.~B.~Thorn and T.~A.~Tran,
``BT worldsheet for supersymmetric gauge theories,''
Nucl.\ Phys.\ B {\bf 649}, 3 (2003)
[arXiv:hep-th/0209102]; ~~~
K.~Bardakci and C.~B.~Thorn,
``An improved mean field approximation on the worldsheet for planar
$\phi^3$ theory,''
Nucl.\ Phys.\ B {\bf 661}, 235 (2003)
[arXiv:hep-th/0212254]; ~~~
C.~B.~Thorn and T.~A.~Tran,
``The fishnet as anti-ferromagnetic phase of worldsheet Ising spins,''
arXiv:hep-th/0307203; ~~~
C.~B.~Thorn, ``Quantum field theory in the language of light-cone
string,'' arXiv:hep-th/0310121; ~~~
A.~Mikhailov, ``Notes on higher spin symmetries,''
arXiv:hep-th/0201019; ~~~
A.~A.~Tseytlin, ``On limits of superstring in $AdS_5 \times S^5$,''
Theor.\ Math.\ Phys.\  {\bf 133}, 1376 (2002) [Teor.\ Mat.\ Fiz.\
{\bf 133}, 69 (2002)] [arXiv:hep-th/0201112]; ~~~
 E.~Sezgin and P.~Sundell,
``Massless higher spins and holography,'' Nucl.\ Phys.\ B {\bf
644}, 303 (2002) [Erratum-ibid.\ B {\bf 660}, 403 (2003)]
[arXiv:hep-th/0205131].
}

\lref\Pisa{
R.~D.~Pisarski and M.~Tytgat,
``Why the $SU(infinity)$ deconfining phase transition might be of  
second order,''
arXiv:hep-ph/9702340.
}

\lref\future{O.~Aharony, J.~Marsano, S.~Minwalla,
K.~Papadodimas and M.~Van Raamsdonk, work in progress.}

\lref\TseytlinGZ{
A.~A.~Tseytlin,
``On limits of superstring in $AdS_5 \times S^5$,''
Theor.\ Math.\ Phys.\  {\bf 133}, 1376 (2002)
[Teor.\ Mat.\ Fiz.\  {\bf 133}, 69 (2002)]
[arXiv:hep-th/0201112].
}
\lref\KarchVN{
A.~Karch,
``Lightcone quantization of string theory duals of free field theories,''
arXiv:hep-th/0212041.
}
\lref\joe{
J.~Polchinski, unpublished.}
\lref\DharFI{
A.~Dhar, G.~Mandal and S.~R.~Wadia,
``String bits in small radius AdS and weakly coupled $N = 4$ super
Yang-Mills theory. I,''
arXiv:hep-th/0304062.
}
\lref\ClarkWK{
A.~Clark, A.~Karch, P.~Kovtun and D.~Yamada,
``Construction of bosonic string theory on infinitely curved anti-de
Sitter space,''
Phys.\ Rev.\ D {\bf 68}, 066011 (2003)
[arXiv:hep-th/0304107].
}
\lref\BianchiWX{
M.~Bianchi, J.~F.~Morales and H.~Samtleben,
``On stringy $AdS_5 \times S^5$ and higher spin holography,''
JHEP {\bf 0307}, 062 (2003)
[arXiv:hep-th/0305052].
}
\lref\deMedeirosHR{
P.~de Medeiros and S.~P.~Kumar,
``Spacetime Virasoro algebra from strings on zero radius $AdS_3$,''
arXiv:hep-th/0310040.
}
\lref\LuciniZR{
B.~Lucini, M.~Teper and U.~Wenger,
``The high temperature phase transition in $SU(N)$ gauge theories,''
arXiv:hep-lat/0307017.
}

\lref\eliezer{
J.~L.~Barbon and E.~Rabinovici,
``Extensivity versus holography in anti-de Sitter spaces,''
Nucl.\ Phys.\ B {\bf 545}, 371 (1999)
[arXiv:hep-th/9805143]; ~~~
J.~L.~Barbon, I.~I.~Kogan and E.~Rabinovici,
``On stringy thresholds in SYM/AdS thermodynamics,''
Nucl.\ Phys.\ B {\bf 544}, 104 (1999)
[arXiv:hep-th/9809033]; ~~~
S.~A.~Abel, J.~L.~Barbon, I.~I.~Kogan and E.~Rabinovici,
``String thermodynamics in D-brane backgrounds,'' 
JHEP {\bf 9904}, 015 (1999)
[arXiv:hep-th/9902058]; ~~~
S.~A.~Abel, J.~L.~Barbon, I.~I.~Kogan and E.~Rabinovici,
``Some thermodynamical aspects of string theory,''
arXiv:hep-th/9911004.
}
\lref\OoguriGX{
H.~Ooguri and C.~Vafa,
``Worldsheet derivation of a large $N$ duality,''
Nucl.\ Phys.\ B {\bf 641}, 3 (2002)
[arXiv:hep-th/0205297].
}

\lref\ChamblinTK{
A.~Chamblin, R.~Emparan, C.~V.~Johnson and R.~C.~Myers,
``Charged AdS black holes and catastrophic holography,''
Phys.\ Rev.\ D {\bf 60}, 064018 (1999)
[arXiv:hep-th/9902170]; ~~~
M.~Cvetic and S.~S.~Gubser,
``Phases of R-charged black holes, spinning branes and strongly
coupled gauge theories,''
JHEP {\bf 9904}, 024 (1999)
[arXiv:hep-th/9902195]; ~~~
A.~Chamblin, R.~Emparan, C.~V.~Johnson and R.~C.~Myers,
``Holography, thermodynamics and fluctuations of charged AdS black holes,''
Phys.\ Rev.\ D {\bf 60}, 104026 (1999)
[arXiv:hep-th/9904197].
}

\lref\rajaraman{R. Rajaraman, ``Solitons And Instantons,''
North Holland Publishing Company, Chapter 6.}

\lref\gw{
D.~J.~Gross and E.~Witten, ``Possible third order phase transition
in the large $N$ lattice gauge theory'',
Phys.\ Rev. \ D{\bf 21}, 446 (1980); ~~~
S.~Wadia,
``A study of $U(N)$ lattice gauge theory in two-dimensions,''
preprint EFI-79/44-CHICAGO; ~~~
S.~R.~Wadia,
``$N = \infty$ phase transition in a class of exactly soluble model 
lattice gauge theories,''
Phys.\ Lett.\ B {\bf 93}, 403 (1980).
}

\lref\Polrev{
A.~M.~Polyakov,
``String theory and quark confinement,''
Nucl.\ Phys.\ Proc.\ Suppl.\  {\bf 68}, 1 (1998)
[arXiv:hep-th/9711002] and references therein.
}

\lref\Polo{
A.~M.~Polyakov,
``String representations and hidden symmetries for gauge fields,''
Phys.\ Lett.\ B {\bf 82}, 247 (1979).
}

\lref\Polt{
A.~M.~Polyakov,
``Gauge fields as rings of glue,''
Nucl.\ Phys.\ B {\bf 164}, 171 (1980).
}

\lref\FotopoulosES{
A.~Fotopoulos and T.~R.~Taylor,
``Comment on two-loop free energy in $N = 4$ supersymmetric Yang-Mills
theory at finite temperature,''
Phys.\ Rev.\ D {\bf 59}, 061701 (1999)
[arXiv:hep-th/9811224].
}

\lref\GubserNZ{
S.~S.~Gubser, I.~R.~Klebanov and A.~A.~Tseytlin,
``Coupling constant dependence in the thermodynamics of $N = 4$
supersymmetric Yang-Mills theory,''
Nucl.\ Phys.\ B {\bf 534}, 202 (1998)
[arXiv:hep-th/9805156].
}

\lref\JurkiewiczIZ{
J.~Jurkiewicz and K.~Zalewski,
``Vacuum structure of the $U(N \to \infty)$ gauge theory on a
two-dimensional lattice for a broad class of variant actions,''
Nucl.\ Phys.\ B {\bf 220}, 167 (1983).
}

\lref\thooft{
G.~'t Hooft,
``A planar diagram theory for strong
interactions,'' Nucl.\ Phys.\ B {\bf 72}, 461 (1974).}

\lref\integrability{
J.~A.~Minahan and K.~Zarembo,
``The Bethe-ansatz for $N = 4$ super Yang-Mills,''
JHEP {\bf 0303}, 013 (2003)
[arXiv:hep-th/0212208]; ~~~
N.~Beisert, C.~Kristjansen and M.~Staudacher,
``The dilatation operator of $N = 4$ super Yang-Mills theory,''
Nucl.\ Phys.\ B {\bf 664}, 131 (2003)
[arXiv:hep-th/0303060]; ~~~
G.~Mandal, N.~V.~Suryanarayana and S.~R.~Wadia,
``Aspects of semiclassical strings in $AdS_5$,''
Phys.\ Lett.\ B {\bf 543}, 81 (2002)
[arXiv:hep-th/0206103]; ~~~
I.~Bena, J.~Polchinski and R.~Roiban,
``Hidden symmetries of the $AdS_5 \times S^5$ superstring,''
arXiv:hep-th/0305116; ~~~
N.~Beisert,
``The complete one-loop dilatation operator of $N = 4$ super Yang-Mills
theory,''
arXiv:hep-th/0307015; ~~~
N.~Beisert and M.~Staudacher,
``The $N = 4$ SYM integrable super spin chain,''
Nucl.\ Phys.\ B {\bf 670}, 439 (2003)
[arXiv:hep-th/0307042]; ~~~
L.~Dolan, C.~R.~Nappi and E.~Witten,
``A relation between approaches to integrability in superconformal
Yang-Mills theory,''
JHEP {\bf 0310}, 017 (2003)
[arXiv:hep-th/0308089]; ~~~
L.~F.~Alday,
``Nonlocal charges on $AdS_5 \times S^5$ and pp-waves,''
arXiv:hep-th/0310146; ~~~
T.~Klose and J.~Plefka,
``On the integrability of large $N$ plane-wave matrix theory,''
arXiv:hep-th/0310232.
}

\lref\AharonyQU{
O.~Aharony and E.~Witten,
``Anti-de Sitter space and the center of the gauge group,''
JHEP {\bf 9811}, 018 (1998)
[arXiv:hep-th/9807205].
}

\lref\PolyakovVU{
A.~M.~Polyakov,
``Thermal properties of gauge fields and quark liberation,''
Phys.\ Lett.\ B {\bf 72}, 477 (1978).
}
\lref\SusskindUP{
L.~Susskind,
``Lattice models of quark confinement at high temperature,''
Phys.\ Rev.\ D {\bf 20}, 2610 (1979).
}
\lref\HagedornST{
R.~Hagedorn,
``Statistical thermodynamics of strong interactions at high-energies,''
Nuovo Cim.\ Suppl.\  {\bf 3}, 147 (1965).
}

\lref\gold{
Y.~Y.~Goldschmidt,
``$1/N$ expansion in two-dimensional lattice gauge theory,''
J.\ Math.\ Phys.\  {\bf 21}, 1842 (1980).
}

\lref\parteig{
A.~Gocksch and R.~D.~Pisarski,
``Partition function for the eigenvalues of the Wilson line,''
Nucl.\ Phys.\ B {\bf 402}, 657 (1993)
[arXiv:hep-ph/9302233].
}

\lref\rajesh{
R.~Gopakumar and C.~Vafa,
``On the gauge theory/geometry correspondence,''
Adv.\ Theor.\ Math.\ Phys.\  {\bf 3}, 1415 (1999)
[arXiv:hep-th/9811131].
}


\lref\ppint{
N.~R.~Constable, D.~Z.~Freedman, M.~Headrick, S.~Minwalla, L.~Motl,
A.~Postnikov and W.~Skiba,
``PP-wave string interactions from perturbative Yang-Mills theory,''
JHEP {\bf 0207}, 017 (2002)
[arXiv:hep-th/0205089]; ~~~
C.~Kristjansen, J.~Plefka, G.~W.~Semenoff and M.~Staudacher,
``A new double-scaling limit of $N = 4$ super Yang-Mills theory and PP-wave
strings,''
Nucl.\ Phys.\ B {\bf 643}, 3 (2002)
[arXiv:hep-th/0205033];~~~
N.~Beisert, C.~Kristjansen, J.~Plefka, G.~W.~Semenoff and M.~Staudacher,
``BMN correlators and operator mixing in $N = 4$ super Yang-Mills theory,''
Nucl.\ Phys.\ B {\bf 650}, 125 (2003)
[arXiv:hep-th/0208178];~~~
N.~R.~Constable, D.~Z.~Freedman, M.~Headrick and S.~Minwalla,
``Operator mixing and the BMN correspondence,''
JHEP {\bf 0210}, 068 (2002)
[arXiv:hep-th/0209002];~~~
D.~J.~Gross, A.~Mikhailov and R.~Roiban,
``Operators with large R charge in $N = 4$ Yang-Mills theory,''
Annals Phys.\  {\bf 301}, 31 (2002)
[arXiv:hep-th/0205066];~~~
R.~de Mello Koch, A.~Jevicki and J.~P.~Rodrigues,
``Collective string field theory of matrix models in the BMN limit,''
arXiv:hep-th/0209155;~~~
N.~Beisert, C.~Kristjansen, J.~Plefka and M.~Staudacher,
``BMN gauge theory as a quantum mechanical system,''
Phys.\ Lett.\ B {\bf 558}, 229 (2003)
[arXiv:hep-th/0212269]; ~~~
J.~Gomis, S.~Moriyama and J.~W.~Park,
``SYM description of SFT Hamiltonian in a pp-wave background,''
arXiv:hep-th/0210153; ~~~
J.~Gomis, S.~Moriyama and J.~W.~Park,
``SYM description of pp-wave string interactions:
Singlet sector and  arbitrary impurities,''
Nucl.\ Phys.\ B {\bf 665}, 49 (2003)
[arXiv:hep-th/0301250];~~~
R.~de Mello Koch, A.~Donos, A.~Jevicki and J.~P.~Rodrigues,
``Derivation of string field theory from the large $N$ BMN limit,''
Phys.\ Rev.\ D {\bf 68}, 065012 (2003)
[arXiv:hep-th/0305042].
}

\lref\pervijay{
V.~Balasubramanian and P.~Kraus,
``A stress tensor for anti-de Sitter gravity,''
Commun.\ Math.\ Phys.\  {\bf 208}, 413 (1999)
[arXiv:hep-th/9902121].
}

\lref\neil{
C.~P.~Burgess, N.~R.~Constable and R.~C.~Myers,
``The free energy of $N = 4$ Super Yang-Mills and the 
AdS/CFT  correspondence,''
JHEP {\bf 9908}, 017 (1999)
[arXiv:hep-th/9907188].
}

\lref\grossreview{
D.~J.~Gross, R.~D.~Pisarski and L.~G.~Yaffe,
``QCD and instantons at finite temperature,''
Rev.\ Mod.\ Phys.\  {\bf 53}, 43 (1981).
}

\lref\thooftunder{
G.~'t Hooft,
``Under the spell of the gauge principle,''
Adv.\ Ser.\ Math.\ Phys.\  {\bf 19}, 1 (1994).
}

\lref\bd{
N.~D.~Birrell and P.~C.~Davies,
``Quantum fields in curved space.''
}

\lref\tHooftHY{
G.~'t Hooft,
``On the phase transition towards permanent quark confinement,''
Nucl.\ Phys.\ B {\bf 138}, 1 (1978).
}

\lref\CabibboIG{
N.~Cabibbo and G.~Parisi,
``Exponential hadronic spectrum and quark liberation,''
Phys.\ Lett.\ B {\bf 59}, 67 (1975).
}

\lref\CollinsKY{
J.~C.~Collins and M.~J.~Perry,
``Superdense matter: neutrons or asymptotically free quarks?,''
Phys.\ Rev.\ Lett.\  {\bf 34}, 1353 (1975).
}

\lref\GattringerFI{
C.~R.~Gattringer, L.~D.~Paniak and G.~W.~Semenoff,
``Deconfinement transition for quarks on a line,''
Annals Phys.\  {\bf 256}, 74 (1997)
[arXiv:hep-th/9612030].
}

\lref\BarbonDI{
J.~L.~Barbon and E.~Rabinovici,
``Closed-string tachyons and the Hagedorn transition in AdS space,''
JHEP {\bf 0203}, 057 (2002)
[arXiv:hep-th/0112173].
}

\lref\maldawil{
J.~M.~Maldacena,
``Wilson loops in large $N$ field theories,''
Phys.\ Rev.\ Lett.\ {\bf 80}, 4859 (1998) [arXiv:hep-th/9803002].}
\lref\reyyee{
S.-J.~Rey and J.~Yee,
``Macroscopic strings as heavy quarks in
large $N$ gauge theories and anti-de Sitter supergravity,''
Eur.\ Phys.\ J.\ C {\bf 22}, 379 (2001)
[arXiv:hep-th/9803001].}

\lref\SathiapalanDB{B.~Sathiapalan, ``Vortices on the string world
sheet and constraints on toral compactification,'' Phys.\ Rev.\ D
{\bf 35}, 3277 (1987); ~~~
Y.~I.~Kogan,
``Vortices on the world sheet and string's critical dynamics,''
JETP Lett.\  {\bf 45}, 709 (1987)
[Pisma Zh.\ Eksp.\ Teor.\ Fiz.\  {\bf 45}, 556 (1987)].
}

\lref\LandsteinerXV{ K.~Landsteiner and E.~Lopez, ``The
thermodynamic potentials of Kerr-AdS black holes and their CFT
duals,'' JHEP {\bf 9912}, 020 (1999) [arXiv:hep-th/9911124].
}

\lref\finn{ D.~Kutasov and F.~Larsen,
``Partition sums and entropy
bounds in weakly coupled CFT,'' JHEP {\bf 0101}, 001 (2001)
[arXiv:hep-th/0009244].
}

\lref\BarbonNW{
J.~L.~Barbon and E.~Rabinovici,
``Remarks on black hole instabilities and closed string tachyons,''
Found.\ Phys.\  {\bf 33}, 145 (2003)
[arXiv:hep-th/0211212].
}

\lref\GreeneCD{ 
L.~A.~Pando Zayas and D.~Vaman,
``Strings in RR plane wave background at finite temperature,''
Phys.\ Rev.\ D {\bf 67}, 106006 (2003)
[arXiv:hep-th/0208066]; ~~~
B.~R.~Greene, K.~Schalm and G.~Shiu, ``On the
Hagedorn behaviour of pp-wave strings and N = 4 SYM theory at
finite R-charge density,'' Nucl.\ Phys.\ B {\bf 652}, 105 (2003)
[arXiv:hep-th/0208163]; ~~~
Y.~Sugawara,
``Thermal amplitudes in DLCQ superstrings on pp-waves,''
Nucl.\ Phys.\ B {\bf 650}, 75 (2003)
[arXiv:hep-th/0209145]; ~~~
R.~C.~Brower, D.~A.~Lowe and C.~I.~Tan, ``Hagedorn
transition for strings on pp-waves and tori with chemical
potentials,'' Nucl.\ Phys.\ B {\bf 652}, 127 (2003)
[arXiv:hep-th/0211201]; ~~~
Y.~Sugawara,
``Thermal partition function of superstring on compactified pp-wave,''
Nucl.\ Phys.\ B {\bf 661}, 191 (2003)
[arXiv:hep-th/0301035]; ~~~
G.~Grignani, M.~Orselli, G.~W.~Semenoff and
D.~Trancanelli, ``The superstring Hagedorn temperature in a
pp-wave background,'' JHEP {\bf 0306}, 006 (2003)
[arXiv:hep-th/0301186]; ~~~
S.~J.~Hyun, J.~D.~Park and S.~H.~Yi,
``Thermodynamic behavior of IIA string theory on a pp-wave,''
arXiv:hep-th/0304239; ~~~
F.~Bigazzi and A.~L.~Cotrone,
``On zero-point energy, stability and Hagedorn behavior of type IIB  
strings on pp-waves,''
JHEP {\bf 0308}, 052 (2003)
[arXiv:hep-th/0306102]; ~~~
L.~A.~Pando Zayas and D.~Vaman,
``Hadronic density of states from string theory,''
Phys.\ Rev.\ Lett.\  {\bf 91}, 111602 (2003)
[arXiv:hep-th/0306107]; ~~~
R.~Apreda, F.~Bigazzi and A.~L.~Cotrone,
``Strings on pp-waves and hadrons in (softly broken) $N = 1$ gauge theories,''
arXiv:hep-th/0307055.
}

\lref\BuchelRE{ A.~Buchel and L.~A.~Pando Zayas, ``Hagedorn vs.
Hawking-Page transition in string theory,'' Phys.\ Rev.\ D {\bf
68}, 066012 (2003) [arXiv:hep-th/0305179].
}

\lref\justin{
J.~R.~David, G.~Mandal and S.~R.~Wadia,
``Microscopic formulation of black holes in string theory,''
Phys.\ Rept.\  {\bf 369}, 549 (2002)
[arXiv:hep-th/0203048].
}

\lref\malda{
R.~Dijkgraaf, J.~M.~Maldacena, G.~W.~Moore and E.~Verlinde,
``A black hole farey tail,''
arXiv:hep-th/0005003.
}

\lref\msv{
J.~Maldacena, M.~M.~Sheikh-Jabbari and M.~Van Raamsdonk,
``Transverse fivebranes in matrix theory,''
JHEP {\bf 0301}, 038 (2003)
[arXiv:hep-th/0211139].
}

\lref\gordon{
K. Furuuchi, E. Schreiber and G. W. Semenoff,
``Five-brane thermodynamics from the Matrix model,''
arXiv:hep-th/0310286.}

\lref\BanksFI{
T.~Banks and E.~Rabinovici,
``Finite temperature behavior of the lattice Abelian Higgs model,''
Nucl.\ Phys.\ B {\bf 160}, 349 (1979); ~~~
E.~H.~Fradkin and S.~H.~Shenker,
``Phase diagrams of lattice gauge theories with Higgs fields,''
Phys.\ Rev.\ D {\bf 19}, 3682 (1979).
}

\lref\martinec{
M.~Li, E.~J.~Martinec and V.~Sahakian,
``Black holes and the SYM phase diagram,''
Phys.\ Rev.\ D {\bf 59}, 044035 (1999)
[arXiv:hep-th/9809061]; ~~~
E.~J.~Martinec and V.~Sahakian,
``Black holes and the SYM phase diagram. II,''
Phys.\ Rev.\ D {\bf 59}, 124005 (1999)
[arXiv:hep-th/9810224].
}

\lref\conifold{
A.~Buchel,
``Finite temperature resolution of the Klebanov-Tseytlin singularity,''
Nucl.\ Phys.\ B {\bf 600}, 219 (2001)
[arXiv:hep-th/0011146]; ~~~
A.~Buchel, C.~P.~Herzog, I.~R.~Klebanov, L.~A.~Pando Zayas and A.~A.~Tseytlin,
``Non-extremal gravity duals for fractional D3-branes on the conifold,''
JHEP {\bf 0104}, 033 (2001)
[arXiv:hep-th/0102105]; ~~~
S.~S.~Gubser, C.~P.~Herzog, I.~R.~Klebanov and A.~A.~Tseytlin,
``Restoration of chiral symmetry: A supergravity perspective,''
JHEP {\bf 0105}, 028 (2001)
[arXiv:hep-th/0102172].
}

\lref\MaldacenaYY{
J.~M.~Maldacena and C.~Nunez,
``Towards the large $N$ limit of pure $N = 1$ super Yang Mills,''
Phys.\ Rev.\ Lett.\  {\bf 86}, 588 (2001)
[arXiv:hep-th/0008001].
}

\lref\mnbh{
A.~Buchel and A.~R.~Frey,
``Comments on supergravity dual of pure $N = 1$ super Yang Mills theory  
with unbroken chiral symmetry,''
Phys.\ Rev.\ D {\bf 64}, 064007 (2001)
[arXiv:hep-th/0103022]; ~~~
A.~Buchel,
``On the thermodynamic instability of LST,''
arXiv:hep-th/0107102; ~~~
S.~S.~Gubser, A.~A.~Tseytlin and M.~S.~Volkov,
``Non-Abelian 4-d black holes, wrapped 5-branes, and their dual  
descriptions,''
JHEP {\bf 0109}, 017 (2001)
[arXiv:hep-th/0108205].
}

\lref\others{
D.~Z.~Freedman and J.~A.~Minahan,
``Finite temperature effects in the supergravity dual of the $N = 1^*$ 
gauge  theory,''
JHEP {\bf 0101}, 036 (2001)
[arXiv:hep-th/0007250]; ~~~
A.~Buchel and J.~T.~Liu,
``Thermodynamics of the $N = 2^*$ flow,''
arXiv:hep-th/0305064.
}

\lref\td{
G.~'t Hooft,
``A two-dimensional model for mesons,''
Nucl.\ Phys.\ B {\bf 75}, 461 (1974); ~~~
S.~Dalley and I.~R.~Klebanov,
``String spectrum Of (1+1)-dimensional large $N$ QCD with adjoint matter,''
Phys.\ Rev.\ D {\bf 47}, 2517 (1993)
[arXiv:hep-th/9209049]; ~~~
G.~Bhanot, K.~Demeterfi and I.~R.~Klebanov,
``(1+1)-dimensional large N QCD coupled to adjoint fermions,''
Phys.\ Rev.\ D {\bf 48}, 4980 (1993)
[arXiv:hep-th/9307111]; ~~~
K.~Demeterfi, I.~R.~Klebanov and G.~Bhanot,
``Glueball spectrum in a (1+1)-dimensional model for QCD,''
Nucl.\ Phys.\ B {\bf 418}, 15 (1994)
[arXiv:hep-th/9311015]; ~~~
I.~I.~Kogan and A.~R.~Zhitnitsky,
``Two dimensional QCD with matter in adjoint representation: 
What does it teach us?,''
Nucl.\ Phys.\ B {\bf 465}, 99 (1996)
[arXiv:hep-ph/9509322].
}

\lref\GregoryVY{
R.~Gregory and R.~Laflamme,
``Black strings and p-branes are unstable,''
Phys.\ Rev.\ Lett.\  {\bf 70}, 2837 (1993)
[arXiv:hep-th/9301052].
}

\lref\BermanAB{
D.~S.~Berman and M.~K.~Parikh,
``Confinement and the AdS/CFT correspondence,''
Phys.\ Lett.\ B {\bf 483}, 271 (2000)
[arXiv:hep-th/0002031].
}

\lref\rotbh{
S.~S.~Gubser,
``Thermodynamics of spinning D3-branes,''
Nucl.\ Phys.\ B {\bf 551}, 667 (1999)
[arXiv:hep-th/9810225]; ~~~
P.~Kraus, F.~Larsen and S.~P.~Trivedi,
``The Coulomb branch of gauge theory from rotating branes,''
JHEP {\bf 9903}, 003 (1999)
[arXiv:hep-th/9811120].
}

\lref\HawkingDP{
S.~W.~Hawking and H.~S.~Reall,
``Charged and rotating AdS black holes and their CFT duals,''
Phys.\ Rev.\ D {\bf 61}, 024014 (2000)
[arXiv:hep-th/9908109].
}

\lref\kleb{
I.~R.~Klebanov and E.~Witten,
``Superconformal field theory on threebranes at a Calabi-Yau  singularity,''
Nucl.\ Phys.\ B {\bf 536}, 199 (1998)
[arXiv:hep-th/9807080]; ~~~
I.~R.~Klebanov and N.~A.~Nekrasov,
``Gravity duals of fractional branes and logarithmic RG flow,''
Nucl.\ Phys.\ B {\bf 574}, 263 (2000)
[arXiv:hep-th/9911096]; ~~~
I.~R.~Klebanov and A.~A.~Tseytlin,
``Gravity duals of supersymmetric $SU(N) \times SU(N+M)$ gauge theories,''
Nucl.\ Phys.\ B {\bf 578}, 123 (2000)
[arXiv:hep-th/0002159]; ~~~
I.~R.~Klebanov and M.~J.~Strassler,
``Supergravity and a confining gauge theory: Duality cascades and  
$\chi$SB-resolution of naked singularities,''
JHEP {\bf 0008}, 052 (2000)
[arXiv:hep-th/0007191].
}


\lref\herbert{
R.~Narayanan and H.~Neuberger,
``Large $N$ reduction in continuum,''
Phys.\ Rev.\ Lett.\  {\bf 91}, 081601 (2003)
[arXiv:hep-lat/0303023]; ~~~
J.~Kiskis, R.~Narayanan and H.~Neuberger,
``Does the crossover from perturbative to nonperturbative physics in QCD 
become a phase transition at infinite $N$ ?,''
Phys.\ Lett.\ B {\bf 574}, 65 (2003)
[arXiv:hep-lat/0308033].
}

\lref\KimSG{
C.~J.~Kim and S.~J.~Rey,
``Thermodynamics of large-$N$ super Yang-Mills theory and AdS/CFT  
correspondence,''
Nucl.\ Phys.\ B {\bf 564}, 430 (2000)
[arXiv:hep-th/9905205].
}

\lref\ReyBQ{
S.~J.~Rey, S.~Theisen and J.~T.~Yee,
``Wilson-Polyakov loop at finite temperature in large $N$ gauge theory and  
anti-de Sitter supergravity,''
Nucl.\ Phys.\ B {\bf 527}, 171 (1998)
[arXiv:hep-th/9803135].
}

\lref\BrandhuberBS{
A.~Brandhuber, N.~Itzhaki, J.~Sonnenschein and S.~Yankielowicz,
``Wilson loops in the large $N$ limit at finite temperature,''
Phys.\ Lett.\ B {\bf 434}, 36 (1998)
[arXiv:hep-th/9803137]; ~~~
A.~Brandhuber, N.~Itzhaki, J.~Sonnenschein and S.~Yankielowicz,
``Wilson loops, confinement, and phase transitions in large $N$ gauge  
theories from supergravity,''
JHEP {\bf 9806}, 001 (1998)
[arXiv:hep-th/9803263].
}

\lref\HalpernFZ{
M.~B.~Halpern,
``On the large N limit of conformal field theory,''
Annals Phys.\  {\bf 303}, 321 (2003)
[arXiv:hep-th/0208150].
}

\lref\spentatoappear{
L.~Alvarez-Gaume, C. Gomez, H. Liu and S. Wadia, to appear.}

\def\my_Title#1#2{\nopagenumbers\abstractfont\hsize=\hstitle\rightline{#1}%
\vskip .5in\centerline{\titlefont #2}\abstractfont\vskip .5in\pageno=0}

\my_Title {\vbox{\baselineskip12pt
\hbox{WIS/29/03-OCT-DPP}
\hbox{\tt hep-th/0310285}}} {\vbox{\centerline{The Hagedorn/Deconfinement
Phase Transition} \vskip 5pt \centerline{in Weakly Coupled Large $N$ Gauge
Theories}}}

\centerline{Ofer Aharony$^{a}$, Joseph Marsano$^{b}$, Shiraz Minwalla$^{b}$,}
\centerline{Kyriakos Papadodimas$^{b}$ and Mark Van Raamsdonk$^{c,d}$}

\medskip

\centerline{\sl $^{a}$Department of Particle Physics, Weizmann Institute of
Science, Rehovot 76100, Israel}
\centerline{\sl $^{b}$Jefferson Physical Laboratory, Harvard University,
Cambridge, MA 02138, USA}
\centerline{\sl $^{c}$Department of Physics, Stanford University,
Stanford, CA 94305, USA}
\centerline{\sl $^{d}$Department of Physics and Astronomy,
University of British Columbia,}
\centerline{\sl Vancouver, BC, V6T 1Z1, Canada}
\medskip


\medskip

\noindent
We demonstrate that weakly coupled, large $N$, $d$-dimensional $SU(N)$
gauge theories on a class of compact spatial manifolds (including $S^{d-1}
\times$ time) undergo deconfinement phase transitions at temperatures
proportional to the inverse length scale of the manifold in question.
The low temperature phase has a free energy of order one, and is characterized
by a stringy (Hagedorn) growth in its density of states. The high
temperature phase has a free energy of order $N^2$. These
phases are separated either by a single first order transition that
generically occurs below the Hagedorn temperature or by two continuous phase
transitions, the first of which occurs at the Hagedorn temperature.
These phase transitions 
could perhaps be
continuously connected to the
usual flat space deconfinement transition in the case of confining
gauge theories, and to the Hawking-Page nucleation of $AdS_5$ black holes
in the case of the $\CN=4$ supersymmetric Yang-Mills theory. We suggest that
deconfinement transitions may generally
be interpreted in terms of black hole
formation in a dual string theory. Our analysis proceeds by first
reducing the Yang-Mills partition function to a $(0+0)$-dimensional
integral over
a unitary matrix $U$, which is the holonomy (Wilson loop) of the gauge field
around the thermal time circle
in Euclidean space; deconfinement transitions are
large $N$ transitions in this matrix integral.

\Date{}  

\centerline{\bf Contents}\nobreak\medskip{\baselineskip=12pt
 \parskip=0pt\catcode`\@=11  

\noindent {1.} {Introduction} \leaderfill{2} \par 
\noindent {2.} {Background} \leaderfill{7} \par 
\noindent \quad{2.1.} {Hagedorn behavior in string theory} \leaderfill{7} \par 
\noindent \quad{2.2.} {Deconfinement transitions in gauge theories} \leaderfill{10} \par 
\noindent \quad{2.3.} {Hagedorn versus deconfinement} \leaderfill{13} \par 
\noindent \quad{2.4.} {Large $N$ deconfinement on compact spaces} \leaderfill{14} \par 
\noindent {3.} {Partition function for free Yang-Mills theory on a compact space} \leaderfill{15} \par 
\noindent \quad{3.1.} {Two matrix harmonic oscillators give Hagedorn behavior for large $N$} \leaderfill{16} \par 
\noindent \quad{3.2.} {Exact partition function for $N=\infty $} \leaderfill{17} \par 
\noindent \quad{3.3.} {Exact partition function for free Yang-Mills theory} \leaderfill{20} \par 
\noindent \quad{3.4.} {Generalization to arbitrary chemical potential} \leaderfill{21} \par 
\noindent \quad{3.5.} {$U(N)$ gauge theories with adjoint matter on $S^3 \times \relax {\fam 0\tenrm I\kern -.18em R}$} \leaderfill{22} \par 
\noindent {4.} {Path integral derivation of the matrix integral and an order parameter} \leaderfill{23} \par 
\noindent \quad{4.1.} {Basic set-up} \leaderfill{23} \par 
\noindent \quad{4.2.} {The integration measure} \leaderfill{24} \par 
\noindent \quad{4.3.} {Evaluation of $S_{eff}$ at one-loop} \leaderfill{25} \par 
\noindent {5.} {Solution of the free Yang-Mills matrix model} \leaderfill{29} \par 
\noindent \quad{5.1.} {Low temperature behavior} \leaderfill{31} \par 
\noindent \quad{5.2.} {Behavior near the transition} \leaderfill{33} \par 
\noindent \quad{5.3.} {High temperature behavior} \leaderfill{34} \par 
\noindent \quad{5.4.} {Exact solution for $T > T_H$} \leaderfill{34} \par 
\noindent \quad{5.5.} {Perturbative expansion slightly above the Hagedorn temperature} \leaderfill{38} \par 
\noindent \quad{5.6.} {Summary of thermodynamic behavior} \leaderfill{41} \par 
\noindent \quad{5.7.} {The Polyakov loop as an order parameter at finite volume} \leaderfill{42} \par 
\noindent \quad{5.8.} {Results for specific theories} \leaderfill{44} \par 
\noindent {6.} {Phase structure at weak coupling} \leaderfill{45} \par 
\noindent \quad{6.1.} {General properties of the effective action} \leaderfill{45} \par 
\noindent \quad{6.2.} {The general form of the effective action in perturbation theory} \leaderfill{46} \par 
\noindent \quad{6.3.} {Possible phase structures at weak coupling} \leaderfill{47} \par 
\noindent \quad{6.4.} {Toy model} \leaderfill{50} \par 
\noindent \quad{6.5.} {Density of states as a function of energy} \leaderfill{54} \par 
\noindent {7.} {Extrapolation to strong coupling and the dual description} \leaderfill{58} \par 
\noindent \quad{7.1.} {Possible phase diagrams for large $N$ Yang-Mills theories} \leaderfill{58} \par 
\noindent \quad{7.2.} {Dual interpretation of the ${\cal N}=4$ SYM thermodynamics at strong coupling} \leaderfill{64} \par 
\noindent \quad{7.3.} {Deconfinement and black holes} \leaderfill{66} \par 
\noindent \quad{7.4.} {Dual description at a general point in the phase diagram} \leaderfill{68} \par 
\noindent {8.} {Discussion and future directions} \leaderfill{70} \par 
\noindent Appendix {A.} {Properties of group characters} \leaderfill{74} \par 
\noindent Appendix {B.} {Counting states in $U(N)$ gauge theories} \leaderfill{76} \par 
\noindent \quad{\hbox {B.}1.} {Counting gauge-invariant states precisely} \leaderfill{76} \par 
\noindent \quad{\hbox {B.}2.} {Evaluating single-particle partition functions on spheres} \leaderfill{77} \par 
\noindent Appendix {C.} {Hagedorn transitions at finite string coupling} \leaderfill{80} \par 
\noindent References \leaderfill{83} \par
\catcode`\@=12 \bigbreak\bigskip}

\newsec{Introduction}

The thermodynamics of large $N$ $SU(N)$ gauge theories is interesting for at
least two different reasons.  On the one hand, these
theories are believed to share many qualitative features with the finite
$N$ non-Abelian gauge theories which are relevant to real-world QCD (or GUTs),
including confinement at zero temperature and
a deconfinement transition as the temperature is increased. On
the other hand, large $N$ gauge theories are believed \refs{\thooft,
\Polrev} to be dual to weakly coupled string theories, which
display Hagedorn behavior (reviewed below) \HagedornST\
associated with
singularities in various thermodynamic quantities. Thus, by studying
large $N$ gauge theory thermodynamics, one may hope to achieve a
better understanding of both deconfinement transitions in gauge theory
and Hagedorn behavior in string theory and, possibly, some
relationship between the two.

Unfortunately, asymptotically free gauge theories in Minkowski space
are strongly coupled except at very high temperatures, so direct
analysis of the thermodynamic behavior usually relies on numerical or
lattice techniques. However, by placing the theory on a compact space,
one obtains a tunable dimensionless parameter $R \Lambda_{QCD}$
(where $R$ is the size of the compact space and $\Lambda_{QCD}$ is the
dynamically generated scale of the gauge theory\foot{We use here, and
throughout the paper, notations
which are appropriate for four dimensional gauge theories. For lower
dimensional gauge theories, $\Lambda_{QCD}$ 
should be replaced by a scale which is an appropriate power 
of the 't Hooft coupling constant $g_{YM}^2 N$.})
which gives back the Minkowski space theory when it is large,
but gives a weakly coupled theory when it is small\foot{A similar
dimensionless parameter appears also in conformal gauge theories like the
$d=4$ $\CN=4$ supersymmetric Yang-Mills theory, which can also be studied
at weak coupling.}. In this
weak coupling limit the thermodynamics can be studied at all
temperatures in perturbation theory, and this analysis is the goal of
the present paper. Specifically, we investigate the thermodynamics of
$d$-dimensional $U(N)$ (or $SU(N)$) gauge theories with arbitrary
adjoint-valued matter fields\foot{Our methods can be easily generalized
to include also matter fields in other representations.}
on compact spatial manifolds, such as
$S^{d-1}$, in the limit of weak coupling.  This theory is simple
enough to analyze exactly in the large $N$ limit; we will find that it
has a surprisingly rich structure with many of the expected features
of strongly coupled gauge theory thermodynamics, including stringy
Hagedorn behavior and a deconfinement transition.

We should start by emphasizing that even at very weak coupling, 
$SU(N)$ Yang-Mills theory on a compact manifold behaves very differently 
from $(N^2-1)$ copies of the $U(1)$ theory. The excitations of
the $U(1)$ theory are arbitrary numbers of  photons; in contrast
(for example) a single $SU(N)$ gluon is not an allowed excitation
on a compact manifold, as gluons source lines of color electric
flux. Due to Gauss' law, the only allowed excitations of an $SU(N)$
theory on any compact manifold are combinations of gluons (and any
other charged particles in the theory) that are grouped into
$SU(N)$ singlets.

As we will see below, the projection onto the singlet sector
introduces effective interactions between the gluons.  It will
turn out that these interactions are negligible at high energies,
but they dominate the dynamics at low energies. As a consequence,
in the $N \to \infty$ limit, the gauge theories under study in
this paper have at least two distinct phases. The lowest
temperature phase (dominated by states with $E/N^2 \ll 1/R$) has a
strongly stringy flavor; it is characterized (see
\refs{\SundborgUE,\PolyakovAF} for the free theory) by a
Hagedorn-like density of states $\rh(E)$ that grows exponentially
with energy, $\rh(E) \propto e^{E/ T_H}$ with $T_H=C/R$, where $C$
is a constant depending on the matter content of the theory and on
the shape of the compactification manifold.
On the other hand, the high temperature
phase (dominated by states with $E/N^2 \gg 1/R$) behaves qualitatively
like a gas of free particles; the free energy $F$ takes the form
$F=N^2f(T)$, where $f(T) \propto T^d$ for $T \gg 1/R$. In some cases
these two phases may be separated by a more mysterious intermediate
temperature phase.

The free gauge theory undergoes precisely one phase transition as a
function of the temperature; this transition is of first order and
it occurs precisely at the Hagedorn temperature $T_H$.
As far as we know, this result was first found by \HallinKM\ in
a specific case, and was derived more generally in a beautiful paper
by Sundborg \SundborgUE.\foot{Sundborg's results do not seem to have
received wide attention; in particular, we learned of his paper only
after we had independently rederived all of its results, which have 
substantial overlap with section 3 and parts of section 5 of our paper.} 
For finite $N$, this phase
transition is smoothed out, as it must be for any theory with a
finite number of degrees of freedom at finite volume.

Turning on a small 't Hooft coupling $\lambda \equiv g_{YM}^2 N$
qualitatively alters the behavior of the free gauge theory. It
turns out that the interacting theory displays one of two possible
behaviors, depending on the sign of a coefficient determined by
two-loop and three-loop vacuum diagrams. When varying the
temperature, this theory either undergoes a single first order
transition {\it below} the Hagedorn temperature, or it undergoes
two continuous phase transitions, one at the Hagedorn temperature
and the second one slightly above it. Note that, in the first
case, the phase transition shields the Hagedorn spectrum of the
theory in the sense that there is no temperature at which string
excitations of arbitrarily high energy dominate the partition
function. In the second case the first of the two phase
transitions is truly Hagedorn-like; the singularities in the
neighborhood of the phase transition directly encode the
high-energy density of states of stringy oscillators.\foot{This is
consistent with the general arguments of
\refs{\CabibboIG,\PisarskiDB}, that a large $N$ second order
deconfinement transition must always be accompanied by a
Hagedorn-like spectrum and occur at the Hagedorn temperature.}

We derive the results described in the previous paragraphs
by reducing the thermal partition function to an integral over a single
unitary $U(N)$ (or $SU(N)$) matrix
\eqn\eqform{Z(\beta)= \int [dU] \exp[-S_{eff}(U)],}
where $U=e^{i \beta \alpha}$ ($\alpha$ is the
zero mode of $A_0$ on ${\cal M} \times S^1$, $\cal M$
is the spatial manifold in question, and $\beta\equiv 1/T$). The mode $\alpha$
is the lightest mode\foot{We
assume that there are no additional zero modes for any of the
fields. This is not always true, since, for instance, the gauge field
may have additional zero modes when the compactification manifold is not
simply connected. We will not discuss such examples here.}
of the gauge theory on ${\cal M} \times S^1$;
$S_{eff}(U)$, the quantum effective action for this light mode,
may be computed by integrating out all other fields in the theory.
This procedure (integrating out the heavy modes)  may be explicitly
performed in perturbation theory, generating an expression for $S_{eff}(U)$
as a power series in the 't Hooft coupling 
$\lambda$. The lowest term in this power series,
corresponding to the free gauge theory, is
particularly easy to compute, either by evaluating one-loop vacuum diagrams or
by explicitly counting gauge-invariant states in free
Yang-Mills theory using a projection onto the singlet sector of the
theory.\foot{This second method was already used to derive similar expressions
in \refs{\HallinKM,\SundborgUE}.} For the free theory
we obtain an expression (of the form \eqform) for
$Z(\beta)$ which is exact, even at finite $N$,
and applies to any gauge theory with any matter content.
The computation of higher order terms is straightforward in principle but
tedious in practice.

Having computed $S_{eff}$ we proceed to evaluate \eqform\ in the usual
manner, by changing the integration variable to the eigenvalues
$e^{i\theta}$ of $U$.  For the adjoint theories we consider, the
resulting effective action in the free limit is simply the sum over a
pairwise potential between the eigenvalues with a temperature-independent
repulsive term
and an attractive term that increases from zero to infinite strength
as the temperature increases from zero to infinity.  In the large $N$
limit, the integral is determined by a saddle point characterized by a
density of eigenvalues $\rho(\theta)$ that minimizes the effective
action. At sufficiently low temperatures, the repulsive term
dominates and $\rho(\theta)$ is constant on the circle.  At high
enough temperatures, the attractive term forces the eigenvalues to
bunch together and $\rho(\theta)$ vanishes outside a narrow interval on the
circle \SundborgUE. In the free theory, these two regimes are 
separated by a first
order transition at the Hagedorn temperature in which the eigenvalue
distribution jumps discontinuously from the uniform distribution to
a sinusoidal distribution.
However, it turns out that depending on the details of the interactions in 
$S_{eff}$, the nature of this transition
may be modified at arbitrarily weak non-zero coupling.

Based on the general form of $S_{eff}$, we show that two classes of
behavior are possible at weak coupling, as noted earlier. In the first
case, we have a single first order phase transition at which the
uniform low temperature distribution jumps discontinuously to a
non-uniform distribution. This is similar to the behavior in the free
theory, but the transition happens strictly below the Hagedorn
temperature for non-zero coupling. The second possibility is that the
eigenvalue distribution evolves continuously as a function of
temperature, changing from a uniform to a non-uniform (but nowhere
vanishing) distribution at a first critical temperature, and then
developing a gap (on which the distribution vanishes) at a second
critical temperature. Consequently, the corresponding theory undergoes
two phase transitions as a function of temperature; the first of these
is a second order transition at the Hagedorn temperature, while the
second is a third order transition closely related to the Gross-Witten
phase transition \gw\ of two dimensional lattice gauge theories.

Which of these two possibilities is realized depends on the sign of a
particular coefficient in $S_{eff}$ which appears to depend on the
details of the field content and interactions of the theory in question. We are
currently involved in a computation of this coefficient for the $d=4$
pure Yang-Mills
theory and for the $d=4$ $\CN=4$ supersymmetric Yang-Mills (SYM) theory;
we hope to report the result of this computation soon \future.

The phase transition at (or near) the Hagedorn temperature has a
natural interpretation in the gauge theory\foot{As we have described above,
in some situations the theory undergoes two phase transitions upon raising
the temperature. The discussion in this paragraph applies to the
first of these.}; it is simply a deconfining
transition.  The low-temperature phase can be thought of as a gas of
singlet `glueballs', with the free energy scaling as $N^0$, while the
high temperature phase, with the free energy scaling as $N^2$, is a
plasma of gluons (and other particles). Indeed, the traditional
deconfinement order parameter, the Polyakov loop in the fundamental
representation ${1\over N}\vev{\tr
(P e^{i \oint A_0})} \simeq {1\over N}\vev{\tr(U)}$, is a good order
parameter for the phase transitions described in this paper.
It is zero in the confining phase, and non-zero in the deconfining
phase\foot{Strictly
speaking, the Polyakov loop is zero in both phases as flux
conservation forbids placing a single quark on a compact space. In \S5.7 we
actually define our order parameter more carefully; it is either
${1\over N}\vev{\tr(U)}$ in an infinitesimal deformation of the gauge
theory corresponding to an infinitesimal Higgsing of the theory, in the
limit that the deformation goes to zero, or we can use ${1\over
N^2}\vev{\tr(U)\tr(U^{\dagger})}$.}.

The phase transitions we discuss in this paper could potentially be
continuously related to several other interesting phase transitions --
at least, the values of all the order parameters are consistent with
such a continuous relation.
In the case of the $3+1$-dimensional
$\CN=4$ $SU(N)$ SYM
theory compactified on a sphere, our analysis applies to the
deconfinement transition at weak 't Hooft coupling; the deconfinement
transition at strong coupling is related \WittenZW\
by the AdS/CFT correspondence \adscft\
to the Hawking-Page transition \HawkingDH\
of gravitational theories on asymptotically anti-de Sitter ($AdS$) spaces. For
asymptotically free gauge theories, our weakly coupled results apply when the
scale $R$ of the compactification manifold is much smaller than the
strong coupling scale $1/\Lambda_{QCD}$. However, when we take
$R\Lambda_{QCD} \gg 1$, our phase transitions 
could
turn into the infinite volume deconfinement transition.

We propose a set of phase diagrams for gauge theories as a function
of coupling ($\lambda$ or $R\Lambda_{QCD}$, respectively) and temperature
that interpolate between our weakly coupled results and the known
strong coupling
behavior. We are led to speculate on a dual interpretation
of deconfinement transitions; in particular,
we conjecture that deconfinement transitions are always
associated with black hole formation in a dual string theory. We argue that
the mysterious intermediate temperature phase associated with second order
deconfinement transitions (if such a phase is realized in some theory)
would be dual to a string theory in a background dominated by a strange
new type of stable black hole.

The paper is organized as follows. We begin in \S2 with a review
of some of the relevant background concerning Hagedorn behavior in
string theory, deconfinement transitions,
and the relation between them. In \S3, we show that large $N$ free
gauge theories have a Hagedorn-like behavior of their spectrum and
provide a first derivation of the matrix model expression for the thermal
partition function of compactified gauge theories by explicitly summing
over gauge-invariant states. In \S4, we derive
the same expression by using a path integral formalism.
In \S5, we describe the solution of this matrix model, exhibit the
phase transition, and both compute and discuss the relevant order parameters.
In \S6, we generalize the path integral derivation of \eqform\ to
interacting gauge theories and determine the general form of $S_{eff}$ in
perturbation theory. Utilizing these results, we discuss the different
possibilities for the phase structure of the theory at weak
coupling. In \S7, we suggest extrapolations of 
our results to strong coupling and
propose a set of possible phase diagrams for gauge theories as a function
of coupling and temperature. We also speculate on the interpretation of 
deconfinement
transitions in terms of black hole formation in a dual string theory.
We end in \S8 with a summary of our results and a
discussion of some possible future directions. Three appendices contain
useful technical results.

\newsec{Background}

\subsec{Hagedorn behavior in string theory}

A single free closed string has an infinite number of vibration modes, each of
which may be excited to arbitrary level. Each vibrational state
 of the string corresponds to a distinct
particle species in space-time. In fact (for strings in Minkowski
space), the number of particle species grows exponentially with their mass.

To be specific, consider a free type II superstring whose worldsheet is the
direct sum of the free SCFT on $\IR^{d-1,1}$ and a compact unitary SCFT
$C'$ with central charge $\hat{c}=10-d$. Let $N(M)$ denote the number of
particle species in this theory with mass less than $M$.
For $M^2$ much bigger than the string tension $1 / 2\pi \apm$,
\eqn\denpart{{dN \over dM}={K e^{2\pi \sqrt{2 \apm} M} \over M^{d}}  +
{ \it subleading},}
where $K$ is a constant. In the zero coupling limit, each particle species
is described by its own free quantum field theory.
The finite temperature partition function $Z$ of such a system of free
quantum fields is easily computed; the contribution
of highly excited string states (particle species of large mass,
$\apm M^2 \gg 1$) is given by\foot{
Note that $d$ in \denpart\  is the number of noncompact directions on
the space  $\IR^{d-1,1} \times C'$. It is possible that
\denpart\ (and the other formulas of this section)  apply to more
general string backgrounds (for instance with a warped space-time
geometry) upon replacing $d$ by $d_{eff}$, the effective number of noncompact
space-time dimensions in such spaces. In particular, $d_{eff}=1$
on a space like $AdS_5$ in global coordinates on which particle
propagation is gapped.}
\eqn\pf{ \ln(Z(T))  =K' T^{{d-1 \over 2}} \int
dM {e^{M ({2\pi \sqrt{2 \apm} - {1\over T}})} \over M^{{d+1 \over 2}} } }
for some constant $K'$.
Note that $Z(T)$ diverges for $T$ larger than the Hagedorn
temperature $T_H ={1 \over {2\pi   \sqrt{2 \apm}}}$. As $T$ approaches this
critical temperature from below, the partition function develops a
singularity
\eqn\pffb{
\ln(Z(T)) \propto \left\{ \matrix{ (\delta T)^{{d -1 \over 2}} \ln(\delta T) &
\qquad d \; {\rm odd} \cr (\delta T)^{{d -1 \over 2}} & \qquad d \; {\rm even}}
\right.
}
where $\delta T= T_H-T$.
\foot{Note that the expectation
value of the energy density
at $T=T_H$ is convergent for $d>3$ but diverges when $d \leq 3$. Note also
that states of very high energy contribute significantly just below the
Hagedorn temperature. Consequently, the properties of the theory at fixed
energy $E \sqrt{\apm}$ (microcanonical ensemble) may differ significantly
from the properties of the theory at any fixed temperature
(canonical ensemble).}

It is useful to understand the origin of the divergence in \pf\ from an
alternate point of view. Recall that, for any system, the
thermal partition function may be computed by the Euclidean partition
function with the time direction compactified on a thermal circle $S_T^1$
of circumference $\beta = 1/T$ (bosons have periodic boundary conditions
and fermions have anti-periodic boundary conditions around a thermal
circle).
For free string theory, this means
that $Z(T)$ in \pf\ can be computed by the string theory
torus partition function on the Euclidean space-time $\IR^{d-1} \times S_T^1
\times C'$.
If we now take the time direction to be in $\IR^{d-1}$, we can reinterpret
$Z(T)$ as the
one-loop contribution to the vacuum energy of string theory on the 9
dimensional space  $\IR^{d-2} \times S_T^1 \times C'$. The one-loop
contribution to the vacuum energy of any theory is determined completely by
its spectrum; in particular it diverges if the spectrum includes a tachyon.
Indeed, superstrings winding a thermal
circle an odd number of times have a tachyonic mode when the size of the circle
is small enough \AtickSI; this is because modular invariance forces the
imposition of the opposite GSO projection, one that projects in the identity
operator, on states of odd winding.
Specifically, the ground state of a superstring that winds once around
the thermal circle has mass
\eqn\massofw{m_W^2={2 \over \apm} \left({1 \over 8 \pi^2 \apm T^2} -1\right),}
and becomes tachyonic for $T > T_H$, providing an alternate
explanation for the divergence of $Z(T)$ for $T > T_H$.

For the theory with strictly vanishing string coupling $g_s$ the
partition function is simply ill-defined at $T > T_H$. However,
for any finite $g_s$, the perturbative analysis of the spectrum
breaks down at sufficiently high energies (of order $1/g_s^2$), so
the behavior as $T \to T_H$ is difficult to analyze directly
(recall that thermodynamics near the Hagedorn temperature receives
non-negligible contributions from states with arbitrarily high
energy). It has been speculated \AtickSI\ that at any finite
coupling, no matter how small, the divergence of the free
$Z(\beta)$ above the Hagedorn temperature may be replaced by a
phase transition to a new high temperature phase. According to
this suggestion, the string mode $W$ winding the thermal circle is
the order parameter for this transition
\refs{\SathiapalanDB,\AtickSI}. The phase transition occurs when
this mode condenses. This could occur either below the Hagedorn
temperature (which is the temperature at which this mode becomes
massless), in which case the transition is of first order, or at
the Hagedorn temperature, in which case the transition is second
order.

Unfortunately, in general we know very little about this conjectured high
temperature `phase' of string theory.
In \AtickSI\ it was argued that, were such a phase transition to occur in
flat space, it must be of first order.  However, thermodynamics and phase
transitions in gravitational systems in flat space
are at best approximate notions (see \AtickSI\ for a nice discussion of
this issue). The high temperature `phase' in flat space seems
likely to be ill defined; in this phase energy densities are
$\CO({1\over g^2})$ and would (since the Jeans instability is triggered at
these energy densities) appear to involve black holes. As the density of
states of Schwarzschild black holes grows even faster than exponential,
the existence of such a phase seems problematic.

The picture is much clearer for type IIB string theory on $AdS_5 \times S^5$.
At small curvatures, the spectrum of single string states in this
theory exhibits a Hagedorn growth, with Hagedorn temperature
$T_H={1\over 2 \pi \sqrt{2} \apm}$ and $d_{eff}=1$ (see \S7.2). However, this
theory (with temperature conjugate to global time) undergoes a first order
phase transition\foot{As the space is effectively
compact, the
phase transition is sharp
only in the limit $g_s \to 0$.} well below its Hagedorn temperature
\refs{\HawkingDH,\WittenZW}. As in flat space, energy densities
in the high temperature phase are $\CO({1\over g^2})$, and the
high temperature phase is dominated by a big black hole sitting at the
center of $AdS$ space. In Euclidean space this black hole may indeed be
thought of as a condensate of winding modes\foot{Recall that the Euclidean
time cycle is contractible in any black hole. Thus the `time
winding number' symmetry, present in global $AdS$, has been spontaneously
broken in the black hole phase, implying that the formation of the black
hole must involve the condensation of winding modes.}, so the
Atick-Witten order parameter analysis seems to apply, at
least qualitatively, to
this situation.\foot{Note, however, that it is not possible to use this
analysis to predict (in this case) the order of the transition, as the
Atick-Witten analysis, applied to string theories with a mass gap, could
be consistent with either a first order or a second order phase transition,
depending on the coupling constants of the theory (see appendix C).}
Note that the entropy of big black holes grows relatively slowly with energy
in $AdS$ space ($S \propto E^{{3\over 4}}$), and thermodynamics is well
defined at all temperatures.

In this paper we demonstrate that the thermodynamics of weakly coupled large
$N$, $SU(N)$ gauge theories at finite volume (for instance gauge theories
on a sphere) has some striking similarities to string thermodynamics on
the weakly curved $AdS$ space described above. The theories we study all
undergo a phase transition as a function of temperature. The low
temperature phase has a Hagedorn growth in the density of states with
$d_{eff}=1$. Finally, the high temperature phase in Euclidean space may
be thought of as a condensate of winding modes.

\subsec{Deconfinement transitions in gauge theories}

In this paper we will study the thermodynamics of gauge theories on compact
manifolds of size $R$. On taking $R$ to infinity our theories reduce to
gauge theories on flat space, whose thermodynamics has been
studied extensively (see \grossreview\ for a review).
In this subsection we review the thermal behavior of
gauge theories on  $\IR^{d-1,1}$. We start with a discussion of confining
gauge theories, and at the end of this section we briefly mention the
situation with the $d=4$ $\CN=4$ supersymmetric Yang-Mills theory.

It is widely believed that pure $SU(N)$ gauge theories in $d=4$ confine
at zero temperature (see \thooftunder\ and \Polrev\ for reviews);
the low energy spectrum of the theory may be thought
of as a Fock space of interacting glueballs.\foot{Similar expectations hold
for gauge theories in lower dimensions, and for four dimensional gauge
theories with a sufficiently small amount of charged matter fields; for
QCD, the $SU(3)$ gauge theory describing the strong interactions,
they are experimentally verified. For simplicity of terminology we will
restrict the discussion of this subsection to pure gauge theories in $d=4$.}
Consequently, it is expected that in such theories
the low temperature ($T \ll \Lambda_{QCD}$) dynamics
may be understood in terms of a sparsely populated thermal bath of
glueballs.
On the other hand, asymptotic freedom permits a reliable
computation of the $T \gg \Lambda_{QCD}$ behavior of such theories\foot{In
infinite volume this statement is far from trivial due to infrared
divergences which affect finite-temperature perturbative computations. These
problems do not appear at finite volume so they will not be relevant for us.},
revealing that this high temperature phase may be understood
as a weakly coupled gas of gluons \refs{\CollinsKY,\PolyakovVU}.
The high and low temperature phases appear qualitatively different; we will
now 
review the argument showing
that they are distinguished by an
order parameter, and so are
separated by a phase transition at some intermediate temperature.

It is a defining property of a confining theory (with no fields in the
fundamental representation of $SU(N)$) that a single external particle in
the fundamental representation (a ``quark'') can be inserted
into this theory only at an infinite cost in energy. Heuristically,
such a quark forms one end of a QCD string which is infinitely long
because it has nowhere else to end. In the low temperature confining phase
this long string also has infinite positive free energy. Let $F_q(T)$
represent the Yang-Mills free energy at temperature $T$ in the presence of
an external quark. It follows from the form of the coupling of an external
quark to the gauge fields that $e^{-F_q(T)/T}=\langle {\cal P}
\rangle$, where ${\cal P}={1 \over N} \tr P\exp[ -\oint A  ]$ is the so
called Polyakov loop (sometimes called the Polyakov-Susskind loop
\refs{\PolyakovVU,\SusskindUP}), the trace of
a Wilson loop around the compactified Euclidean thermal time circle.
Thus, $\langle {\cal P} \rangle=0$ in the low temperature confining phase.

On the other hand, asymptotic freedom permits reliable computations
that establish $\langle{\cal P} \rangle  \neq 0$ at high enough temperatures.
Thus, $\langle {\cal P} \rangle$ constitutes an order parameter that sharply
distinguishes the low temperature confining phase from the high temperature
deconfined phase. From a low temperature point of view, the transition
that separates these two phases is associated with the condensation of
flux tubes whose effective free energy is driven negative at high enough
temperatures (when the energy of these strings is overcome by the entropy of
their vibrations).

In the large $N$ limit (with fixed 't Hooft coupling, or equivalently
fixed $\Lambda_{QCD}$) the deconfinement
phase transition has yet another order parameter \ThornIV . The confined phase is
dominated by gauge-invariant bound states and so its free energy $F(T)$
scales like $N^0$ at large $N$. On the other hand, the
deconfined phase is described by free gluons, and consequently
its free energy scales as $N^2$ at large $N$. Thus, in the large $N$
limit, $\lim_{N\to \infty} F(T)/N^2$ constitutes a second order parameter
for deconfinement; like the Polyakov loop this new order parameter vanishes in
the confined phase but is finite in the deconfined phase.

Next, we turn to a discussion of the order of the deconfinement phase
transition. To study the phase transition
one may generate an effective action for the
Polyakov loop $\cal P$ by integrating all other fields out of the path
integral that generates the Yang-Mills free energy.
$\cal P$
is a complex scalar field on $\IR^3$; note also that ${\cal P}
\rightarrow e^{{2 \pi i \over N}} {\cal P} $
is a symmetry of the $SU(N)$ Yang-Mills path integral \tHooftHY\
(it is generated by gauge
transformations that are single valued only up to an element of the
center of the gauge group $\IZ_N$; such gauge transformations act trivially
on the dynamical fields of the pure Yang-Mills theory),
which is spontaneously broken in
the deconfinement phase transition.

The effective action for ${\cal P}$ completely determines the nature and
properties of the deconfinement transition. If this transition turns out to
be first order then its properties depend on the details of the theory
under consideration. However, if the deconfinement transition is of
second order then the phase transition point has universal behavior; it
must be described by a $\IZ_N$-invariant fixed point of a complex scalar
field on $\IR^3$ \refs{\SvetitskyGS,\YaffeQF}. Only a small number of fixed
points with the required properties are known. For $N=2$ the conditions above
uniquely pick out the Wilson Fischer fixed point (the $d=3$ Ising model).
For $N=3$
no attractive fixed point with the symmetries listed above
is known. For $N \geq 4$ the only known fixed point is
$U(1)$ invariant (all operators of the form $\ph^{k N}$ for integer $k$ and
$N \geq 4$ are irrelevant at this fixed point, hence the enhanced
symmetry).

The arguments of the paragraph above apply to all confining
$SU(N)$ gauge theories with adjoint matter.  In the rest of this
subsection, however,  we will describe what is known about the
deconfinement transition of the pure gauge theory on $\IR^{3,1}$.
Since this phase transition is a
strong coupling phenomenon, we cannot study it in perturbation
theory, and most knowledge about it comes from lattice simulations.
Lattice simulations indicate that the $SU(2)$ deconfinement
phase transition is second order and is indeed in the Ising model
universality class. The $SU(3)$ deconfinement phase
transition is first order, in agreement with the predictions above. However,
the $SU(N)$ deconfinement phase transitions for $N \geq 4$
(at least for $N=4, 5, 6$) also appear to be first order;
in particular the $U(1)$ invariant ($d=3$ XY fixed point) appears not to
be attained in pure Yang-Mills\foot{However, it is entirely possible
that this beautiful fixed point describes the phase transition of some
other confining gauge theory -- say, for instance,
$\CN=1$ supersymmetric Yang-Mills theory. The critical exponents of this
fixed point imply that the singular piece of the free Energy scales like
$F\propto (\d T)^{2.08}$ (where $\d T =T-T_c$) upon approaching the phase
transition, implying $d_{eff} \approx 5.16$, see \pffb. It would be
fascinating to find a stringy interpretation or confirmation of
this result.} (see, for instance, \LuciniZR, but see also \Pisa).

As the final topic in this subsection we review the thermodynamics of the
$\CN=4$ SYM theory in flat space. We focus on the limit
$N \to \infty$ with $\lambda =g^2_{YM} N$ held fixed. As this theory is
conformal (and from large $N$ counting) the free energy density $F$ must
take the form $F=-{\pi^2 \over 6} N^2 f(\lambda) T^4$.
Perturbative computations \FotopoulosES\ establish that, at small
$\lambda$, $f(\lambda)=1-{3 \lambda \over 2 \pi^2 }$, while computations
using the AdS/CFT
correspondence \GubserNZ\ show that, at large $\lambda$,
$f(\lambda)={3\over 4}+{45\over 32}
{\zeta(3)\over (2 \lambda)^{{3\over 2}}}$. It is
generally assumed 
(see, e.g., \KimSG)
that $f(\lambda)$ is analytic on the positive real
line (i.e. that one encounters no phase transitions in extrapolating the
$\CN=4$ SYM coupling from zero to infinity; more about this in
\S7). In summary,
the thermodynamics of the $\CN=4$ SYM theory in flat
space is qualitatively rather boring. The same theory has much more
interesting thermodynamics on a sphere as was pointed out in 
\refs{\WittenZW,  \SundborgUE}; we will discuss this issue in great
detail below.

\subsec{Hagedorn versus deconfinement}

't Hooft has argued \thooft\ that it should be possible to recast any
gauge theory, in the limit of large $N$ with fixed $g_{YM}^2 N$,
as a string theory with string coupling $g_s \propto 1/N$. If we accept
this conjecture, then it is reasonable to expect that the gauge theory deconfinement transition has
a dual description as a stringy thermal transition of the sort
described in \S2.1\foot{As we have commented in \S2.1, this
transition may occur below the Hagedorn temperature.}.
Indeed, stringy and gauge thermodynamics have many points
of similarity \refs{\ThornIV,\PisarskiDB,\AtickSI};
in each case the low temperature phase is a gas of
weakly interacting thermally populated stringy particles whose free
energy is $\CO(1)$ in the relevant coupling constants (${1\over N}$ or
$g_s$). The phase transition is driven by a condensation of these
particles, and is marked by an order parameter that winds around the
thermal time circle (Polyakov loop or Euclidean winding string).  The
high temperature phase has a free energy that scales like the inverse
square of the relevant coupling.

If we accept that stringy thermodynamics can be understood in terms of
 deconfinement, it follows that the relation $\rho(E) \sim \exp(E / T_H)$
must break down in the energy range  which is relevant for the high temperature
deconfined phase. We can explain this
breakdown as follows. In the high temperature phase of asymptotically
 free gauge theories, the density of
states is approximately that of a $3+1$ dimensional free field theory
with $N^2$ degrees of freedom on a space of volume $V$,
$\rho(E) \sim \exp(E^{3/4} N^{1/2}
V^{1/4})$. A crude estimate of the energy at the point of transition
 between the two phases
may be obtained by equating these two formulas for $\rho(E)$, yielding
$E \sim N^2 V T_H^4$. Note that this energy is proportional to
$N^2 \sim 1/g_s^2$. Gravitational effects are large at these
energies for every $g_s$, no matter how small, explaining the failure of the
free string estimate.

So far we lack a quantitative demonstration that large $N$ gauge
thermodynamics
exhibits sharp stringy features\foot{A notable exception is the case of 
two dimensional gauge theories,
which one can explicitly solve in the large $N$ limit in many cases
(see, for instance, \refs{\td,\KutasovGQ}). 
In this limit one obtains a Hagedorn-like spectrum with Regge trajectories.
Moreover, there is also a stringy description of two dimensional QCD 
\tdstring.}.
The deconfinement transition in flat space occurs at strong coupling
and has so far resisted efforts at quantitative analysis. In this paper we
will make progress on this important problem by analyzing confining gauge
theories on compact spaces\foot{The AdS/CFT correspondence, which
establishes a duality between a gauge theory on a sphere and a string
theory, motivates this study.} of size $R$ rather than flat space, taking
$R \Lambda_{QCD} \ll 1$. The thermal behavior of such a theory may be
computed reliably at all values of the temperature; we will demonstrate
that it displays stringy features.

\subsec{Large $N$ deconfinement on compact spaces}

As we have argued at the end of the previous subsection, there are several
good motivations to study the thermodynamics of Yang-Mills theories on
compact manifolds. The AdS/CFT correspondence forces us to address this
problem. In a confining theory the finite size of the manifold
cuts off the running of the
coupling at the scale $1/R$, introducing a dimensionless
coupling constant into
the problem. 

In the case of a conformal theory like the $\CN=4$ SYM theory
the new scale permits nontrivial temperature dependence of
thermodynamic quantities. In this subsection we will take a first look at
the thermal behavior of gauge theories on such spaces.

Clearly, the thermodynamical
properties of an $SU(N)$ gauge theory on a compact space of size $R
\gg 1 / \Lambda_{QCD}$ closely resemble those of the theory
in flat space. This resemblance cannot be perfect at finite $N$, however, as
the transition between confined and deconfined behavior
must be smooth rather than sharp for any system with a
finite number of degrees of freedom. In the
$N \to \infty$ limit, on the other hand, the number of degrees
of freedom is infinite even at finite volume, and the deconfinement
transition mimics its flat space counterpart more closely;
in particular it remains sharp.

Note that the Polyakov loop is no longer a good order parameter in the
finite volume theory because its expectation value vanishes for
kinematical reasons (Gauss' law makes it impossible to put a single
fundamental quark on a compact manifold,
independently of the phase the theory is
in). However, as we will explain in \S5.7, it is possible to define related
order parameters, either by looking at the norm of the Polyakov loop
or by introducing an infinitesimal amount of fundamental
matter. Alternately, one can use the second order parameter discussed
in \S2.2 above; the high and low temperature phases are
sharply distinguished by the fact that the free energy scales as
$\CO(N^2)$ and $\CO(1)$, respectively, in these phases.
It is still reasonable to write the effective Landau-Ginzburg free
energy for this theory as
the theory of a complex scalar field ($\cal P$) on the
compact space,
and we will use a variant of this description in our analysis later
in this paper.

We have argued above that a large $N$  confining gauge theory,
compactified on a space of size $R \gg 1/\Lambda_{QCD}$, undergoes
a phase transition at
a temperature of order $\Lambda_{QCD}$. In this paper we will study the
opposite limit $ R \Lambda_{QCD} \ll 1$. The gauge theory is weakly coupled
in this limit (recall that the gauge coupling stops running at scale
${1\over R}$ on a compact space),
and the thermodynamics may reliably be computed
at all values of the temperature. We will find that the phase transition
persists in this limit at a temperature of order
$\CO({1\over R})$. In the case of the $d=4$ $\CN=4$ SYM theory on $S^3$,
the AdS/CFT correspondence has established \refs{\WittenZW, \HawkingDH}
that, at strong coupling $\lambda$, the theory undergoes a phase transition
at a temperature of $\CO({1\over R})$. We will establish that this phase
transition continues to occur at weak coupling $\lambda$ at a
temperature of the same order.

The weak coupling nature of the phase transitions we
describe in this paper
(at $R \Lambda_{QCD} \ll 1$ for confining theories, and at $\lambda \ll 1$ for
the $\CN=4$ SYM theory) allows us to analyze them in detail. Quite remarkably
we will find sharp signatures of stringy thermodynamics in these theories.
In particular, we will demonstrate that the low temperature phase is always
characterized by a Hagedorn growth in the density of states
(see also \refs{\SundborgUE,\PolyakovAF}).
We view this result as evidence for
the existence of string duals for weakly coupled gauge theories on
compact manifolds.

\newsec{Partition function for free Yang-Mills theory on a compact space}

We now proceed to directly analyze the thermodynamics of weakly
coupled Yang-Mills theory on a compact space, beginning in this
section (and the next two)
with the free theory. For simplicity, we restrict here to
spaces (such as $S^{d-1}$) for which all modes of the various fields are
massive. In most of this section we will discuss the specific case
of a $U(N)$ gauge theory with fields in the adjoint representation,
but it is easy to generalize our arguments to more general theories. 
In particular,
our exact results for the partition function in \S3.3 will
be completely general.
As usual, the behavior of thermodynamic quantities in the
canonical ensemble is governed by the partition function
\eqn\defpart{
Z(\beta) = \sum_{physical \; states} x^{E_i} = \int \rho(E) x^E dE,
}
where $\rho(E)$ is the density of states and we define
$x\equiv e^{-\beta}=e^{-1/T}$. As we discussed earlier, it is essential to
keep in mind that the constraint from Gauss' law for a gauge theory on a
compact space implies that the physical states over which we sum must
be gauge-invariant.

\subsec{Two matrix harmonic oscillators give Hagedorn behavior for large $N$}

By expanding the fields into modes on the compact space
(e.g. spherical harmonics on $S^3$), our field theory may be viewed as
a quantum mechanical system with infinitely many degrees of freedom
(one for each mode). In the free theory, these degrees of freedom are
decoupled harmonic oscillators, each in some representation of the
gauge group. Before embarking on our detailed analysis of this theory,
we present a simple demonstration that even restricting to two such
oscillators in the adjoint representation of $U(N)$ yields a theory
that displays Hagedorn-like behavior in the large $N$ limit\foot{
Hagedorn-like behavior in free large $N$ systems was also observed
in \HalpernFZ.}.

Thus, we consider two such modes, each with unit energy, created by
operators $A^\dagger$ and $B^\dagger$ in the adjoint of
$U(N)$. Physical states are gauge-invariant and correspond to traces
of products of these operators acting on the Fock space
vacuum. Single-trace states of energy $E$ are specified by a series of
$E$ $A^\dagger$'s and $B^\dagger$'s inside the trace. The number of such
states satisfies
\eqn\naivebnd{
2^E/E < n(E) \leq 2^E,
}
where the upper bound arises since for each of the $E$ positions in the
trace we may choose $A^\dagger$ or $B^\dagger$, and the lower bound
comes because cyclicity of the trace equates a given state to at most
$E-1$ others. In the large $N$ theory all of these states are independent, so
the density of independent single-trace states is Hagedorn-like :
\eqn\naiveden{
\rho(E) = {{\partial n} \over {\partial E}} \sim E^\alpha e^{\beta_H E},
}
where the Hagedorn temperature in this case is $T_H = 1/\beta_H = 1/\ln(2)$.

For finite $N$, we cannot have Hagedorn behavior at all energies,
since the high temperature behavior must be that of a field theory
(or quantum mechanics in the toy example described above)
with finitely many fields. The departure from Hagedorn behavior at
high energies comes about because the oscillators have only finitely
many gauge-invariant degrees of freedom (of order $N^2$), so that only
a number of traces of order $N^2$ give independent states\foot{For
example, for a single $N\times N$ matrix, $\tr(A^{N+1})$ may be expressed
as a combination of products of traces of lower powers of $A$.}.
Thus, at energies of order
$N^2$, trace relations will cut off the exponential growth of states,
resulting in a cross-over to field theory (or quantum mechanics)
behavior. In the limit of
large $N$, this cross-over becomes a sharp transition, as we will see
explicitly in the next sections.

\subsec{Exact partition function for $N=\infty$}

We now show that the exact partition function of our free gauge theory
in the strict $N=\infty$ limit (where no trace relations exist) may be
obtained by simple counting arguments, following
\refs{\SundborgUE,\PolyakovAF}.

We consider a model with $m \geq 2$ matrix-valued bosonic harmonic
oscillators; the number $m$ can be finite or infinite, as in the example of a
compactified $d$-dimensional gauge theory. Let the $i^{th}$
oscillator have energy $E_i$. We encode the spectrum of oscillators in
a ``single-particle partition function'' $z(x) \equiv \sum_{i}
x^{E_i}$ where the sum goes over all oscillators\foot{To be precise,
this is the single-particle partition function for the $U(1)$ theory,
since in general the single oscillator states are non-physical.}. Examples
of such single-particle partition functions for compactified $3+1$
dimensional theories will be provided in \S3.5 below.

Following the discussion above, the partition function of single-trace
states with $k$ oscillators is given by
\eqn\Zir{ Z_{k}={z(x)^k \over k} + (positive)}
where the factor of $1/k$ compensates for cyclicity in the trace, and
the $+ (positive)$ acknowledges that this is an over-compensation in cases
where the trace breaks up into repeated sequences of
oscillators. Since in the large $N$ limit a single-trace state can
have any number of oscillators, we need to sum this result over $k$ to
find a partition function for single-trace states,
\eqn\Zirr{Z_{ST}=\sum_{k=1}^{\infty} Z_k = -\ln (1-z(x)) + (positive).}
Actually, it is not too difficult to promote \Zirr\ to an exact
expression which correctly accounts for repetitions inside the
trace. This careful counting utilizes Polya theory and was already performed
in \refs{\SundborgUE,\PolyakovAF}. For the convenience of
the reader we reproduce it in appendix B. The exact result in the
large $N$ limit is
\eqn\sumex{Z_{ST} = -\sum_{q=1}^{\infty} {\ph(q) \over q} \ln( 1-z(x^q)),}
where $\ph(q)$ represents the number of positive
integers which are not larger than $q$ and are relatively prime to $q$.
Note that the first term in \sumex\ is precisely the term we explicitly
wrote down in \Zirr.

Using \sumex, we may now write the full partition function of
our model, summing over states with arbitrarily many traces. The space
of multi-trace states is simply the Fock space of single-trace states,
so the full multi-trace partition function $Z$ is easily obtained from
$Z_{ST}$ (taking into account the fact that the single-trace states
behave as identical bosonic particles),
\eqn\multipart{\eqalign{\ln(Z) &= \sum_{n=1}^{\infty}
{1\over n} Z_{ST}(x^n) = \cr
&= -\sum_{n=1}^{\infty} {1\over n} \sum_{q=1}^{\infty}
{\ph(q)\over q} \ln(1 - z(x^{qn})) = \cr
&= -\sum_{k=1}^{\infty} {1\over k} (\sum_{q | k} \ph(q)) \ln(1 - z(x^k)) = \cr
&= -\sum_{k=1}^{\infty} \ln(1 - z(x^k)), \cr}}
where we have used the formula $\sum_{q|k} \ph(q) = k$.
It is easy to generalize this result to include also fermionic oscillators;
denoting by $z_B$ the single-particle partition function
of the bosonic oscillators,
and by $z_F$ the single-particle partition function
of the fermionic oscillators, we
obtain
\eqn\bfmultipart{\ln(Z) = -\sum_{k=1}^{\infty}
\ln(1 - z_B(x^k) + (-1)^k z_F(x^k)).}
The inclusion of fermions does not change the qualitative behavior, so
it is sufficient to assume that all oscillators are bosonic for the purposes of qualitative
discussions.

Equation \bfmultipart\ gives the exact partition function in the
strict $N=\infty$ theory. It is not correct for the finite $N$ theory
because we have assumed all traces to be independent.\foot{When the
number of fields in a trace is of order $N^2$ trace identities relate various
states that are assumed to be different in \multipart. For example,
any $2\times 2$ Hermitian matrix $M$ obeys $\tr(M^3)= \half(
3\tr(M) \tr(M^2) - \tr(M)^3)$.}
However, as long as the thermodynamics is dominated by
states for which the number of oscillators in a trace is small
compared to $N^2$ (which will certainly be true for low enough
temperatures), trace relations will be unimportant and \bfmultipart\
should give a very good approximation to the correct behavior.

In order to understand the dynamics of \multipart\ we note some obvious
properties of the ``single-particle partition function'' $z(x)$, namely that
$z(0)=0$, $z(x)$ is a monotonically increasing
function of $x$, and that (since $m \geq 2$) $z(1) > 1$.
It follows that the equation $z(x)=1$ has a unique
solution (which we will denote by $x=x_H$) for $0<x<1$.
If we denote the corresponding inverse temperature by $\beta_H$
 ($e^{-\beta_H}=x_H$), then $\ln(Z(\beta))$ is well defined for
 $\beta >\beta_H$, but diverges like $-\ln (\beta-
\beta_H)$ as $\beta $ tends to $\beta_H$ from above.
For $\beta< \beta_H$ (high temperatures) $Z(\beta)$ in \multipart\
is ill defined. All these features are
reproduced by the formula
\foot{Let
\eqn\rept{\ln(Z)=\int dE \rho'(E) e^{-\beta E}}
and
\eqn\repth{ Z_{ST}= \int dE \rho_{ST}(E)e^{-\beta E}.}
Then, $\rho'(E) \sim  \rho_{ST}(E) \sim e^{\beta_H E}/E$ at high energies.
Note that $\rept$ and \repth\ match \pf\ and \denpart, respectively, upon
setting $d=d_{eff}=1$ in those formulas.}
\eqn\rep{Z(\beta)=\int dE \rho(E) e^{-\beta E},}
where $\rho(E) \approx e^{\beta_H E}$ at high
energies. Consequently, our system of oscillators has Hagedorn-like
thermodynamics with the inverse Hagedorn temperature $\beta_H$. Note that
$z(x)=2x$ for the two oscillator model of the previous subsection, yielding
$x_H=1/2$ and
$\beta_H=\ln(2)$ in agreement with the discussion in that subsection.

The fact that $Z(\beta)$ diverges for $T > T_H = \beta_H^{-1}$
indicates that $T_H$ is a limiting temperature if $N$ is set strictly equal
to infinity. Of course, this cannot be correct at any finite $N$, no matter
how large. Indeed, a system of $m N^2$ harmonic oscillators
(or even a compactified field theory for which $m$ is infinite but there
is a finite number of oscillators below any fixed energy) can certainly
be heated to any temperature.  It must instead be the case that at
high temperatures ($T > T_H$), the condition for the validity of
\bfmultipart\ (namely, that the number of fields contained
in the typical state is much smaller than $N^2$) fails;
we will see in \S5 that this is indeed the
case. Thus, in the high temperature phase, trace relations invalidate
the analysis above, and traces no longer provide a useful basis for
$U(N)$ singlets in the high temperature phase. This is true at any finite
$N$ and so it is also true in the large $N$ limit of principal interest to us.
In the next subsection we therefore turn to a more generally
applicable method of dealing with the $U(N)$ singlet condition.

\subsec{Exact partition function for free Yang-Mills theory}

In this section, we will derive an exact expression for the partition
function of free Yang-Mills theory with arbitrary gauge group and
matter content, on
any compact space for which all modes of the various fields are
massive. As we have discussed, the basic excitations of such a theory
are single-particle states, corresponding to oscillators which arise
by expanding the various fields into physical modes on the space.  The
general physical state is obtained by acting with arbitrary
collections of these oscillators on the Fock space vacuum, with the
constraint that the total state should be a singlet of the gauge
group.

To obtain an explicit expression for the partition function, suppose
we have bosonic modes with energies $E_i$ in representations $R_i$ of
the gauge group and fermionic modes with energies $E'_i$ in
representations $R'_i$ of the gauge group.
\foot{Prior to this subsection we have
assumed that all modes are in the adjoint representation.
We relax that assumption in this subsection.}
Then, the partition function
may be expressed as a sum over the occupation numbers of all modes,
with a Boltzmann factor $e^{-\beta E}$ corresponding to the total energy, and a
numerical factor counting the number of singlets (physically allowed
states) in the corresponding product of representations,
\eqn\partone{\eqalign{
Z(x) =& \sum_{n_1=0}^{\infty} x^{n_1 E_1} \sum_{n_2=0}^{\infty}
x^{n_2 E_2} \cdots
\sum_{n'_1=0}^{\infty} x^{n'_1 E'_1} \sum_{n'_2=0}^{\infty}
x^{n'_2 E'_2} \cdots \times \cr & \{
{\rm \# \; of \; singlets \; in \;} sym^{n_1}(R_1)\otimes
sym^{n_2}(R_2) \otimes \cdots \otimes anti^{n'_1}(R'_1) \otimes
anti^{n'_2}(R'_2) \otimes \cdots \}. }}
Note that in order to correctly account
for particle statistics, we must symmetrize (antisymmetrize)
representations corresponding to identical bosonic (fermionic) modes. An
explicit expression for the group theory factor may be obtained by using
the fact that the number of singlets in a product of representations
is simply the integral over the group manifold of the product of
characters for these representations, as reviewed in appendix A. Thus,
we find
\eqn\parttwo{\eqalign{
Z(x) =& \int [dU] \prod_i \left\{ \sum_{n_i=0}^{\infty} x^{n_i E_i}
\chi_{sym^{n_i}(R_i)}(U) \right\} \prod_i \left\{ \sum_{n'_i=0}^{\infty}
x^{n'_i E'_i} \chi_{anti^{n'_i}(R'_i)}(U) \right\}. }}
The sums in brackets are
generating functions for the characters of symmetrized
(antisymmetrized) products of arbitrarily many copies of a given
representation $R_i$ ($R'_i$). As we show in appendix A, these
turn out to have simple expressions given by equation (A.8). Using this
equation, we find that
\eqn\partthree{\eqalign{
Z(x) =& \int [dU] \exp \left\{ \sum_{i;m=1}^{\infty} {1\over m} x^{m E_i}
\chi_{R_i}(U^m) \right\} \exp \left\{ \sum_{i;m=1}^{\infty}
{(-1)^{m+1} \over m} x^{m E'_i} \chi_{R'_i}(U^m)\right\}. }}
Finally, since the sums over
$i$ in the exponentials are simply sums over the single-particle
modes, it is convenient to define bosonic and fermionic
single-particle partition functions for each representation $R$,
counting the single-particle states in this representation without
the degeneracy coming from the dimension of the representation and
without any gauge-invariance constraints,
\eqn\singpart{
z_B^R(x) = \sum_{R_i = R} x^{E_i}, \qquad z_F^R(x) = \sum_{R'_i =
R} x^{E'_i}.}
With these definitions, we may express our final result
for the partition function as \SundborgUE\
\eqn\partfour{
Z(x) = \int [dU] \exp \left\{ \sum_R \sum_{m=1}^{\infty}
{1 \over m} \left[z_B^R(x^m) +
(-1)^{m+1} z_F^R(x^m) \right] \chi_{R}(U^m) \right\}. }
Thus, the partition
function for an arbitrary free field theory (with no zero modes) on a
compact space may be expressed as a single-matrix integral, with the action
determined in a simple way by the single-particle modes of the theory.

\subsec{Generalization to arbitrary chemical potential}

Before considering specific gauge theories, we note that the result
\partfour\ may be easily generalized to the case of a non-zero chemical
potential. For a set of commuting charges $Q^a$ which commute with the
Hamiltonian, we may introduce chemical potentials $\mu_a \equiv
\ln(q_a)$ simply by replacing $x^{E_i} \to x^{E_i} \prod_a q_a^{Q^a_i}$
in \partone, where $Q_i^a$ is the $a^{th}$ charge of the $i^{th}$ mode.
The resulting partition function is
\eqn\partfive{\eqalign{
Z(x,\{q_a\}) =& \int [dU] \exp \left\{ \sum_R \sum_{m=1}^{\infty} {1 \over m}
\left[z_B^R(x^m,\{q_a^m\}) + (-1)^{m+1} z_F^R(x^m,\{ q_a^m \}) \right]
\chi_{R}(U^m) \right\}, }}
where the generalized single-particle partition functions are now given by
\eqn\singpart{
z^R(x,\{q_a\}) = \sum_{R_i = R} x^{E_i} \prod_a q_a^{Q^a_i}.
}

\subsec{$U(N)$ gauge theories with adjoint matter on $S^3 \times \IR$}

For most of the remainder of this paper, we will focus on $U(N)$ gauge
theories on a sphere with adjoint matter. In this case, using
$\chi_{adj}(U) = \tr(U) \tr(U^{\dagger})$ (where we denote by $\tr$
without a subscript the trace in the fundamental representation),
we obtain the unitary matrix model
\eqn\partsix{\eqalign{
Z(x) =& \int [dU] \exp \left\{ \sum_{m=1}^{\infty} {1 \over m} (z_B(x^m) +
(-1)^{m+1} z_F(x^m)) \tr(U^m) \tr((U^\dagger)^m) \right\}. }}
It is straightforward to work out explicit expressions for the single
particle partition functions for various types of fields on a sphere
of unit radius.
In appendix B (see also \refs{\SundborgUE,\PolyakovAF}) we find that for
$(3+1)$-dimensional scalars, vectors, and chiral
fermions on $S^3 \times \IR$,
\eqn\sthreefields{
z_{S4}(x) = {x+ x^2 \over (1-x)^3}, \qquad z_{V4}(x) = {6x^2 - 2x^3
\over (1-x)^3}, \qquad z_{F4}(x) = {4 x^{3 \over 2} \over (1-x)^3}.
}
In a $3+1$ dimensional free $U(N)$ gauge theory with $n_S$ scalar fields,
$n_V$ vector fields and $n_F$
chiral fermions (all in the adjoint representation), we should use
in \partsix\ $z_B(x) = n_S z_{S4}(x) + n_V z_{V4}(x)$, $z_F(x) =
n_F z_{F4}(x)$.
Note that all of these functions (and the single-particle partition
functions in general) increase monotonically from $z=0$ at $x=0$ (zero
temperature), and in all dimensions above $0+1$ they
diverge at $x=1$ (infinite temperature)\foot{For a $d$
dimensional field theory compactified on $S^{d-1}$, $z(x)$ diverges as
$2{\cal N}^{dof}/(1-x)^{d-1}$, where ${\cal N}^{dof}$
is the number of single-particle
degrees of freedom.}.

For $SU(N)$ gauge theories, the only difference (beyond the fact that $U$
must have determinant one) is that every
$\tr(U)\tr(U^{\dagger})$ in \partsix\ should be replaced everywhere by
$(\tr(U)\tr(U^{\dagger})-1)$, the character of the adjoint
representation in $SU(N)$.

\newsec{Path integral derivation of the matrix integral and an
order parameter}

In the previous section we computed the partition function of free
Yang-Mills theory on a compact space $\cal M$ by counting gauge-invariant
states. However, $Z(T)$ may also be computed by the vacuum path
integral of Euclidean Yang-Mills theory\foot{We thank Nima Arkani Hamed for
emphasizing the utility of this approach to us, and for several extremely
enjoyable and productive conversations on related issues.}
on ${\cal M} \times S_T^1$ where the circumference of the thermal
circle $S_T^1$ is $\beta=1/T$.
In this section we will rederive the expression \partsix\ for the
partition function by computing this path integral for gauge theories on
an $S^3$ of unit radius.

Though it gives an identical final result for the partition function,
the path integral derivation provides a physical interpretation for
the unitary matrix $U$ as the Wilson loop of the gauge field (averaged
over the compact space) around
the thermal circle. Consequently, ${1\over
N}\tr(U)$ is precisely the Polyakov loop operator described in \S2.2;
a standard order parameter for deconfinement. We will see in the next
section that ${1\over N}\tr(U)$ is also the natural order parameter
for a large $N$ phase transition of the matrix model \partsix.

A further advantage of the path integral derivation is that it
generalizes easily to the calculation of the large $N$ partition
function $Z(T, \lambda)$ at weak 't Hooft coupling $\lambda \equiv
g_{YM}^2 N$ when the theory is no longer free. In \S6.2 we analyze
 the general structure of $Z(T, \lambda)$ at small $\lambda$. An explicit
calculation of the weak-coupling partition function, in specific gauge
theories,  is postponed to a future paper \future.

\subsec{Basic set-up}

We work in the gauge
\eqn\gf{\del_i A^i=0,}
where $i=1,2,3$ runs over the sphere coordinates,
and $\del_i$ are covariant
derivatives. The choice \gf\ fixes the gauge freedom only partially; 
it leaves spatially
independent but time dependent gauge transformations unfixed. We fix
this residual gauge-invariance with the condition
\eqn\gff{\p_t \alpha(t)=0,}
where
\eqn\defalpha{\alpha \equiv {1\over \omega_3} \int_{S^3} A_0,}
where $\omega_3$ is the volume of $S^3$.

The mode $\alpha$ will play a special role in what follows because it
is the only zero mode (mode whose action vanishes at quadratic
order) in the decomposition of Yang-Mills theory into Kaluza-Klein
modes on $S^3 \times S^1$. Consequently, $\alpha$ fluctuations are
strongly coupled at every value of $\lambda$, including in the limit
$\lambda \to 0$. In particular, they cannot directly be integrated out in
perturbation theory.

In order to proceed with the perturbative evaluation
of the partition function we will adopt a two step procedure. In the
first step, discussed in this section, we integrate out all
non-zero modes\foot{As $\alpha$ is a zero mode this is a standard Wilsonian
procedure. Note the parallel with the discussions in \S2.2 and in
appendix C.}
and generate an effective action $S_{eff}(\alpha)$ for
$\alpha$. $S_{eff}(\alpha)$ is non-trivial even at zero coupling and, as we
will discuss in \S6, it is
further modified perturbatively in $\lambda$.
Once we have obtained $S_{eff}(\alpha)$, we must then proceed to perform the
integral over $\alpha$ in order to obtain $Z(T)$. This is the subject
matter of \S5.

\subsec{The integration measure}

In this subsection we will define $S_{eff}(\alpha)$ more carefully.
Recall that the Yang-Mills free energy may be written as
\eqn\intaft{e^{-\beta F}= \int d \alpha \int \CD A
\Delta_1  \Delta_2  e^{-S_{YM}(A, \alpha)},}
where $\Delta_1$ is the Fadeev Popov determinant conjugate to \gf,
$\Delta_2$ is the Fadeev Popov determinant conjugate to \gff, and
$S_{YM}$ is the Yang-Mills action. It is not difficult to explicitly
evaluate $\Delta_2$ (see later in this subsection) and verify that
it is independent of $A$. Consequently, \intaft\ may be rewritten as
\eqn\intafto{e^{-\beta F}=\int d\alpha \Delta_2
\exp[-S_{eff}(\alpha)], }
where
\eqn\intaftt{\exp[-S_{eff}(\alpha)]=
\int \CD A \Delta_1 \exp[-S_{YM}(A, \alpha)].}
In the rest of this subsection, we will explicitly evaluate
$\Delta_2$ and so determine the effective measure of integration \intafto.

It follows from \gff\ that
\eqn\fpdet{\Delta_{2 }=\det\,'\left(\partial_0-i\left[\alpha,\ast\right]
\right),}
where the prime asserts that the determinant is over non-zero modes of
$A_0$. Denoting by $\lambda_i$ ($i = 1,\cdots,n$) the eigenvalues of
$\alpha$, and choosing a convenient basis of matrix functions whose
time-dependence is given by $\exp(2\pi i n t / \beta)$,
the determinant is easily evaluated as the product
\eqn\fpdett{\eqalign{\Delta_{2}&=\prod_{n\ne 0}\prod_{i,j}\left[\frac{2\pi
in}{\beta}-i\left(\lambda_i-\lambda_j\right)\right]=\cr
&=\left(\prod_{m\ne 0}\frac{2\pi i m}{\beta}\right)\prod_{i,j}
\frac{2}{\beta\left(\lambda_i-\lambda_j\right)}
\sin\left(\frac{\beta\left(\lambda_i-\lambda_j\right)}{2}\right).}}
Notice that up to an overall constant,
\eqn\comb{d\alpha \Delta_{2}= [dU],}
where
\eqn\msreign{d\alpha=\prod_i d\lambda_i
\prod_{i<j}\left(\lambda_i-\lambda_j\right)^2}
is the left-right invariant integration measure over Hermitian matrices
$\alpha$, and
\eqn\unitmeas{[dU]=\prod_i d\lambda_i \prod_{i<j}
\sin^2\left(\frac{\beta\left(\lambda_i-\lambda_j\right)}
{2}\right)}
is the left-right invariant integration measure in the integral
over the unitary matrices
\eqn\udefined{U\equiv e^{i\beta\alpha}.}
We will see in the next section at one loop order and argue generally
in \S6\ that $S_{eff}$ may be regarded as a function of $U$ rather
than $\alpha$, so that \intafto\ may be written as
\eqn\intaftp{e^{-\beta F}= \int [dU] e^{-S_{eff}(U)}}
where $S_{eff}(U)$ is defined in \intaftt.

\subsec{Evaluation of $S_{eff}$ at one-loop}

The path integral in \intafto\ may be evaluated diagrammatically,
generating an expansion of $S_{eff}(U)$ in powers of the gauge coupling.
In this
subsection we use this method to evaluate $S_{eff}(U)$ to lowest order in
$\lambda$ ($\CO(\lambda^0)$). This evaluation is rather simple as only
one-loop graphs in the expansion of \intafto\ contribute to $S_{eff}$
at this order.

We first consider the pure gauge theory. The Fadeev-Popov determinant for
gauge-fixing \gf\ is
\eqn\fpd{ \det{\del_i D^i } = \int \CD c \CD {\bar c}
e^{- {\bar c} \del_i D^i c},}
where $D^i$ denotes a gauge covariant derivative
\eqn\gaugecovariant{D_i c=\del_i-i[A_i,c]}
and $c$ and ${\bar c}$ are complex ghosts.
Consequently, the action of the free gauge theory may be written as
\eqn\fymact{ e^{-S_{eff}(U)}
=\int \CD A_i \CD A_0 \CD c \CD {\bar c}
\delta( \p_i A^i)
\exp \left[- \int \Tr \left( \half A_i (\tilde{D}_0^2 +\del^2) A^i +
\half A_0 \del^2 A_0 + {\bar c} \del^2 c\right)\right],}
where
\eqn\deft{\tilde{D}_0 X\equiv \p_0 X -i[\alpha, X].}

To proceed further we note that any vector field on the sphere may be
decomposed as
\eqn\decomp{A_i= \p_i \ph + B_i}
where $\p_i B^i=0$. The integral over $\ph$ in \fymact\ is easily performed
using the $\delta$ function, yielding $1/ \sqrt{\det'(\p_i \p^i) }$ where the
derivatives act on scalar functions on $S^3$ and the prime denotes omission
of the zero mode. The integral over $A_0$ yields the identical factor
(the zero mode is $\alpha$ which is not integrated over). The integral over
the ghosts, on the other hand, evaluates to $\det'(\p_i \p^i)$. These three
factors cancel nicely, so that \fymact\ simplifies to
\eqn\fymactn{ e^{-S_{eff}(U)}
=\int \CD B_i
\exp\left[-\half \int \Tr \left( B_i (\tilde{D}_0^2 +\del^2) B^i
\right)\right].}
Thus,
\eqn\sefflndet{S_{eff}= {1\over 2}\ln(\det(-\tilde{D}_0^2 -\del^2 )),}
where the operator acts on the space of divergenceless vector functions on
the sphere, i.e. the space of vector functions spanned by the vector
spherical harmonics. Consequently
\eqn\sefffin{S_{eff}={1\over 2}\sum_{\Delta} n(\Delta) \ln
\det(-\tilde{D}_0^2 +\Delta^2),}
where $\Delta^2$ are the eigenvalues of the Laplacian acting on vector
spherical
harmonics on the compact manifold and $n(\Delta)$ is the degeneracy of each
eigenvalue. When the compact space is an $S^3$ of unit radius we have
$\Delta=h+1$ (for integer $h \ge 0$) and $n(\Delta)=2h(h+2)$ (see the counting of vector
spherical harmonics in appendix B).

The determinant of $(-\tilde{D}_0^2 +\Delta^2)$ is easily
evaluated by passing to Fourier space in the time direction,
yielding the infinite product \eqn\prodeval{{\rm det}_{U(N)}
\left[ \prod_{n=-\infty}^{\infty}
\left(\frac{4\pi^2n^2}{\beta^2}+\frac{4\pi
n}{\beta}\alpha+\alpha^2+\Delta^2\right) \right].} The infinite
product of matrices may be rewritten as
\eqn\prdfctr{\eqalign{\left(\alpha^2+\Delta^2\right)&\left[\prod_{m\ne
0}\frac{\beta}{2\pi m}\right]^{-2}\left[\prod_{n\ne
0}\left(1+\frac{\beta\alpha}{\pi
n}+\frac{\left(\alpha^2+\Delta^2\right)\beta^2}{4\pi^2n^2}\right)\right]
=\cr &=\left(\alpha^2+\Delta^2\right)\left[\prod_{m\ne 0}
\frac{\beta}{2\pi m}\right]^{-2}\left[\prod_{n=1}^{\infty}
\left(1-\frac{\beta^2(\alpha+i\Delta)^2}{4\pi^2n^2}\right)
\left(1-\frac{\beta^2(\alpha-i\Delta)^2}{4\pi^2n^2}\right)\right]=\cr
&=\left[\prod_{m\ne 0}\frac{\beta}{2\pi m}\right]^{-2}
\left(\frac{4}{\beta^2}\right)\sin\left[\frac{\beta(\alpha+i\Delta)}{2}\right]
\sin\left[\frac{\beta(\alpha-i\Delta)}{2}\right]=\cr
&=\frac{2}{\beta^2}\left[\prod_{m\ne 0}\frac{\beta}{2\pi
m}\right]^{-2}
\left[\cosh(\beta\Delta)-\cos(\beta\alpha)\right]=\cr &={\cal
N}e^{\beta\Delta} (1-e^{-\beta \Delta+i\beta \alpha})(1-e^{-\beta
\Delta -i \beta \alpha}),}} where
 \eqn\evaln{{\cal N}=\left({1 \over \beta^2} 
\left[\prod_{m\ne 0}\frac{\beta}{2\pi m}\right]^{-2} \right).}

The divergent factor ${\cal{N}}$ is a constant independent of both 
$\Delta$ and $\alpha$, and we set it to unity to reproduce the free
energy of the harmonic oscillator. Thus
\eqn\deteval{\ln( \det(-\tilde{D}_0^2 +\Delta^2) ) = \Tr \left(
\beta \Delta + \ln( 1-e^{-\beta \Delta+i\beta \alpha}) +
\ln(1-e^{-\beta \Delta -i \beta \alpha}) \right),} where the trace
is over $N^2 \times N^2$ dimensional matrices and $\alpha$ acts in
the adjoint representation.  Expanding the logarithms in a power
series, summing over $\Delta$, and passing from the adjoint to the
fundamental ($\Tr_{adj}(e^{i n \beta \alpha}) \rightarrow \tr(U^n)
\tr(U^{-n})$) and using \sefflndet\ we obtain 
\eqn\effactv{S_{eff}= {1\over 2}\beta N^2
\sum_{\Delta} \Delta n(\Delta) - \sum_{n=1}^{\infty}
\frac{z_V(x^n)}{n} \tr(U^n) \tr(U^{-n}). } For the case of an
$S^3$ of unit radius the first term (appropriately regularized) is
equal to ${11\over 120}\beta N^2$, where ${11 \over 120}$ is the
Casimir energy $\sum_0^\infty h(h+1)(h+2)$ for a vector field on
the unit sphere.\foot{See equation (64) of \pervijay. This
equation, and the other Casimir energies in this subsection, may
be justified as follows. On a sphere of radius $R$ we wish to
compute $\sum_{m=1}^\infty (m^2-1)(m/R)$, where $E=m/R$ is the
energy of the mode in question. We regulate this sum by
multiplying the summand with a cutoff function $f(E/\Lambda)$,
where $f$ is a smooth function such that $f(0)=1$, $f'(0)=0$ and
$f(\infty)=0$. We now evaluate the regulated sum using the
Euler-MacLaurin formula \eqn\eulmac{\half F(0) + F(1) + F(2)
\ldots =\int_0^\infty F(x) dx +\zeta(-1) F'(0) +\zeta(-3) {F'''(0)
\over 3!}+ \cdots} with $F(m)=(m^2-1)(m/R)f({m \over R \Lambda})$,
and we find \eqn\regsum{{\Lambda^4 R^3 }\int_0^{\infty} x^3 f(x) -
{\Lambda^2 R} \int_0^{\infty} x f(x) +{-\zeta(-1)+\zeta(-3) \over
R} + \CO({1 \over \Lambda^2}).} In order that the vacuum energy of
the theory be zero in flat space (a necessary condition, for
instance, for conformal invariance), the two divergent terms above
must be cancelled by counterterms; indeed, counterterms of the
form $a \Lambda^4\int \sqrt{g}$ and $ b \Lambda^2 \int \sqrt{g}
{\cal R}$ (with suitable values for $a$ and $b$) achieve this
cancellation. The remaining finite piece of the energy is $11/120
R$. Finally, note that we have been careful to choose a `general
coordinate invariant' regulator; one that cuts off modes in a
manner that depends only on their proper energy. It would be
incorrect,  for instance,  to regulate the summand above with the
smoothing function $f((m-1)/\Lambda R)$; in this regulation scheme
theories with the same Lagrangian and same $\Lambda$ but different
$R$ are not identical in the UV. See chapter 6 of \bd\ for an
interesting and extensive discussion of related issues.}

It is a simple matter to generalize this calculation to include the
contributions from free conformally coupled
scalar and spinor fields. When the compact manifold is $S^3$,
the contribution to
$S_{eff}$ from a single additional scalar field is
\eqn\change{\d S_{eff}= \half \ln(\det(-\tilde{D}_0^2 -\del^2+1 )),}
where the operator acts on scalar fields on $S^3 \times S^1$ and the
constant piece of the operator is a consequence of the ${\cal R} \ph^2$
term in the
Lagrangian for conformally coupled scalars.
As above, this determinant is easily evaluated to yield
\eqn\effacts{S_{eff}= {{1 \over 240}\beta N^2} - \sum_{n=1}^{\infty}
\frac{z_S(x^n)}{n} \tr(U^n) \tr(U^{-n}) ,}
where ${1 \over 240} = \half \sum_{n=0}^\infty (n+1)^3$.
A Weyl spinor field on $S^3$ contributes
\foot{We take the spinor field
to be anti-periodic on the $S^1$ in order to compute the trace of
$e^{-\beta H}$ rather than the trace of $(-1)^F e^{-\beta H}$.}
\eqn\effactf{\eqalign{S_{eff}=&-\ln(\det(-{\tilde \p}_0 \gamma^0 -\pslash)) 
=\cr
=& {{17\over 960} \beta N^2} - \sum_{n=1}^{\infty}
\frac{(-1)^{n+1} z_F(x^n)}{n} \tr(U^n) \tr(U^{-n}) ,}}
where ${17\over 960}=-\sum_{n=1}^{\infty} n(n+1) (n+\half)$.

Thus, we have rederived the expression for
the free partition function \partsix, but this time with a physical
interpretation for the matrix $U$ as the {\it holonomy around the
time circle}, and for ${1\over N}\tr(U)$ as a Wilson loop around the time
circle, or a Polyakov loop (in the fundamental representation).

\newsec{Solution of the free Yang-Mills matrix model}

In this section we proceed to directly analyze the unitary matrix
model \partsix\ for the exact partition function of free $U(N)$ Yang-Mills
theory in order to extract the thermodynamic behavior.

We begin by recalling that any unitary matrix model with gauge-invariant
action and measure may be rewritten entirely in terms of the
eigenvalues of the unitary matrix, which must lie on the unit
circle. Denoting these eigenvalues by $\{e^{i \alpha_i}\}$ (with
$-\pi < \alpha_i \leq \pi$), we may
rewrite the partition function \partsix\ in terms of the eigenvalues
by the replacements
\eqn\replace{
\int [dU] \to \prod_i \int_{-\pi}^{\pi} [d \alpha_i]
\prod_{i<j} \sin^2\left({{\alpha_i - \alpha_j}\over 2}\right);
\qquad \tr(U^n) \to
\sum_j e^{i n \alpha_j} \; .
}
With only adjoint matter, we find from \partsix\ that the
effective theory for the eigenvalues is governed entirely by a pairwise
potential,
\eqn\partsixn{
Z(x) = \int [d \alpha_i] e^{-\sum_{i \ne j} V(\alpha_i - \alpha_j)}
}
where
\eqn\pairpotl{\eqalign{
V(\theta) &= -\ln|\sin(\theta/2)| - \sum_{n=1}^\infty {1 \over
n}\left[z_B(x^n) + (-1)^{n+1}z_F(x^n)\right]\cos(n \theta) =\cr &= \ln(2) +
\sum_{n=1}^\infty {1 \over n}(1- z_B(x^n) - (-1)^{n+1}z_F(x^n))\cos(n
\theta) \; .}}
In the first line, the first term coming from the measure is a
temperature-independent repulsive potential, while the remaining terms
provide an attractive potential which increases from zero to infinite
strength as the temperature increases from zero to infinity.\foot{To
see that the second term in the potential is always attractive, note
that we may rewrite it as
\eqn\vsecond{
V_2(\theta) = {1 \over 2} \int dE \left\{ \rho_B(E) \ln(1 - 2x^E
\cos(\theta)+x^{2E}) - \rho_F(E) \ln(1 + 2x^E \cos(\theta)+x^{2E})
\right\},
}
where $\rho_B$ and $\rho_F$ give the single-particle density of states
for bosonic and fermionic modes . This potential is always attractive
since for any value of $E$ and $x$ the integrand is decreasing in the
interval $\theta \in [-\pi,0]$ and increasing in the interval $\theta \in
[0,\pi]$.} This
suggests that at low temperatures, the minimum action configurations
will correspond to eigenvalues distributed evenly around the circle,
while at high temperatures, the eigenvalues will tend to bunch up.

For finite $N$, the partition function receives contributions from all
configurations of eigenvalues and depends smoothly on the temperature.
On the other hand, it is well known that in the large $N$ limit, the
matrix model partition function is dominated by the minimum action
configurations, with the exact leading and subleading terms in the
$1/N$ expansion of the free energy given, respectively, by the minimum
of the action and by the Gaussian integral about the minimum action
configuration. When this minimum action configuration (or its
derivatives) changes abruptly as a function of temperature, we may
have a phase transition in the large $N$ limit (see, for example, \gw ).
We will now see that exactly
this behavior occurs and leads to a phase transition for our model.

We begin in \S5.1 by analyzing the low temperature phase. In \S5.2 we
analyze the behavior near the phase transition temperature, and in
\S5.3 we analyze the high temperature limit. In \S5.4 we write down
the solution of the matrix model in the high temperature phase, and
in \S5.5 we provide a perturbative expansion of this solution around
the phase transition temperature. In \S5.6 we summarize the thermodynamical
behavior of the theory. In \S5.7 we discuss how the
Polyakov loop may be used as an order parameter at finite volume, and
we end in \S5.8 with an application of our results to some interesting
$3+1$ dimensional gauge theories.

\subsec{Low temperature behavior}

Consider first the theory at low temperatures, where the effective
potential between the eigenvalues is dominated by the repulsive
term. In this case, we expect that the pairwise repulsion drives the
eigenvalues to spread uniformly around the circle. In fact,
since there is no difference between displacing any individual
eigenvalue in the uniform distribution to the left and displacing it
to the right, the uniform eigenvalue distribution will always be a
stationary point of the action for a pairwise potential. To see when
this stationary point is a minimum, it is convenient to introduce an
eigenvalue distribution $\rho(\theta)$ proportional to the density of
eigenvalues $e^{i\theta}$ of $U$
at the point $\theta$. Note that $\rho$ must be everywhere
non-negative, and we may choose its normalization so that
\eqn\rhonorm{
\int_{-\pi}^\pi d \theta \rho(\theta) = 1 \; .
}
With this definition, the effective action for the eigenvalues becomes
\eqn\distact{\eqalign{
S[\rho(\theta)] &=
N^2  \int d \theta_1 \int d \theta_2 \rho(\theta_1) \rho(\theta_2)
V(\theta_1 - \theta_2) = \cr
  &= {N^2 \over 2 \pi} \sum_{n=1}^\infty |\rho_n|^2 V_n (T),
}}
where in the second line, we have defined $\rho_n\equiv \int d\theta
\rho(\theta) \cos(n\theta)$ and $V_n \equiv \int d\theta V(\theta)
\cos(n\theta)$ to be the
Fourier modes on the circle of $\rho$ and $V$, respectively,
and we assume without loss
of generality that the eigenvalue distribution is symmetric around
$\theta=0$. From the
latter expression, it is clear that the uniform distribution ($\rho_{n
\ge 1} = 0$) will be an absolute minimum of the potential as long as
all $V_n$ are positive. From \pairpotl\ we see that
\eqn\potmodes{
V_n = {2 \pi \over n} (1- z_B(x^n) - (-1)^{n+1}z_F(x^n)),
}
so the uniform distribution is an absolute minimum if and only if
\eqn\unicond{
z_B(x^n) + (-1)^{n+1} z_F(x^n) < 1 }
for all $n$. But, since the single-particle partition functions are
monotonically increasing, and $0 \le x < 1$, the $n=1$ condition is
always strongest, and the uniform distribution will be stable for
temperatures $T < T_H = -1/\ln(x_H)$ where $x_H$ is the solution to
\eqn\defxh{
z(x_H) \equiv z_B(x_H) + z_F(x_H) = 1 \; ;  }
note that this is precisely the Hagedorn temperature we discussed in \S3.2.
As long as we have more than a single quantum-mechanical
mode, equation \defxh\ always has a
unique solution with $0 < x < 1$, so the uniform distribution becomes
an unstable extremum beyond some finite temperature $T_H$ determined
by the single-particle partition function.

For $T < T_H$, we may now evaluate the free energy at leading and
subleading order in $N$. Since $\tr(U^n)$ vanishes for any $n \ge 1$
for the uniform distribution, the classical value for the action
\partsix\ (and
thus the leading ${\cal O}(N^2)$ contribution to the free energy)
vanishes. The first non-zero contribution to the free energy thus
arises from the Gaussian integral around this configuration. From the
quadratic action \distact,\potmodes,
we see that the leading contribution to the partition function is
\foot{The `Jacobian' for the `change of variables'
from $\lambda_i$ to $\rho_n$ appears to be irrelevant at every
order in the $1/N$ expansion, see \gold.}
\eqn\largeNpart{
Z(x) = \prod_{n=1}^\infty {1 \over 1-z_B(x^n)-(-1)^{n+1}z_F(x^n)},
}
where the normalization has been arbitrarily
fixed by choosing the vacuum free energy to vanish, $Z(x=0)=1$ (more
physically, we could choose $Z(0)$ to account for the Casimir energy of
the vacuum, see the previous subsection).
This is precisely the same as the result \bfmultipart\ that we found
above by counting the states with $E \ll N^2$; we see that this is
indeed the leading large $N$ behavior as long as $T < T_H$.

This partition function diverges at $T = T_H$, as should be expected,
since this is the point where the quadratic action \distact\ develops
an unstable direction. The corresponding divergence in the free energy
$F = -T \ln(Z)$ goes like
\eqn\divfree{
F \to T_H \ln(T_H - T)
}
as $T$ approaches $T_H$ from below. By the general discussion in
\S2, we may conclude that this divergence is associated with a
Hagedorn density of states,
\eqn\newhag{
\rho(E) \propto E^0 e^{\beta_H E} \; ,
}
as we also found by counting states in \S3.
Thus, just as for perturbative string theory, we have a Hagedorn
divergence which is associated with an unstable mode in the Euclidean
path integral; note that the power of the energy is also the same as
in string theory at finite volume.
In this case, the ``tachyon'' causing the divergence is the lowest ``momentum''
fluctuation mode $\rho_1$ of the eigenvalues. In the next subsection,
we will be
able to see explicitly what the endpoint of this tachyon condensation
brings as we raise the temperature beyond the Hagedorn temperature.

\subsec{Behavior near the transition}

At the temperature $T=T_H$, $V_1$ vanishes and the mode $\rho_1$ (the
lowest momentum fluctuation in the eigenvalue distribution) becomes
massless. Since the action is quadratic, this corresponds to an
exactly flat direction in the potential, and the corresponding family
of minimum action configurations are given (up to overall translations
of $\theta$) by
\eqn\flatdir{
\rho(\theta) = {1 \over 2 \pi}(1+t \cos(\theta)), \qquad \qquad 0 \le t \le 1,
}
which interpolate between the uniform distribution and a sinusoid
vanishing at a single point. For $T > T_H$, the quadratic form $S$ has
negative coefficients, so all minimum action configurations must lie
at the boundary of configuration space, at a point where a hyperboloid
of constant $S$ lies tangent to this boundary. Since this boundary is
provided by the positivity condition $\rho(\theta) \geq 0$,
these minimum action eigenvalue
distributions necessarily vanish on a subset of the circle.

In the limit of small, positive $\Delta T = T-T_H$, the action
contour $S=0$ is a cone with opening angle going to zero. The
contours of smaller $S$ are hyperboloids inside this cone, so the
minimum action configuration lies on the boundary of the
configuration space inside the cone. Finally, it is easy to see
that the region of allowed $\rho$'s is convex (if points
$\{\rho_n\}$ and $\{\rho'_n\}$ are in the region then all points
$\{t \rho_n + (1-t) \rho'_n\}$ are in the region for $0 \le t \le
1$). Thus, the boundary region interior to the cone is a
simply-connected neighborhood of the $t=1$ configuration in
\flatdir\ whose size goes to zero as $T \to T_H$. We conclude that
the minimizing configuration changes continuously from the $t=1$
configuration in \flatdir\ for $T$ in some interval above $T_H$.

The leading behavior of the free energy in this vicinity of $T_H$ is
then given by evaluating the $V_1$ term in the action \distact\ on the
$t=1$ configuration in \flatdir , since the effects of changing the
eigenvalue distribution come in at higher orders.

Thus, the leading result for the minimum action is
\eqn\minact{
S^{min}_{T > T_H}(x) = {N^2 \over 8 \pi} (T-T_H) \; V_1'(T_H) + \dots
}
from which we may obtain the free energy as $F = T S^{min}$. The
leading behavior near the transition is therefore given by
\eqn\leadingfree{
\lim_{N \to \infty} {1 \over N^2} F_{T \to T_H}(T) =
\left\{ \matrix{ 0 & \qquad T < T_H \cr
-{1 \over 4} (T-T_H) z'(x_H){x_H \over T_H} & \qquad T > T_H} \right.
}
so we have a first order phase transition at the Hagedorn temperature. Note
also that the behavior is characteristic of a deconfinement
transition, since the free energy is ${\cal O}(1)$ below the
transition and ${\cal O}(N^2)$ above the transition. This will be
verified in \S5.7 below when we discuss the behavior
of a second order parameter for confinement, related to the Polyakov loop.

\subsec{High temperature behavior}

For general values of $T > T_H$, we will not be able to write down
an exact expression for the free energy, though we describe in the
next subsection a formal solution that may be used to evaluate the
free energy to arbitrary accuracy. However, the theory simplifies
again in the large temperature limit, where the potential becomes
strongly attractive\foot{The high temperature regime and its
relation to string theory was discussed for uncompactified
theories in \PolchinskiTW\ in a very similar language to the one
we are using here, and this discussion was applied to two
dimensional QCD in \KutasovGQ.}. In this case, the action is
minimized by a tightly clustered configuration of eigenvalues that
approaches a delta function in the limit of infinite temperature.
Thus, we should have $\rho_n=1$ to leading order, and \eqn\hights{
S = {N^2 \over {2\pi}} \sum_{n=1}^{\infty} V_n(T \gg 1). } The
high temperature limit of the single-particle partition functions
depends only on the space-time dimension $d$ and the number of
physical polarizations ${\cal N}^{dof}$ of the fields in the
theory, \eqn\limofzi{ z_i(x) \to 2 {\cal N}^{dof}_i T^{d-1} +
{\cal O}(T^{d-2}) \; . } Using \potmodes, we find that the high
temperature limit of the free energy is \eqn\highF{ F(x \to 1) =
-2 N^2 T^d \zeta(d) \left[{\cal N}^{dof}_B + \left(1 - {1 \over
2^{d-1}}\right){\cal N}^{dof}_F \right] + {\cal O}(T^{d-1}). } The
limiting free energy density here coincides with that of the flat
space theory, which should be expected since in the free field
analysis we are dealing with a scale-invariant theory, so taking
the dimensionless temperature $TR$ to infinity is equivalent to
taking the limit of large volume at fixed temperature (see
\refs{\neil, \finn,\LandsteinerXV} among others for the high
temperature expansion of the free energy for the $U(1)$ theory).

\subsec{Exact solution for $T > T_H$}

Having understood the qualitative features of the $T>T_H$
behavior, we now describe an exact solution for the matrix model at
arbitrary temperatures above the transition.  Just beyond the Hagedorn
temperature, we have seen that the eigenvalue distribution becomes
sinusoidal, vanishing at $\theta = \pi$ as the lowest momentum
fluctuation mode condenses. As the temperature is increased further,
the attractive term in the pairwise potential continues to increase in
strength, so the eigenvalues will become increasingly bunched
together, occupying only a finite interval $I = [-\theta_0, \theta_0]$
on the circle (we arbitrarily choose the middle of this interval to be
at $\theta=0$ for convenience).

The precise distribution may be determined by the condition that a
single additional eigenvalue added in the interval $I$ experiences no
net force from the other eigenvalues,
\eqn\stabcond{
\int_{-\theta_0}^{\theta_0} V'(\alpha - \theta) \rho(\theta) d\theta =0,
\qquad \qquad \alpha \in I.
}
Substituting in our potential \pairpotl, we obtain
\eqn\equione{
\int_{-\theta_0}^{\theta_0} \cot \left( {\alpha - \theta \over 2} \right)
\rho(\theta) d\theta = 2 \sum_{n=1}^{\infty} a_n \rho_n \sin(n \alpha),
}
where on the right hand side,
$a_n \equiv z_B(x^n) + (-1)^{n+1} z_F(x^n)$,
and we have used the fact that the eigenvalue
distribution should be symmetric about $\theta = 0$ and therefore be
orthogonal to $\sin(n \theta)$ for all $n$.
As before, $\rho_n$ are the moments of
the eigenvalue distribution,
\eqn\pmoments{
\rho_n = \int_{-\pi}^\pi \rho(\theta) \cos(n \theta) d\theta.}

To solve for
$\rho(\theta)$, it is convenient treat the moments $\rho_n$ as
independent variables, and self-consistently
solve for $\rho(\theta)$ and $\rho_n$
together using the two equations \equione\ and \pmoments.
Fortunately, the equations \equione\ provide the equilibrium
conditions for a matrix model with action
\eqn\auxact{
S = N \sum_{n=1}^{\infty} {a_n \rho_n \over n}
(\tr(U^n) + \tr(U^\dagger {}^n)),}
that has been solved exactly in \JurkiewiczIZ. The solution for $\rho(\theta)$
is\foot{In the language of \JurkiewiczIZ, the solution we are looking for is of
type $A_1$.}
\eqn\psoln{
\rho(\theta) = {1 \over \pi} \sqrt{\sin^2 \left({\theta_0 \over 2} \right) -
\sin^2 \left({\theta \over 2} \right)} \sum_{n=1}^\infty Q_n \cos((n -
{ 1\over 2}) \theta)
}
for $-\theta_0 < \theta < \theta_0$ and $\rho(\theta)=0$ otherwise, where
\eqn\Qdef{
Q_n \equiv 2 \sum_{l=0}^\infty a_{n+l} \rho_{n+l} P_l ( \cos(\theta_0))
}
and the angle $\theta_0$ that bounds the eigenvalue distribution is
determined by the condition
\eqn\Qcond{
Q_1 = Q_0 + 2 \; .
}
In the formula \Qdef\ for $Q_n$, the $P_l$ are Legendre polynomials, defined by
\eqn\defleg{
\sum_{l=0}^{\infty} P_l(x) z^l = (1-2 xz + z^2)^{-{ 1\over 2}}.
}
We may now eliminate $\rho(\theta)$ to obtain a linear system of
equations for the $\rho_n$ from \pmoments\ and \Qcond . To describe
these, we define a matrix $R$ and a vector $A$ whose elements are
polynomials in $s^2 \equiv \sin^2(\theta_0/2)$,
\eqn\qmdef{\eqalign{
R_{ml} &\equiv a_l \sum_{k=1}^l
(B^{m+k-{1 \over 2}}(s^2) + B^{|m-k + {1 \over 2}|}(s^2))
P_{l-k}(1-2s^2), \cr
A_m &\equiv a_m(P_{m-1}(1-2s^2) - P_m(1-2s^2)),
}}
where
\eqn\defofb{
B^{n-{1 \over 2}}(s^2) = {1 \over \pi} \int_{-\theta_0}^{\theta_0}
d \theta \sqrt{\sin^2 \left({\theta_0 \over 2} \right) -
\sin^2 \left({\theta \over 2} \right)} \cos((n - { 1\over 2}) \theta)
}
are polynomials defined by the generating function
\eqn\genofb{
\sum_{n=0}^\infty B^{n+{1\over 2}}(x) z^n = {1 \over 2 z}
(\sqrt{(1-z)^2 + 4 z x} + z - 1).
}
The determining equations \pmoments\ and \Qcond\
for the eigenvalue moments and the angle $\theta_0$
are then simply
\eqn\effqm{
R \vec{\rho} = \vec{\rho}; \qquad \qquad \vec{A} \cdot \vec{\rho} = 1.
}
Thus, the vector of moments of the eigenvalue distribution must be
an eigenvector of the matrix $R$ with eigenvalue 1, normalized so that
its dot product with $A$ is 1 (ensuring that the eigenvalue
distribution integrates to 1). The condition
\eqn\givestheta{
\det(R-1) = 0
}
that the matrix $R$ has such an eigenvector determines the angle
$\theta_0$. Note that while this equation is a complicated function of
$\theta_0$, it is linear in each of the coefficients $a_n$, so it is
convenient to parameterize the solution in terms of $(
\sin(\theta_0/2), \{ a_{n>1} \})$, solving for $a_1$ in terms of these
variables using \givestheta. With \givestheta\ satisfied, the
explicit solution for the unit eigenvector may be given as
\eqn\soln{
\rho = M^{-1} e_1,
}
where $M$ is a matrix obtained by replacing the first row of the
singular matrix $(1-R)$ with the vector $A$, and $e_1 = (1,0,0,\dots)$.

This solution is rather formal for the general case when $R$ is an
infinite matrix. However, in the case when $a_{n>k}$ vanishes for some
$k$, the matrix $M$ takes the form
\eqn\formofm{
M = \left( \matrix{ M_{k \times k} & 0 \cr L & 1} \right),
}
so the explicit solution may be obtained by inverting a finite matrix,
\eqn\solofrho{
\vec{\rho} = \left( \matrix{ M_{k \times k}^{-1} & 0 \cr
-LM_{k \times k}^{-1} & 1} \right) e_1.
}
Note that once we evaluate the moments $\rho_{n \le k}$, the full
eigenvalue distribution is given by \psoln\ since $Q_{n>k} = 0$ in
this case. Similarly, only the upper $k \times k$ submatrix of $R$
contributes to the condition \givestheta\ determining $\theta_0$.

In our model, the coefficients $a_n$ die off with $n$ (exponentially
for $x \ll 1$, and like a power law as $x \to 1$)
so we may obtain an arbitrarily good approximation to the exact
eigenvalue distribution by truncating such that $a_{n>k} = 0$ for
sufficiently large $k$.  It turns out that even restricting to $k=1$
retains the same qualitative behavior of the model.\foot{In
this case, the auxiliary theory \auxact\ is exactly the model studied
in \gw. The truncation to $k=1$ was also considered in \SundborgUE.} 
In this case, the angle $\theta_0$ is determined
by
\eqn\toydet{
0 = \det(R_{1 \times 1} - 1) = a_1(2s^2 - s^4) - 1,
}
while $\rho_1$ is determined by
\eqn\forrhoone{
\rho_1 = M_{1 \times 1}^{-1} \cdot  1 = (2 a_1 s^2)^{-1}.
}
The full distribution is then determined from $\rho_1$ using \psoln\
(noting that $Q_{n>1} =0$). Parameterizing the solution by $s =
\sin(\theta_0/2)$, the final result is
\eqn\toydist{
\rho(\theta) = {1 \over \pi \sin^2 \left({\theta_0 \over 2} \right)}
\sqrt{\sin^2 \left({\theta_0 \over 2} \right) -
\sin^2 \left({\theta \over 2} \right)}  \cos\left({\theta \over 2} \right),
}
with $\theta_0$ determined in terms of $a_1$ by \toydet , or explicitly
\eqn\solforone{
\sin^2\left({\theta_0\over 2}\right) = 1 - \sqrt{1-{1 \over a_1(T)}}.
}
As $T$ increases from $T_H$ to infinity, $a_1$ increases from 1 to
infinity, so $\theta_0$ decreases from $\pi$ to 0, such that the
eigenvalue distribution \toydist\ eventually approaches a delta
function. Using \toydist, we may evaluate the free energy in the
approximation $a_{n>1}=0$ and find that
\eqn\truncfree{
-{1 \over N^2} \ln(Z_{trunc}) = -\left({1 \over 2 s^2} + {1 \over 2} \ln(s^2) -
{1\over 2}\right).
}
Near the transition, we find
\eqn\nearf{
{F\over N^2} = -{T_H \over 4} (a_1 - 1) + {\cal O}((a_1-1)^2) =
-{T_H\over 4} (T - T_H) {{\del a_1} \over {\del T}} \bigg|_{T=T_H} +
{\cal O}((T-T_H)^2),
}
which gives the same leading behavior \leadingfree,
characteristic of a first order transition, as we found in the full model.

\subsec{Perturbative expansion slightly above the Hagedorn temperature}

Using the general solution above, it is also possible to determine
explicitly the exact behavior near the transition as a perturbation
expansion in $(T-T_H)$. To determine $\theta_0$ as a function of
temperature, it is convenient to define $\Delta \equiv a_1-1$ and $\epsilon
\equiv \cos^2(\theta_0/2)$, and to expand $R = R_0 + \delta R$ where
\eqn\defszero{
R_0 \equiv {\rm diag}(a_1, a_2, a_3, \dots),
}
so that $\delta R$ will be small for small $(T-T_H)$. We may then
solve the condition \givestheta\ perturbatively for $\Delta$ as a function
of $\epsilon$ by writing
\eqn\pgt{
0 = \det(1 - (1-R_0)^{-1} \delta R) }
and expanding the determinant out in terms of traces. This expansion
in powers of $\delta R$ may further be expanded as a power series in
$\epsilon$ and $\Delta$, and finally solved perturbatively for
$\Delta$ as a power series in $\epsilon$. Using the fact that
\eqn\Sexp{
\delta R_{ml} = (-1)^{m+l+1} l^2 m a_l \epsilon^2 + {\cal O} (\epsilon^3)\; ,
}
and that the lowest power of $\Delta$ appearing in \pgt\ is $1/\Delta$ (from
$(1-R_0)^{-1}_{11}$), it follows that $\Delta$ may be obtained to
order $\epsilon^{2k+1}$ by keeping terms to order $\delta R^k$ in \pgt.
Explicitly, we find that the leading terms are
\eqn\anglefn{\eqalign{
\Delta &= \epsilon^2 + \epsilon^4(1- \sum_{n=2}^{\infty} {n^3 a_n
\over 1-a_n}) + {\cal O}(\epsilon^5) \; .
}}
When $a_{n>1} = 0$,
this correctly reduces to the leading orders of the truncated model result
\solforone
\eqn\truncres{
\Delta = \sum_{n \ge 1} \epsilon^{2n}.
}

To find perturbative expressions for the moments of the eigenvalue
distribution near the transition, we use the general result \soln,
expanding $M=M_0 + \delta M$ with
\eqn\defmzero{
M_0 = \left( \matrix{A_1 & A_2 & A_3 & \cdots \cr & 1-a_2 & & \cr & &
1-a_3 & \cr & & & \ddots} \right) \qquad M_0^{-1} = \left( \matrix{{1
\over A_1} & {-A_2 \over A_1(1- a_2)} & {-A_3 \over A_1(1-a_3)} &
\cdots \cr & {1 \over 1-a_2} & & \cr & & {1 \over 1-a_3} & \cr & & &
\ddots} \right),
}
so that the explicit perturbative solution is
\eqn\pertsol{
\rho = M_0^{-1}(1+\delta M M_0^{-1})^{-1} e_1 =
(M_0^{-1} - M_0^{-1} \delta M M_0^{-1} + \dots) e_1 \; .
}
Note that 
\eqn\deltaml{
\delta M_{ml} = \left\{ \matrix{ - \delta R_{ml} & \qquad \qquad m > 1, \cr 
0 & \qquad \qquad m=1,} \right.  
}
so from \Sexp\ we see that $\delta M$ starts at order $\epsilon^2$. Thus,
we may obtain the eigenvalue
moments to order $\epsilon^{2k-1}$ by keeping the first $k$ terms
here. Using the first two terms shown and the expansions of $a_1$ and
$A_n$ in powers of $\epsilon$, we find
\eqn\corrected{\eqalign{
\rho_1 &= (M_0^{-1})_{11}+ \sum_{n=2}^{\infty} (M_0^{-1})_{1n} \delta R_{n1}
(M_0^{-1})_{11} + {\cal O}(\epsilon^4) =\cr
&=  {1 \over 2}  + {\epsilon \over 2}   + {\epsilon^2 \over 2}
\sum_{n=2}^{\infty}  {n a_n \over 1-a_n} + {\cal O}(\epsilon^3), \cr
\rho_{n>1} & =(M_0^{-1})_{nn} \delta R_{n1} (M_0^{-1})_{11} +
{\cal O}(\epsilon^4) =\cr & = {\epsilon^2 \over 2} (-1)^n{ n \over 1 - a_n}
+ {\cal O}(\epsilon^3). \cr }}
The result for $\rho_{n>1}$ seems
at first somewhat surprising since it appears that the moments
increase without bound as $n$ is increased. However, it turns out that
the leading terms \Sexp\ in the expansion of $\delta R_{n1}$ in $\epsilon$
provide a good approximation only for $\epsilon \ll 1/n^2$. To obtain
an approximation valid for all $n$ at some fixed value of $\epsilon$,
we should use the complete expression $\delta R_{n1} = a_1(B^{n+{1
\over 2}}(1-\epsilon) + B^{n - {1 \over 2}}(1-\epsilon))$, which does
fall off for large $n$. Alternately, we note that the perturbative
expansion of the difference between the moments of the exact
distribution and those of the truncated $k=1$ model is well behaved for large
$n$,
\eqn\momdif{\eqalign{
\delta \rho_1 = \rho_1 -\rho_1(a_{n>1}=0) &=  {\epsilon^2 \over 2}
\sum_{m=2}^{\infty} {m a_m \over 1-a_m} + {\cal O}(\epsilon^3), \cr
\delta \rho_n = \rho_{n>1} - \rho_{n>1}(a_{n>1}=0) & = {\epsilon^2 \over 2}
(-1)^n{ n a_n \over 1 - a_n} + {\cal O}(\epsilon^3), }}
due to the exponential decay of $a_n$ for large $n$. 

Using the expressions \corrected\ for $\rho_n$, we may finally
evaluate the free energy from \distact\ and \potmodes . To avoid
subtleties associated with the apparent growth in the moments
\corrected\ for large $n$, it is again convenient to express the
result as a correction to the truncated model free energy.
The final result, taking into account all corrections to the truncated
model, is 
\eqn\goodfree{\eqalign{
-{1 \over N^2} \ln(Z) &= \sum_{n=1}^\infty {1 \over n} (1-a_n) 
(\rho_n (a_{n>1}=0) + \delta \rho_n)^2 =\cr
&= -{1 \over N^2} \ln(Z_{trunc})+ {\epsilon^4
\over 4} \sum_{n=2}^{\infty} {n(n^2-1) a_n \over 1-a_n} +
{\cal O}(\epsilon^5)=\cr
&=-{\epsilon^2 \over 4}- {\epsilon^3 \over 3} - \epsilon^4 \left({3 \over
8}-{1 \over 4} \sum_{n=2}^{\infty} {n(n^2-1) a_n \over 1-a_n}\right) + {\cal
O}(\epsilon^5), }}
where $Z_{trunc}$ is the partition function \truncfree\ of the $k=1$ 
truncation computed earlier in this section. 
\foot{ While
the perturbation expansion about the $k=1$
truncated model in the second line of \goodfree\ 
is perfectly well behaved, the naive evaluation of
$\ln(Z)$ using \corrected\ yields the divergent expression
\eqn\freenaive{\eqalign{
-{1 \over N^2} \ln(Z) &= \sum_{n=1}^\infty {1 \over n} (1-a_n)
\rho_n^2 =\cr &=-{\epsilon^2 \over 4}- {\epsilon^3 \over 2} +
{\epsilon^4 \over 4} \sum_{n=2}^{\infty} n -{\epsilon^4 \over 2}+ {\epsilon^4
\over 4} \sum_{n=2}^{\infty} {n(n^2 - 1) a_n \over 1-a_n} + \cdots. }}
The divergence in the third term of  \freenaive\ is unphysical; it is a
consequence of the (apparent) linear growth in the moments observed 
in \corrected. As we noted earlier, this linear growth is actually cut off 
beyond $n \sim 1/\sqrt{\epsilon}$, where it is replaced by a
decaying behavior. Thus, the divergent sum in
\freenaive\ should be cut off at  $n=C/\sqrt{\epsilon}$, resulting in 
a contribution that scales like $C^2 \epsilon^3 /8$; 
a finite contribution to a lower order of perturbation theory. Indeed, it is
easily verified that the (finite) $\epsilon^3$ term in \freenaive\
disagrees with the exact $\ep^3$ term in the expansion of the 
solution \truncfree\ 
\eqn\ztruncn{
-{1 \over N^2} \ln(Z_{trunc}) = -{\epsilon^2 \over 4}- {\epsilon^3 \over 3}-
{3\epsilon^4 \over 8} + {\cal O}(\epsilon^5),}
as our analysis above would suggest. We conclude that the 
divergent sums at higher orders in the naive perturbation theory 
conspire to give finite contributions 
$\epsilon^3/6 + \epsilon^4/8 + \cdots$. A systematic understanding of
this naive perturbation theory would be an intricate task; one that
may, however, be avoided by choosing the correct starting point for
the perturbation expansion, as we have done above. The smarter
perturbation expansion \goodfree\ is 
entirely divergence free; this
follows since the $a_n$ decay exponentially with $n$, whenever $x <
1$.} 
If desired, the free energy 
may be rewritten in terms of $\Delta$ using \anglefn,
or explicitly in terms of temperature using
\eqn\deltabyt{
\Delta = a_1(T) - a_1(T_H) = a_1'(T_H)(T-T_H) +
{1 \over 2} a_1''(T_H)(T-T_H)^2 + \cdots.
}

\subsec{Summary of thermodynamic behavior}

\fig{Free energy as a function of temperature in free Yang-Mills theories.}
{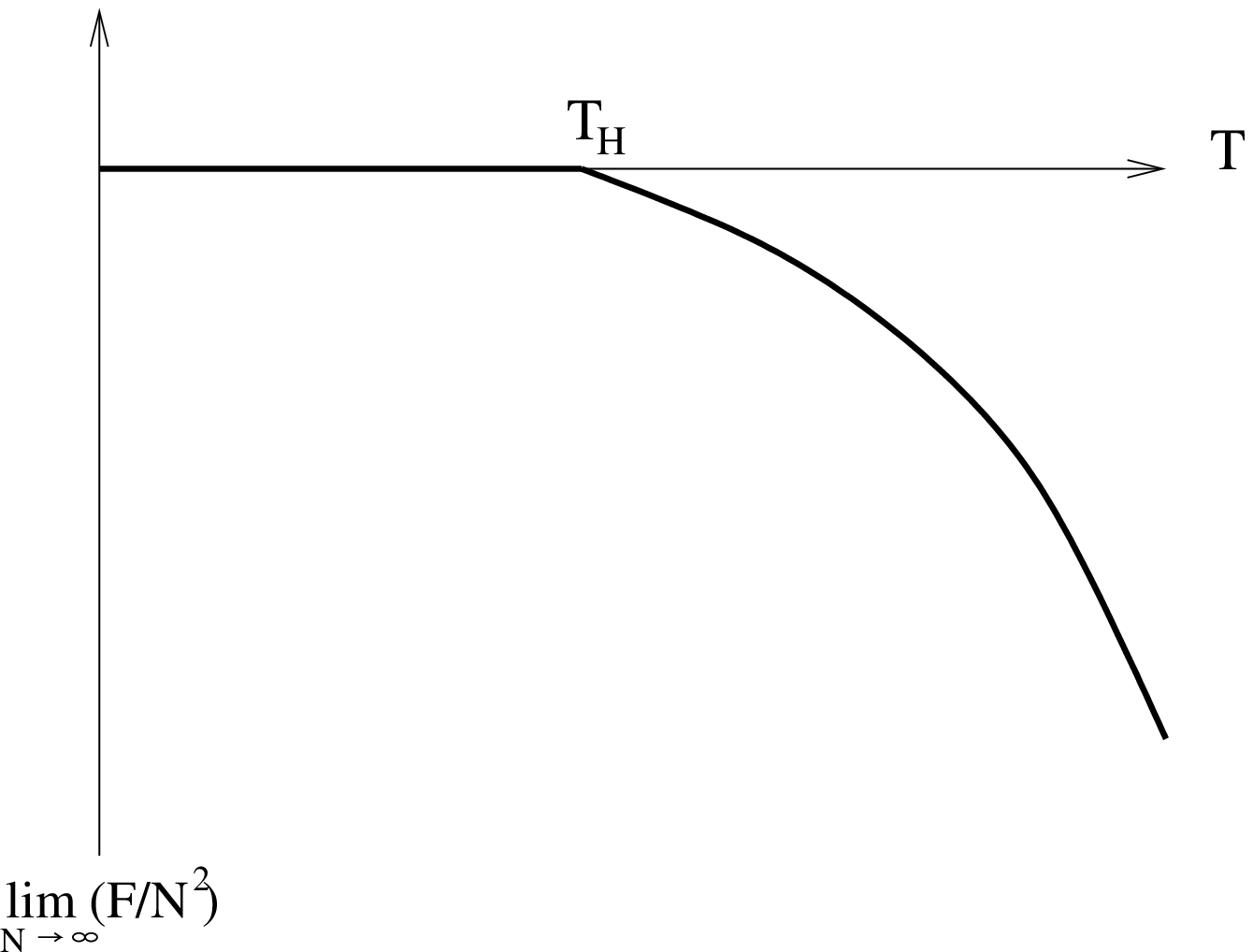}{2.75truein}
\figlabel{\freefr}

We are now in a position to present a reasonably detailed picture of
the behavior of the free energy of the free gauge theory at all
temperatures, depicted in figure \freefr.

We find that the uniform eigenvalue distribution provides an absolute
minimum of the matrix model potential for all temperatures below some
$T_H$ determined by the single-particle partition function as
\eqn\defth{
z(T_H) = 1 \; .
}
The free energy in this regime, given by \largeNpart , is ${\cal
O}(1)$ and decreases monotonically to a logarithmic divergence
\divfree\ at $T=T_H$, characteristic of a Hagedorn density of states
\newhag.

As $T$ increases past $T_H$, the uniform eigenvalue distribution
develops an unstable mode which condenses to give a pure sinusoid
distribution \flatdir\ for $T$ just above $T_H$, and a distribution with a gap
for $T > T_H$.  For small $(T-T_H)$, the ${\cal O}(N^2)$ free energy
decreases linearly in $T-T_H$, so the first derivative of the free
energy is discontinuous and we have a first order phase transition. The free
energy as a function of temperature may be written as a perturbation
expansion in $T-T_H$ with the leading terms given in \goodfree.

In the high temperature limit, the pairwise potential between
eigenvalues becomes strongly attractive, and the eigenvalue
distribution approaches a delta function. In this regime, the free
energy asymptotes to \highF\ which corresponds to the expected
flat-space free-energy density. Using the formal exact solution of the
previous section, it should be possible to work out the high
temperature behavior as a perturbation expansion in $1/T$, as we did
explicitly for the behavior near the transition.

\subsec{The Polyakov loop as an order parameter at finite volume}

In the theory we analyzed above, the Polyakov loop $\langle {1\over
N}\tr(U) \rangle$ vanishes in both the high temperature and the low
temperature phases.  However, the reason for this vanishing is rather
different in the two phases. The low temperature phase is governed by
a single saddle point (uniform distribution of eigenvalues); the
Polyakov loop vanishes on this saddle point. On the other hand the
high temperature phase is governed by a one parameter set of saddle
points (related to each other by the $U(1)$ rotation $U \rightarrow
e^{i \alpha } U$, which acts on the eigenvalue distribution by
shifting $\theta$ periodically). The value of $\langle {1\over
N}\tr(U) \rangle$ is non-zero when evaluated on each saddle point
individually but vanishes upon integrating over $\alpha$ (averaging
over all saddle points\foot{Witten \WittenZW\ has described a rather
similar phenomenon in the bulk dual to the strongly coupled $\CN=4$
Yang Mills.}), since the phase varies over a complete
circle\foot{For $SU(N)$ theories this is not an integral but a sum
over $N$ discrete saddle points, but it has the same effect.}. To
obtain an order parameter which correctly distinguishes the two
phases, it is therefore sufficient to use the squared magnitude of the
Polyakov loop, $\langle |{\cal P}|^2 \rangle = \lim_{N \to \infty}
\langle {1 \over N^2} |\tr(U)|^2 \rangle$.

This order parameter may be given a nice physical interpretation by
the following alternative definition. Define
$\langle {1\over N}\tr(U) \rangle_{\epsilon}$
as the expectation value of $\langle
{1\over N}\tr(U) \rangle$ upon perturbing the effective
action in \partsix\ by the infinitesimal term ${N \epsilon}
(\tr(U) + \tr(U^\dagger)$).  Note that this perturbation breaks the $U(1)$
symmetry (or the $\IZ_N$ symmetry for $SU(N)$);
in the presence of this perturbation, the high temperature phase
is also
governed by a single saddle point\foot{Of course at any finite $N$ there
is some finite $\epsilon_0(N)$ that is so small that when $\epsilon \leq
\epsilon_0(N)$ the path integral receives significant contributions from
the entire manifold of `almost' saddles. However $\epsilon_0(N) \rightarrow
0$ as $N \to \infty$. Consequently, if we take the $N \to \infty$ limit
first, only a single saddle point contributes to the path integral for any
$\epsilon$ no matter how small.}. Consequently, the order parameter
\eqn\op{{\cal P}^{eff}= \lim_{\ep \to 0} \lim_{N \to \infty} \langle
{1\over N}\tr(U) \rangle_{\epsilon} }
is zero in the low temperature phase but non-zero in the high temperature
phase\foot{Note that this construction
is rather similar to the definition of the order
parameter of a ferromagnet in 3 dimensions. Naively, the magnetization of
the ferromagnet vanishes due to $SO(3)$ symmetry in both the high and low
temperature phases. However if we define $|M|_\ep$ to be the magnetization
in the presence of a small magnetic field in the $z$ direction, then
$\lim_{\ep \to 0} \lim_{V \to \infty} |M|_\ep$ (where $V$ is the
volume) is the standard order parameter for the ferromagnet.}. It is
easy to show that in the large $N$ limit, $\langle |{\cal P}|^2
\rangle =|{\cal P}^{eff}|^2$, so the two definitions of the order
parameter are equivalent.

The construction presented in the previous paragraph may be
understood physically as follows.
As discussed in \S2.4, $\langle \tr(U) \rangle$
vanishes in both
phases for purely kinematical reasons; Gauss' law does not allow the
introduction
of a quark on a compact space. However, the perturbation  described in the
previous paragraph effectively adds a condensate (or plasma) of classical
quarks and anti-quarks to the theory; in the presence of such a condensate
the constraint from Gauss' law is circumvented (recall that flux lines
decay in a Higgs phase). In the high temperature phase the additional
classical quark has finite free energy even in the limit that the classical
Higgsing is taken to zero. On the other hand, in the low temperature phase
the Polyakov loop vanishes as this Higgsing is taken to zero.

The analysis above shows that ${\cal P}^{eff}=1/2$ immediately after
the phase transition, and it goes up to ${\cal P}^{eff}\to 1$ at high
temperatures. In fact, ${\cal P}^{eff}$ is exactly the variable
$\rho_1$ discussed in \S5.4, so the exact behavior of the order
parameter is given formally by the first component of $\vec{\rho}$ in
\soln, while the behavior near the transition is given by
\pertsol, or explicitly to leading orders by \corrected.

\subsec{Results for specific theories}

Before closing this section, we explicitly apply our results to two
interesting
$3+1$ dimensional theories, pure
Yang-Mills theory and the ${\cal N} = 4$ SYM theory. Using
the single-particle partition functions given in \sthreefields , we
may write down explicit expressions for the transition temperature
using $z(x_H)=1$, for the behavior of the free energy near the
transition using \goodfree , and for the high temperature behavior
using \highF .

For pure $d=3+1$, $U(N)$ (or $SU(N)$)
Yang-Mills theory on an $S^3$ of unit radius, 
the single-particle partition functions are 
given by
\eqn\zfree{
z_B(x)={6x^2-2x^3  \over (1-x)^3}, \qquad \qquad z_F(x) = 0.
}
Solving $z(x_H)=1$, we find that the
Hagedorn temperature is given by $x_H = 2 - \sqrt{3}$, $T_H = -1 /
\ln(2-\sqrt{3}) \simeq 0.759326$. The free energy below
the Hagedorn temperature may be obtained from the partition function 
$Z$ in \bfmultipart\ using \zfree\ and recalling that $x = \exp(-1/T)$. 
Just above the phase transition, the free 
energy is given by
\eqn\frnznt{
{1 \over N^2}F_{{\cal N} = 0}(T \to T_H^+) =
-0.9877(T-T_H)-3.004(T-T_H)^{3 \over 2}
-5.980(T-T_H)^2 + {\cal O}((T-T_H)^{5 \over 2}),
}
while the high temperature free energy is
\eqn\frnzht{
F_{{\cal N} = 0}(T \gg T_H) \to -{ \pi^2 \over 45} N^2 T^4 V_{S^3},
}
where $V_{S^3} = 2 \pi^2 R^3$ is the volume of the sphere (restoring a
general value for the $S^3$ radius).

For the $d=3+1$ $\CN=4$ SYM theory on an $S^3$ of unit radius, 
the single-particle partition 
functions are
\eqn\zmax{
z_B(x)={6x +12x^2 - 2x^3 \over (1-x)^3}, \qquad \qquad z_F(x)= {16 x^{3\over
2} \over (1-x)^3},
}
so from $z_B(x_H)+z_F(x_H)=1$ we find that the Hagedorn temperature
is given by\foot{Curiously, $x_H$ for the ${\cal N}=4$ theory is
exactly the square of $x_H$ for the pure Yang-Mills theory, and $T_H$ in
$\CN=4$ SYM is precisely half of $T_H$ in pure Yang-Mills. This
arises from the even more peculiar fact that the single-particle
partition functions are related by $z_{{\cal N}=0}(x) = z_{{\cal
N}=4}(x^2)$. }
$x_H = 7 - 4\sqrt{3}$, $T_H = -1 / \ln(7-4\sqrt{3}) \simeq
0.379663$.
The free energy below the Hagedorn temperature may be obtained from 
the partition function 
$Z$ in \bfmultipart\ using \zmax, with
$x=\exp(-1/T)$. The free energy immediately above the phase
transition temperature is given by
\eqn\frnfnt{
{1\over N^2} F_{{\cal N} = 4}(T \to T_H^+)=
-0.9877(T-T_H)-4.248(T-T_H)^{3 \over 2} -11.696(T-T_H)^2 +
{\cal O}((T-T_H)^{5 \over 2}),
}
while the high temperature free energy is
\eqn\frnfht{
F_{{\cal N} = 4}(T \gg T_H) \to -{\pi^2 \over 6} N^2 T^4 V_{S^3}.
}

\newsec{Phase structure at weak coupling}

In this section we turn to an analysis of the phase transition at
non-zero 't Hooft coupling\foot{Some of the key ideas underlying
this section and the next one arose in discussions with
R. Gopakumar. We thank him in particular for emphasizing the qualitative 
difference between the two 
scenarios depicted in figure 4, and for forcing us to understand the
order parameter in a clearer fashion.}.  
We will see that an arbitrarily small non-zero
coupling qualitatively changes the behavior of the transition. In
\S6.1 we deduce the most general form of
$S_{eff}(U,\lambda)$ allowed by gauge invariance. In \S6.2 we analyze
the structure of $S_{eff}(U, \lambda)$ in
perturbation theory. In \S6.3 we study the structure of the
deconfinement transition (discussed in detail for the free theory in the
previous section) at weak coupling. In \S6.4 we verify the general analysis
of \S6.3 in an exactly solvable toy model. We end in \S6.5 with a description
of the implications of our results for the microcanonical ensemble.

\subsec{General properties of the effective action}

For any value of the coupling constant,
gauge invariance imposes tight constraints on the form of the effective action
$S_{eff}(\alpha)$ \intaftt\ for the zero mode of the gauge field;
$S_{eff}(\alpha)$ should be
invariant under all space-time gauge
transformations that
\item{(1)} Are single valued (up to an element of the center of $U(N)$)
on ${\cal M} \times S^1$,
\item{(2)} Preserve the gauge-fixing conditions \gf\ and \gff .

We will restrict attention to gauge transformations $U(t)$ that are
independent of the position on the compact space. Under such a transformation
\eqn\transa{A_0 \rightarrow V(t) A_0 V^\dagger(t) -i \p_t V V^\dagger,}
so that
\eqn\transal{\alpha \rightarrow V(t) \alpha V^\dagger(t) -i \p_t V V^\dagger.}
$V(t)$ obeys the condition (2) above (and \transal\ makes sense) only when
the right-hand side of \transal\ is independent of time.

Constant (time independent) gauge transformations clearly satisfy the
requirements (1) and (2) above. Consequently, $S_{eff}(\alpha)$ is invariant
under $\alpha \rightarrow V \alpha V^{\dagger}$. We may use this invariance to
diagonalize $\alpha$. Once this has been done, we consider the further
gauge transformations
\eqn\gttt{V(t)= e^{i t D},}
where $D$ is a diagonal matrix, whose eigenvalues are all
integral multiples of $2 \pi / \beta$. $\alpha$
transforms under the gauge transformation generated by \gttt\
as $\alpha \rightarrow \alpha+D$. This implies that $S_{eff}(\alpha)$ is
invariant under separate shifts of any of the eigenvalues of $\alpha$
by multiples of
$2 \pi/\beta$. It follows from these invariances that $S_{eff}$ is really a
function only of $\tr(U^n)$ (for all $n$)
where $U = e^{i \beta \alpha}$ (see \udefined)
is the zero mode holonomy around the time circle.

Finally, consider $V(t)=e^{{2  \pi i k t \over
\beta N}}$ where $k$ is an integer if the gauge group is $SU(N)$, and
arbitrary for $U(N)$.
$V(t)$ obeys the single-valuedness condition (1), as
$e^{{2 \pi i k \over N}}$ belongs to the center of the gauge group. Under the
gauge transformation generated by $V(t)$,
$\alpha \rightarrow \alpha + {2 \pi k \over N \beta}$.
Consequently, $S_{eff}$ should be invariant under $U \rightarrow e^{{2 \pi i k
\over N}} U$, for all integers $k$ if the gauge group is $SU(N)$, and for
any $k$ for $U(N)$. In the limit $N \to \infty$ the two cases coincide, and
$S_{eff}(U)$ must be invariant under $U \rightarrow e^{i \theta }U$
for arbitrary $\theta$. Putting everything together, we conclude that
$S_{eff}$ may depend on $U$ only in combinations of the form
\eqn\basictr{
\tr(U^{n_1}) \cdots \tr(U^{n_k}) \tr(U^{-n_1-\dots - n_k}) \; .
}

\subsec{The general form of the effective action in perturbation theory}

In large $N$ perturbation theory, $S_{eff}$ is generated by planar
vacuum diagrams obtained by integrating out the massive modes. In this
calculation, $\alpha$ is a background field, which appears
diagrammatically via external line insertions on the index loops of
diagrams. Planar diagrams at order $\lambda^{k-1}$ include $(k+1)$ index
loops, each of which leads to a trace in the final result (which may
contain arbitrarily many factors of $\alpha$). From the discussion of the 
previous subsection, the only terms with $(k+1)$ traces allowed by gauge
invariance take the form \basictr . It follows that the most general
form of the planar contribution to $S_{eff}$ in perturbation theory is
\eqn\pertpar{\eqalign{S_{eff}(U) = N^2 [ &\sum_n  m_n^2(x) |u_n|^2
+ \lambda \sum_{m,n} F_{m,n}(x) (u_m u_n u_{-n-m} +c.c.) \cr
& + \lambda^2  \sum_{m, n, p} F_{m,n,p}(x) (u_m u_n u_p u_{-m-n-p} +c.c.)
+ \cdots] \; ,}}
where we have defined $u_n \equiv \tr(U^n)/N$ (note that $u_n^* =
u_{-n}$ and $u_0=1$). This agrees, of course, with the quadratic form
of the one-loop effective action that we found in sections 3 and
4. Also, the $u_n$ coincide with the variables $\rho_n$ used in \S5
for eigenvalue distributions which are 
symmetric about $\theta=0$, so the values
of $m_n^2$ in the free field theory may be read off from equation \potmodes.

\subsec{Possible phase structures at weak coupling}

Recall from \S5 that, at $\lambda=0$, $u_1$ is massless at the phase
transition temperature $T_H$, while the other $u_n$'s are all massive.
Consequently, $u_1$
will continue to be the lightest mode in the vicinity of the phase
transition also at weak coupling. Thus, for analyzing the phase transition
at weak coupling, it is
useful to obtain an effective action for $u_1$ by integrating
out the $u_n$'s with $n > 1$ \foot{As we are interested in the large $N$
limit, it is sufficient to integrate out these modes classically.}.
It follows from \pertpar\ that the leading large $N$ terms in the resulting effective action are of the
form
\eqn\sefffa{S_{eff}(u_1)={N^2} \left( m_1^2(x, \lambda) |u_1|^2
+ \sum_{n=2}^{\infty} \lambda^{2n-2} B_n(x,\lambda) |u_{1}|^{2n}
\right),}
where $m_1^2$ and the $B_n$'s are functions of the temperature and power 
series in $\lambda$ starting (generically)
from a $\lambda^0$ term.

In particular, to $\CO(\lambda^2)$ we have
\eqn\compseffu{
S_{eff}(u_1)= N^2 \left( m_1^2 |u_1|^2 + b |u_1|^4 \right),}
where $b(x)=B_2(x,0) \lambda^2$ is a function of temperature which is
generically non-zero. It is not difficult to compute $B_2(0)$ 
starting from \pertpar.
The only terms in \pertpar\ that contribute are
\eqn\fenymsp{
S'_{eff}(U) = N^2 \left( m_1^2 |u_1|^2 + m_2^2 |u_2|^2  + \lambda I
(u_{-2} u_1^2 + u_{2} u_{-1}^2)
+ \lambda^2 A  |u_{1}|^4 \right).}
At this order,
$u_2$ may be integrated out from \fenymsp\ by setting it to be equal to its
classical value $u_2=-I\lambda u_1^2 / m_2^2$ \foot{Thus
setting the eigenvalue distribution to
\eqn\evfu{ \rho(\theta)={1 \over 2 \pi}\left(
1+2|u_1| \cos (\theta+\alpha) - {2\lambda I \over m_2^2}
|u_1|^2 \cos(2 \theta + 2 \alpha)
\right)}
for some $\alpha$.
},
yielding \compseffu\ with $b=(A-I^2/m_2^2)\lambda^2$.
Note that because the eigenvalue density \evfu\ has to be non-negative
everywhere, our expressions for the effective action \sefffa\ and
\compseffu\
are valid only for $u_1$ such that $|u_1| \leq \half + \CO(\lambda)$.

As in \S5, the one-loop contribution to the free
energy coming from integrating over $u_1$ diverges at the temperature
$T_H(\lambda)$ at which $m_1^2(x,\lambda)$ goes to zero and $u_1$ becomes
massless. For any value of the coupling, 
at leading order in the distance from this temperature, we have
$m_1^2 \simeq K(T_H(\lambda)-T)$ for some positive constant $K$.
This divergence signals a Hagedorn behavior of the
single-particle spectrum of the theory with Hagedorn temperature $T_H$
(computable in perturbation theory), so
this behavior persists (at least in the microcanonical ensemble) 
even at non-zero 't Hooft coupling.

As the saddle point
at $u_1=0$ is unstable for $T >T_H$, the theory described by \compseffu\
clearly undergoes a phase transition (to another saddle point) at some $T \leq
T_H$. Whether this phase transition occurs at $T<T_H$ or $T=T_H$ depends on
whether the value of $b$ (defined in \compseffu) at the Hagedorn temperature
$T_H$ is positive or negative, as we now argue in detail \foot{See
\Pisa\ for a similar discussion in a related context.}. The formulas in
the remainder of this section are all correct only to leading order in $b$
(or in $\lambda$).

\fig{Plots of $S_{eff}(u_1)$ as a function of $u_1$ for small positive $b$,
in units of $N^2 b$, at several
values of $m_1^2$ (from top
to bottom) :
$m_1^2=0$ (Hagedorn temperature and first phase transition temperature),
$m_1^2 =-{b\over 4}$, $m_1^2=-{b\over 2}$
(second phase transition temperature), and $m_1^2 =-{3b\over  4}$.}
{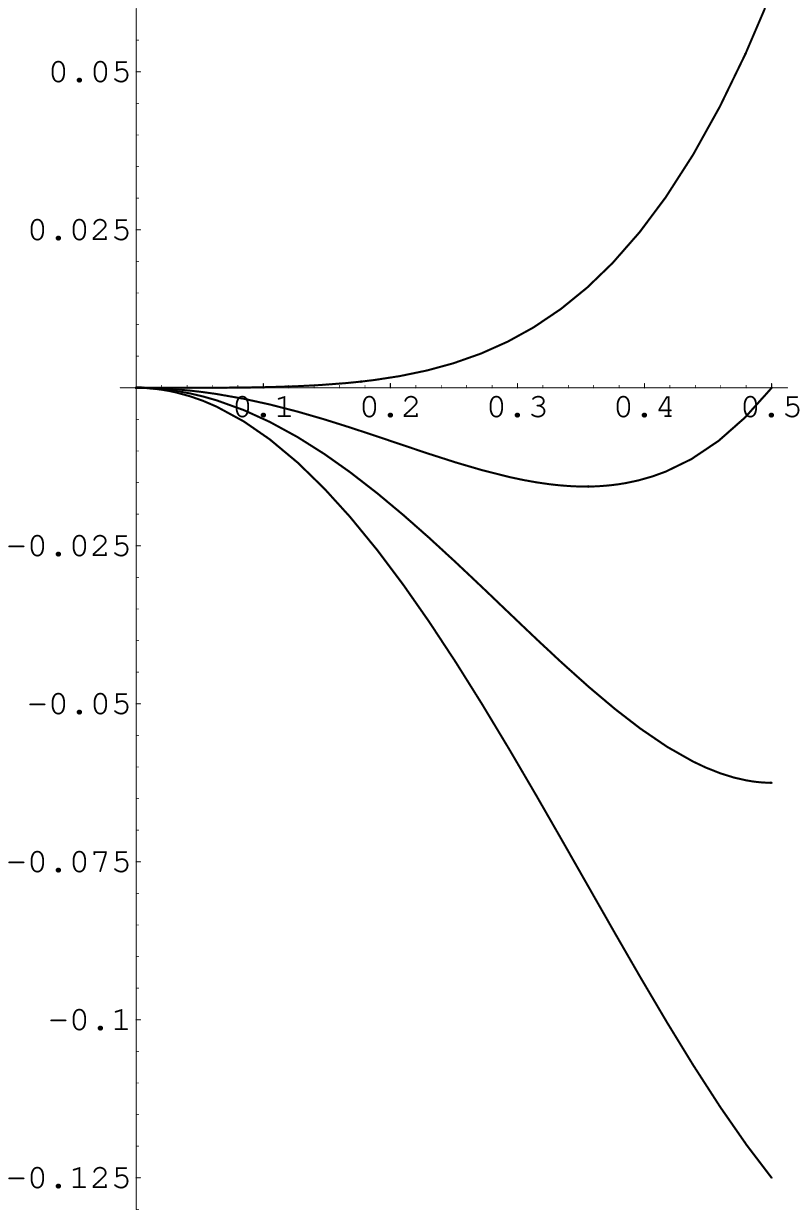}{1.75truein}
\figlabel{\secplots}
First, consider the case $b > 0$. For $T<T_H$, $u_1=0$, corresponding 
to the uniform eigenvalue distribution,
is clearly a
global minimum of the effective action \compseffu.
For $T>T_H$, however, $u_1=0$
is unstable and $S_{eff}$ is minimized at $|u_1|^2=|m_1|^2 / 2  b$. The
value of the effective action at this minimum is
$S_{eff} = - N^2 |m_1|^4 / 4  b $,
which is of order $(T-T_H)^2$, and so the phase transition at $T=T_H$ is
of second order. As the temperature is raised above $T_H$, the eigenvalue 
distribution smoothly becomes non-uniform until we reach $|u_1|^2 = 1/4$; 
this occurs at  $m_1^2 = -b / 2$. At this point, 
$\rho(\theta)$ vanishes at some $\theta$ and
we have reached the boundary of the space of eigenvalue distributions and the
edge of the validity of \compseffu. As the temperature is further raised
the eigenvalue distribution develops a gap on the circle, and the
theory undergoes a further phase transition similar to the Gross-Witten
transition \gw.
This second phase transition (at $T=T_H+{b \over 2 K}$) is
of third order. The behavior above
this second transition temperature is no longer captured by the effective
action \sefffa, and the full action is required to analyze it.

\fig{Plots of $S_{eff}(u_1)$ as a function of $u_1$ for small
negative $b$, in units of $N^2 |b|$, at
several values of $m_1^2$ (from top to bottom) : $m_1^2={3|b|\over
4}$, $m_1^2 ={|b|\over 2}$ (new phase nucleated), $m_1^2={|b|\over 4}$
(first order phase transition temperature), $m_1^2={|b|\over 8}$,
and $m_1^2=-{|b|\over 8}$. }{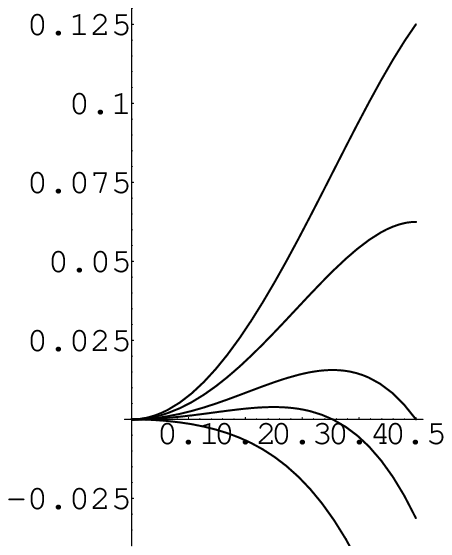}{2.0truein}
\figlabel{\firstplots} 
In the case where $b$ is negative,
$S_{eff}$ develops a new local minimum at $|u_1| = \half$ (the
boundary of our order parameter space) when $m_1^2 < |b| / 2$;
note this happens below the Hagedorn temperature. When $m_1^2 >
|b| / 4$, the free energy at this new local minimum is positive,
and the saddle point $|u_1|=\half$ is disfavored compared to
$u_1=0$. However, when we raise the temperature to $m_1^2<|b| /
4$, the free energy at $|u_1|=\half$ becomes negative, and so this
saddle dominates over the $u_1=0$ saddle point. Consequently, at
$T= T_H-{|b| \over 2 K}$, $|u_1|$ jumps discontinuously
from zero to $\half$ and the theory undergoes a first order phase
transition. When the temperature is further raised, the eigenvalue
distribution develops a break on the circle and the theory is no
longer described by \sefffa. This behavior is qualitatively
similar to that of the free theory which we analyzed in \S5, except
that the phase transition now happens below the Hagedorn temperature $T_H$.

If $b$ vanishes exactly then the higher order terms in \sefffa\ are
required for analyzing the phase transition, but generically $b$ will
not vanish at the Hagedorn temperature. In our discussions above we
assumed that $b$ was constant in a range of temperatures near the Hagedorn 
temperature; this is consistent at weak coupling 
since the range of temperatures which is
relevant for the discussions above is of order $\lambda^2$, so the changes
in $b$ within this range are of higher order in $\lambda$.

\subsec{Toy model}

The varieties of behavior described in the previous subsection may be 
illustrated using a
simple toy example, given by the matrix model
\eqn\ntm{Z(\beta)= \int [dU] \exp\left[ -\left( |\tr(U)|^2 (m_1^2 -1) + b
|\tr(U)|^4 / N^2 \right)\right].}
Based on the previous discussion, we expect this toy model to give a
good picture of the behavior of the full theory for small $\lambda$
near the phase transition, since the $u_n$'s for $|n|>1$ are all small there.
Readers who believe our general discussion above may skip directly to the
next subsection.

Our toy model may be solved exactly using the methods of \S5.4. 
Since the action in \ntm\
depends only on $\tr(U)$ and not on traces of higher powers of $U$,
the possible forms of the
saddle point eigenvalue distribution for the integral \ntm\ (up to an
overall shift in $\theta$) are the same as those in \gw\ (and in our
truncated model in \S5). The possible distributions may be parametrized by a
positive number $g$, and they are given
by one of the two expressions
\eqn\gwsoln{\eqalign{\rho(\theta)&=\frac{1}{2\pi}
\left(1+\frac{2}{g}\cos(\theta)\right)
\qquad\qquad
g\ge 2,\cr
\rho(\theta) &=\frac{2}{\pi g}\cos\left(\frac{\theta}{2}\right)
\sqrt{\frac{g}{2}-\sin^2\frac{\theta}{2}}\qquad g \le 2,}}
where for the matrix model \ntm\ we have
\eqn\gdet{g^{-1}=u_1(1-m_1^2 -2 b u_1^2)}
with
\eqn\htsc{u_1 = {\tr(U)\over N}= \int\,d\theta\,\rho(\theta)\cos(\theta).}
Note that $u_1$ is real since we assumed in \gwsoln\ 
that the eigenvalue distribution
is symmetric about $\theta=0$.

We search for simultaneous solutions to \gwsoln, \gdet\ and \htsc. It
is easy to check that the constant distribution $u_1=0$ ($g= \infty$,
$\rho(\theta) = 1/2\pi$) is always
a solution; we call this solution Phase I.  The free energy in Phase I
is zero. From the discussion of \S5 we know that
this solution is a local minimum when $m_1^2 > 0$.

An additional solution (Phase II) is obtained by
using the first distribution in \gwsoln\ (with $g \geq 2$), upon
solving  \gdet\ and \htsc\ for this case. This gives
\eqn\cdet{u_1=(1-m_1^2)u_1-2 b u_1^3,}
with the solution
\eqn\cpt{u_1^2=-{m_1^2 \over 2 b}.}
Obviously, this only makes sense if $m_1^2 / b \leq 0$; consistency
of the solution also requires
\eqn\tvalid{g^{-2}= \left| {m_1^2 \over {2 b}} \right|
\leq {1\over 4},}
so this phase exists and is a local minimum
for $b > 0$ and
$-b/2 \leq m_1^2 \leq 0$ (such a phase exists also
for $m_1^2 > 0$ and $b < 0$, but in that case its free energy is positive
and so this phase is sub-dominant compared to Phase I).
The free energy $F = - T \ln(Z)$ evaluated
on this saddle point is 
\eqn\freeenII{F(T) = - {N^2 T m_1^4 \over {4 b}},}
which is quadratic in $(T-T_H)$ near the Hagedorn temperature (where
$m_1^2 \propto T_H-T$).

To find solutions using the second distribution in \gwsoln\ (with $g
\leq 2$), we must once again solve \gdet\ and
\htsc\ with
the appropriate distribution, and we find
\eqn\selfconstwnd{u_1=1-{g\over 4} =
1-{1\over 4 (u_1(1-m_1^2)-2 u_1^3 b)}.}
Given the solution for $u_1$, the free energy evaluated on this saddle point is
\eqn\feos{F(T) = N^2 T \left[
(m_1^2-1) u_1^2 + b u_1^4 +{1\over 4} - \half \ln
\left( 2 (1-u_1) \right) \right] }
(the first two terms come from the potential in \ntm\ evaluated on the saddle,
while the remaining two terms come from the measure).

The exact solution to the quartic equation \selfconstwnd\ is
easily obtained, though it is not particularly illuminating. 
It may be checked that there are no physical solutions for 
$m_1^2 > -{b \over 2}$, while for $m_1^2 = -{b \over 2}$ we find $u_1=\half$. 
The discussion
of the previous subsection suggests (and we will demonstrate this below)
that this solution represents the phase transition point from Phase II to 
Phase III
when $b>0$, and a phase nucleation point (the creation of a new phase at
the boundary) when $b<0$.

To understand the behavior near this value of $m_1^2$, 
we analyze the solutions to \selfconstwnd\ in
the neighborhood of this special point. For this purpose we introduce a new
variable $y \equiv -m_1^2-{b \over 2}$ which is linear in $\d T$, the
increment over the phase transition (or nucleation) temperature. We also
introduce a variable $\delta \equiv {b \over 2}(4u_1^2-1)$ that measures the
deviation of $u_1$ from its value at the phase transition
(or nucleation) point.
In terms of these variables, \selfconstwnd\ may be rewritten as
\eqn\selpro{2 \sqrt{1+ {2 \over b} \delta}\left(1 - 
{1 \over 2} \sqrt{1+ {2 \over b} \delta}\right)(1+y-\delta)=1.}
The explicit solution for $y$ in terms of $\delta$ follows immediately
from this; this solution may be 
expanded in a power series in $\delta$ and inverted to obtain
\foot{The following three formulas are only correct for $b>0$; for $b<0$
there is another solution to \selpro\ which gives the leading contribution
\spentatoappear. We thank  L. Alvarez-Gaume, C. Gomez, H. Liu and S. Wadia
for pointing this out to us.}
\eqn\xsoln{\delta=y-{y^2 \over b^2}+
 {2 +b \over b^4}  y^3 -\left({5\over b^6}+{5\over b^5}+{9 \over 4 b^4} 
\right)y^4 + \CO(y^5).}
Using our definition $\delta \equiv {b \over 2}(4u_1^2-1)$ we now find 
\eqn\sfss{u_1=\half \left( 1+{1\over b}y -{2+b  \over {2 b^3}}y^2 +
{4+4b+b^2\over 2 b^5} y^3 + \CO(y^4)
\right),}
and inserting this into \feos\ we obtain
\eqn\feoss{\eqalign{F(T)&=N^2 T \left[-{b \over 16} -{y\over 4 }
-{y^2 \over 4 b} + {1\over 6 b^3} y^3 + \CO(y^4)\right]=\cr
&=N^2 T \left[- {1\over 4b}(y+{b\over 2})^2 +{1\over 6 b^3} y^3 +
\CO(y^4) \right]= \cr
&=N^2 T \left[- {m_1^4 \over 4b} + {1\over 6 b^3} y^3 + \CO(y^4)\right].}}

In summary, Phase I exists for all values of $m_1$ and $b$. Its free energy
is zero. Phase II exists when $0<-m_1^2/b \leq 1/2$. Its free energy is
$F=-N^2Tm_1^4/4b$. Finally, Phase III exists whenever there is a solution
with $u_1 \geq 1/2$ to \selfconstwnd\ and has a free energy given by \feos\ 
(for $b>0$ this phase exists when $-m_1^2/b \geq 1/2$, and for $m_1^2$
close to $-b/2$ it has a free energy given by \feoss). Recall that
$m_1^2=K(T_H-T)$ where $K$ is positive.

It is now clear that the dynamics of our toy model near the Hagedorn
temperature $m_1^2(T_H)=0$ depends sensitively on
the sign of $b$ at this temperature. 
First consider $b >0$. For $T<T_H$ ($m_1^2 >0$),
Phase I with $u_1=0$ is the only saddle point, and it
has vanishing free energy. As $T$ rises above $T_H$ (for small negative
$m_1^2$) a new phase (Phase II) comes into existence. Simultaneously
Phase I becomes unstable, and the theory executes a second order
phase transition into Phase II at $T=T_H$. Finally, at a still higher
temperature ($m_1^2=- b / 2$) Phase II evolves continuously into Phase III;
this transition is of third order, due to \freeenII, \feoss, and the fact
that $y$ in \feoss\ is linear in
$\delta T$. As the temperature is raised further,
the theory remains in Phase III.

On the other hand, if $b< 0$, Phase I is the only saddle point for
$m_1^2 \geq -b / 2$. At this temperature (below the Hagedorn
temperature) a new phase comes into existence. As the temperature
is further raised, this new phase splits into two different saddle
points (Phase II and Phase III). Phase II has positive free energy
whenever it exists, and is always unstable. Phase III is locally
stable; however its free energy is positive in the neighborhood of
$m_1^2 = -b / 2$, and so (at this point) this saddle is
sub-dominant compared to Phase I. Upon further raising the
temperature, $F(T)$ of Phase III becomes negative at some positive
value of $m_1^2$, which may be obtained by solving
\selfconstwnd\ and \feos\ simultaneously for $u_1$ and $m_1^2$;
note that this happens below the Hagedorn temperature. Finally,
as the temperature is raised above the Hagedorn temperature, Phase
II disappears and Phase I becomes unstable, so the theory remains
in Phase III at all higher temperatures.

In conclusion, the exact solution of the toy model \ntm\ verifies, in
complete detail,  the general results of the previous subsection.

\subsec{Density of states as a function of energy}

In the previous subsections we discussed the canonical partition function as a
function of temperature for Yang-Mills theories at weak coupling. We
found three qualitatively different classes of $Z(\beta)$ depending on
whether  $b$ (see \compseffu) is negative, zero or positive. In this
subsection we will qualitatively describe the corresponding Yang-Mills
theories in the microcanonical ensemble, seeing
how the density of states depends on the energy in each of these
three different classes of models.

In order to make contact with the canonical ensemble discussed
in previous subsections, we will find it convenient to characterize the
microcanonical density of states $S(E)=\ln(\rho(E))$
in a rather peculiar way. We will find it convenient to plot the logarithm of
the effective temperature $\ln(T(E))=-\ln (\del S(E)/\del E)$ as
a function of $\ln(E)$ \foot{We thank R. Gopakumar for suggesting this to us.};
such plots were used for very similar purposes in \BarbonDI.

\fig{Plot of $\log(T)$ as a function of $\log(E)$ (for energies of order
$N^2$, in the microcanonical
ensemble), for $\lambda=0$ and for $\lambda>0$ with $b>0$ and $b<0$.}
{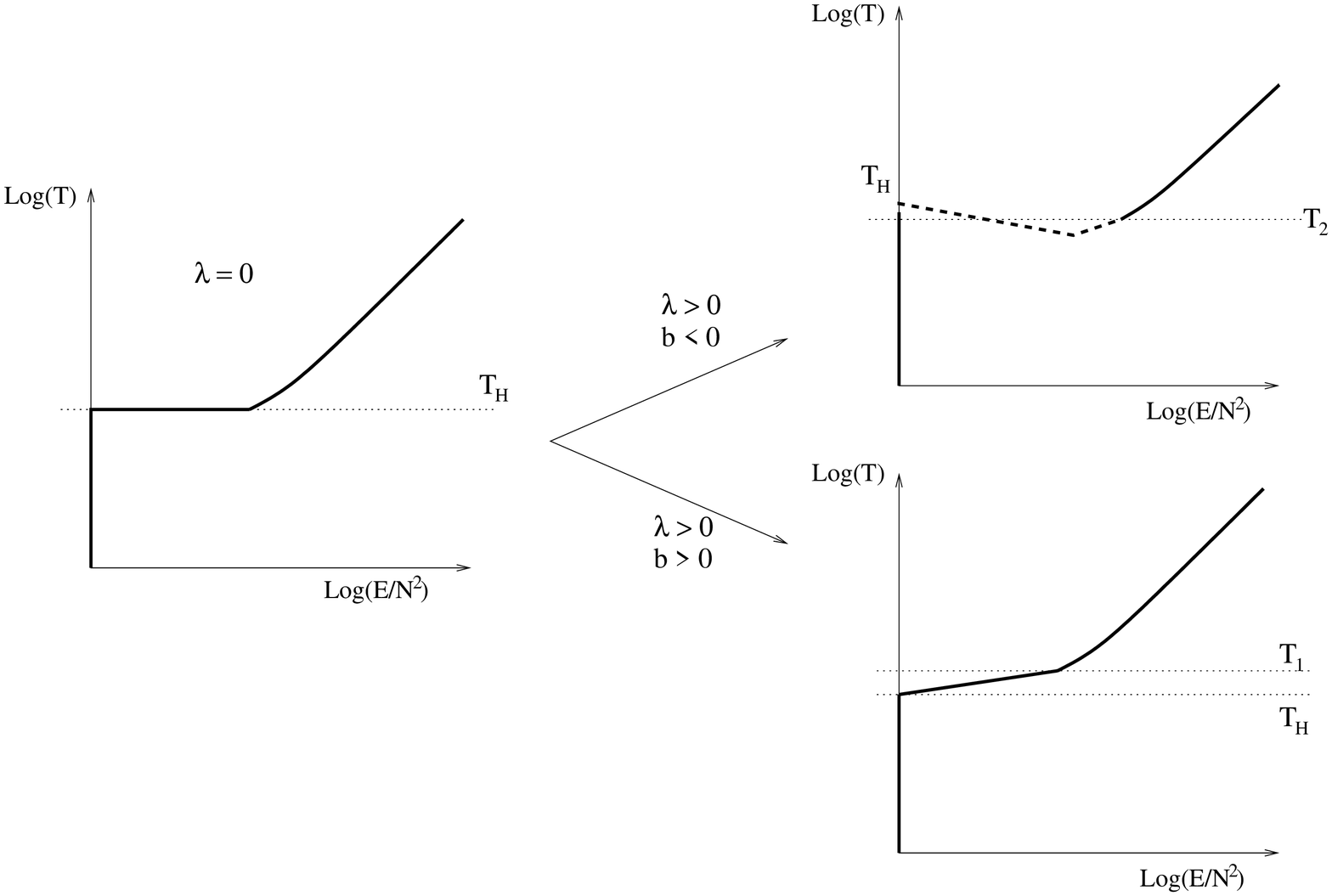}{5.0truein}
\figlabel{\ensquared}

The behavior of the density of states (or ${\partial S \over \partial
E}$) as a function of temperature may be deduced directly from the
behavior of $F(T)$ using the relation $E = \partial_\beta (\beta
F)$. However, this information is incomplete where there are first
order phase transitions, since these correspond to temperatures at
which the energy jumps discontinuously in going from one phase to
another (non-zero latent heat). For the case $\lambda = 0$, we have
seen that the first order transition is a Hagedorn transition, so we
may conclude that for microcanonical energies intermediate between the
two phases, the density of states is Hagedorn, corresponding to a
horizontal line in the graph of $\ln(T)$ as a function of $\ln(E)$. 
Thus, the $F(T)$ behavior
depicted in figure \freefr\ leads to the microcanonical behavior shown
on the left in figure \ensquared . For small $\lambda > 0 $, the
density of states should be a small perturbation of this behavior, and
we may deduce from the results of \S6.3 that the qualitatively
different behavior obtained for $b>0$ and $b<0$ corresponds to whether
the flat Hagedorn region in the $\lambda=0$ plot tilts upwards
($b>0$), resulting in two continuous phase transitions at $T_H$ and
some higher temperature $T_1$, or downwards ($b<0$), preserving the
first order transition (dashed lines in figure 4 indicate points not 
accessible in the canonical ensemble).

The plots in figure \ensquared\ correspond to energies of order $N^2$,
and they are drawn 
in the large $N$ limit where the canonical ensemble exhibits sharp
transitions and the boundaries between phases in the microcanonical
ensemble are distinct. To understand in more detail the phases
available in the microcanonical ensemble at general energies
(including energies of order one), 
it is convenient to stretch out the low energy
regions of the three diagrams for large but finite N, as we do in
figures 5, 6, and 7 below.\foot{Strictly speaking, the boundaries between
phases are smoothed out when $N$ is finite, but for large enough $N$
these should look sharp as shown.}

\fig{Plot of $\log(T)$ as a function of $\log(E)$ (in the microcanonical
ensemble) when $b>0$.}
{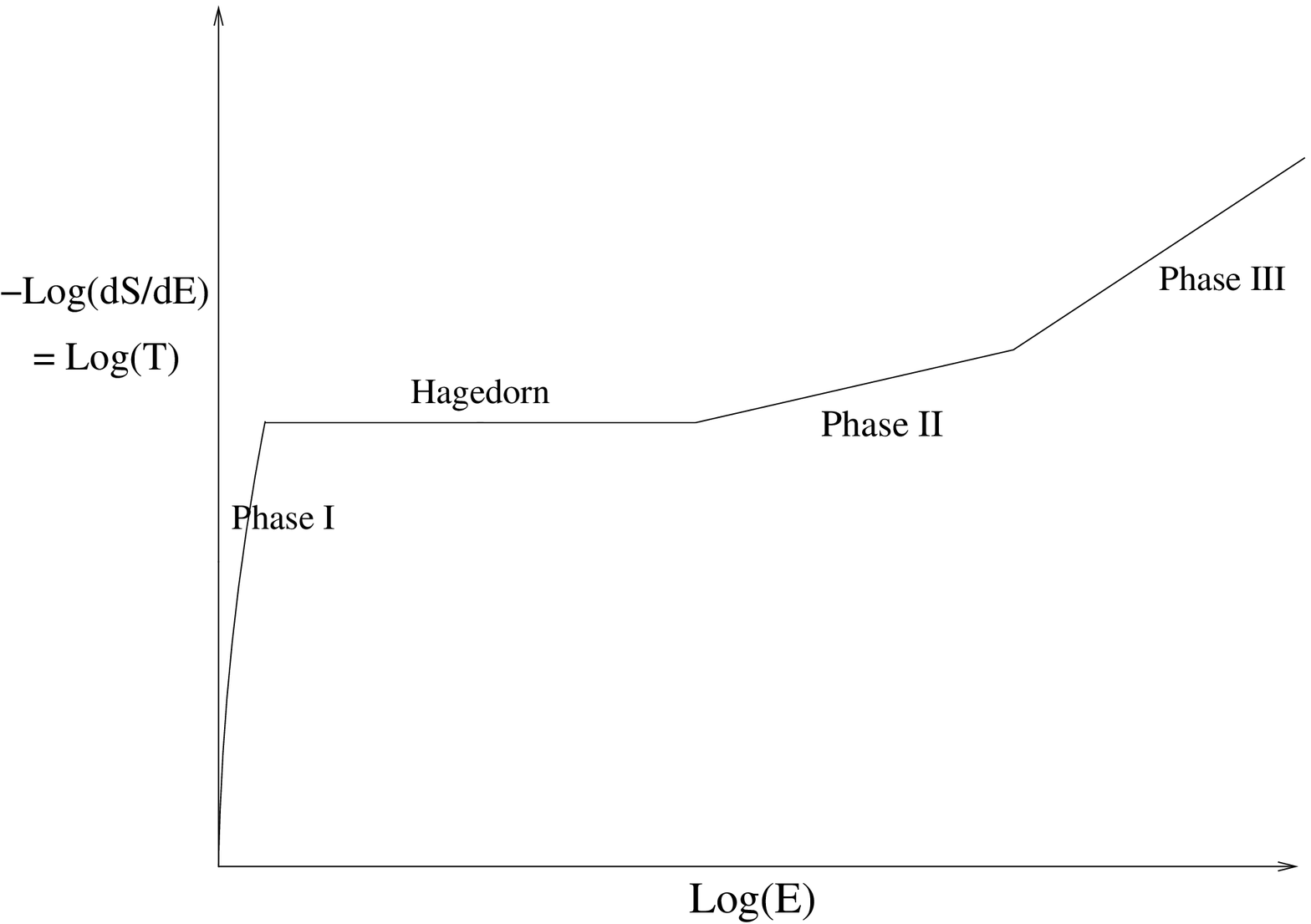}{4.0truein}
\figlabel{\posbmicro}
In the case $b>0$ this plot is displayed in figure \posbmicro.
As is apparent from the figure,
at any given temperature the theory has exactly
one phase available to it; this is consistent with the behavior described
by the series of Landau-Ginzburg diagrams in figure \secplots\ and with the
analysis of the toy model of the previous section.
For $T<T_H$ the theory is in Phase I. At $T_H$,
the theory undergoes a second order Hagedorn phase transition, emerging
into Phase II (the energy after the transition is of order $N^2$).
At a still higher temperature, the theory undergoes yet
another continuous phase transition into Phase III.

Note that  Phase I corresponds to $u_1=0$,
Phase II corresponds to the minimum of (for instance) the second plot in
figure \secplots,
and Phase III is at the boundary of the configuration space of
eigenvalue distributions.

\fig{Plot of $\log(T)$ as a function of $\log(E)$ (in the microcanonical
ensemble) in the free Yang-Mills theory.}
{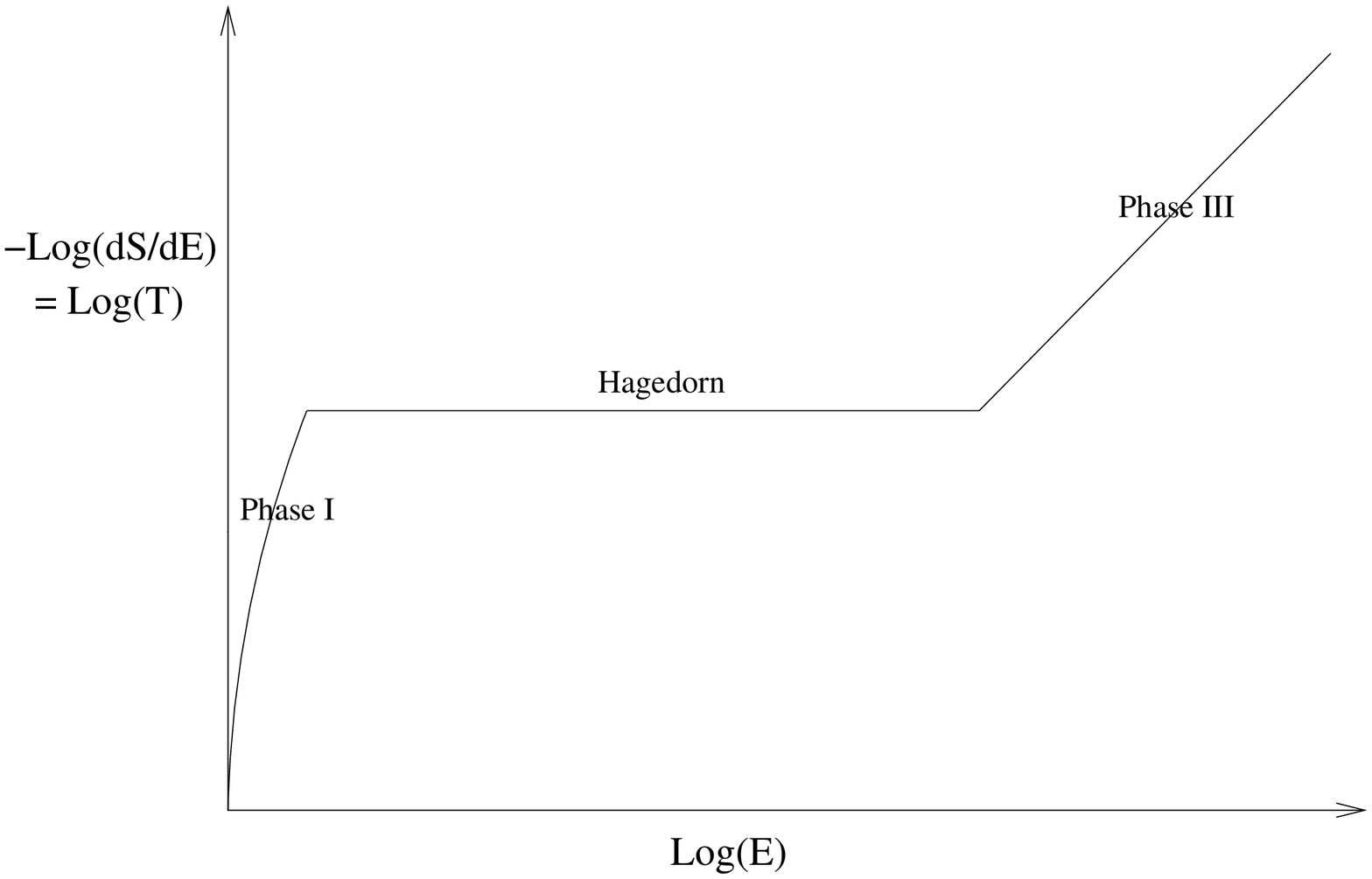}{4.0truein}
\figlabel{\zerobmicro}
The density of states of the free theory may be obtained from the limit
$b \to 0$ of the
previous discussion; the relevant plot takes the form shown in figure
\zerobmicro.
As is apparent from this figure, the free 
theory makes a phase transition (of first
order) directly from Phase I to Phase III at the Hagedorn
temperature, in agreement with the analysis of \S5.

\fig{Plot of 
$\log(T)$ as a function of $\log(E)$ (in the microcanonical ensemble)
when $b<0$.}
{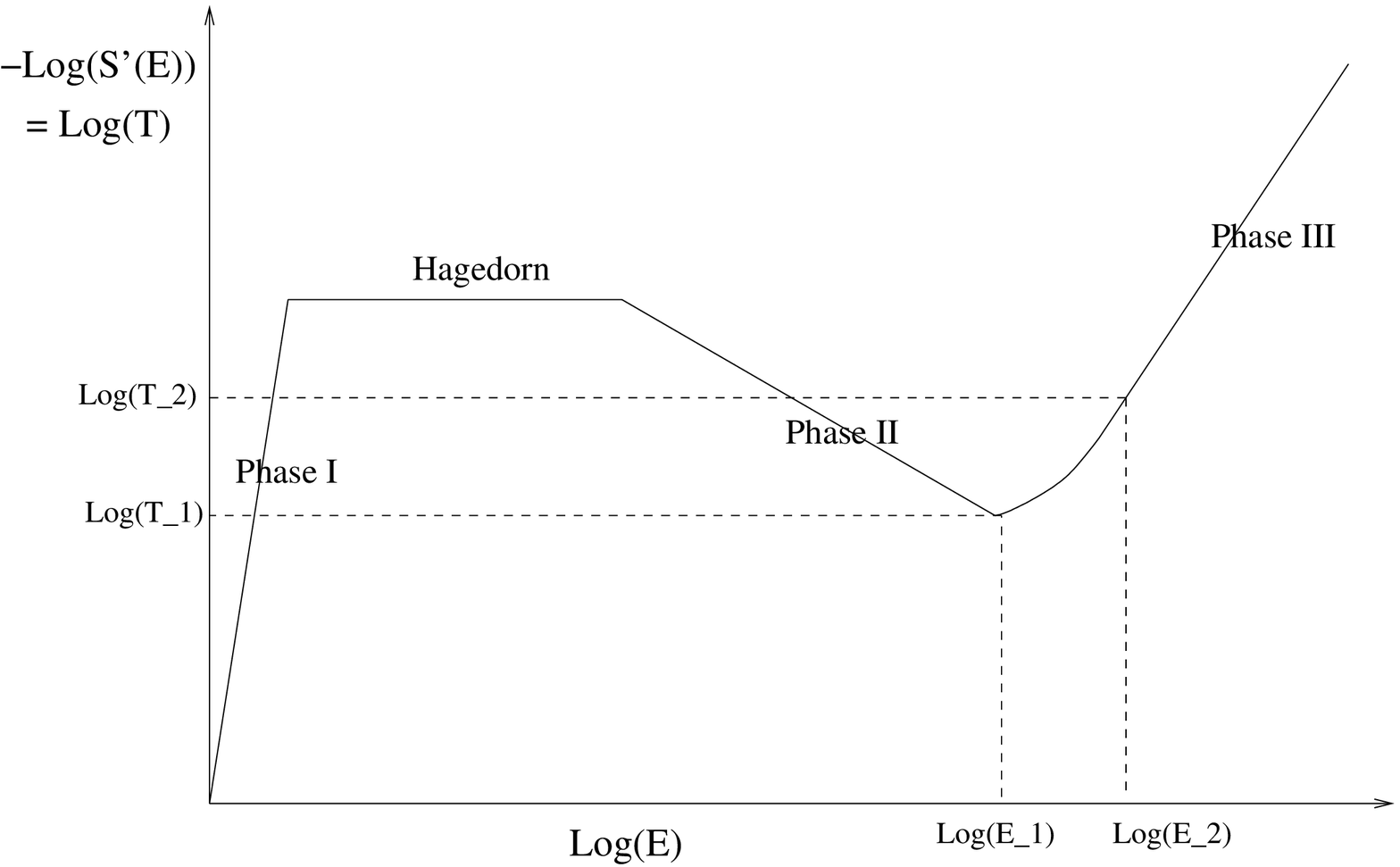}{4.0truein}
\figlabel{\negbmicro}
When $b<0$, we believe\foot{As noted above, when a theory undergoes a first 
order
transition, $Z(\beta)$ is insufficient to reproduce the full
density of states of
the theory. In order to obtain figure \negbmicro\ 
we use one additional assumption,
that the
unstable saddle points appearing in the Landau Ginzburg diagram
of figure \firstplots\ (see also
the toy model of the previous subsection) may be interpreted as an
unstable phase (of negative specific heat) at the corresponding
temperature, in the microcanonical ensemble.}
that the density of states is characterized by the
plot displayed in figure \negbmicro.
The thermodynamics induced by the density of states plotted in
figure \negbmicro\
is consistent with the behavior described by the Landau Ginzburg
plots in figure \firstplots. In particular, the five graphs shown in figure
\firstplots\
(from top to bottom) should be taken to represent $S_{eff}(u_1)$ for
$T<T_1$, $T=T_1$, $T=T_2$, $T_2<T< T_H$ and $T>T_H$, respectively. Phase
I lies at $u_1=0$ in figure \firstplots.
Phase II is the unstable maximum in the
3rd and 4th graphs in figure \firstplots,
and Phase III is the saddle point at the
boundary $u_1=\half$. Note that, according to both figures \negbmicro\ and
\firstplots,
Phase III exists only for $T>T_1$, and Phase II exists only
for $T_1 <T< T_H$.

In ending this section we would like to emphasize that adding
interactions has two qualitatively different effects on the spectrum
of weakly coupled Yang-Mills theories. First, single-trace states
involving a number of fields which is much smaller than $N$ 
pick up corrections to their energy, starting at 
first order in $\lambda$ (for conformal field theories on $S^{d-1}$ these
corrections are equivalent to the anomalous dimensions of the single-trace
operators). These corrections result in a 
renormalization of $m_1^2(T)$, and hence of the Hagedorn temperature. 
However, the physics of the phase transition (and, in particular,
the crucial sign of the coefficient $b$) is governed by a different
effect, which is the deviation from Hagedorn-type behavior in states which
involve a number of fields of order $N^2$. \foot{When the number of 
fields is of order $\sqrt{N}$ or larger, single-trace and multi-trace
states mix, and one cannot distinguish between them. This was 
highlighted recently in the study of the effects of non-planar graphs in the 
Yang-Mills/plane wave duality \ppint.} This effect appears even in
the free Yang-Mills theory, and it appears to be modified in a
qualitatively important fashion by coupling constant effects at 
$\CO(\lambda^2)$, as shown in figure 4. These modifications determine the sign of $b$ and the 
nature of the phase transition at finite coupling.  

\newsec{Extrapolation to strong coupling and the dual description}

In this section we will present possible extrapolations of the results
of section 6 to strong coupling. We will also discuss the possible
interpretation of these results in a stringy dual description.  
Much of this section is rather speculative; we will present 
some questions and outline possible answers.

\subsec{Possible phase diagrams for large $N$ Yang-Mills theories}

Consider the deconfining transition of a $d$-dimensional confining large $N$
Yang-Mills theory on a compact space of size $R$, at large values of
$R \Lambda_{QCD} \to \infty$. As we reviewed in \S2.2,
this transition may be either of first order
(as seems to be the case for pure Yang-Mills theory) or of
second order (in which case, for $d=4$
it must be in the universality class of the $d=3$ XY model).

\fig{The simplest phase diagram for a compactified
confining theory with negative
$b$ and a first order phase transition at $R=\infty$ .
Only solid lines represent phase transitions.}
{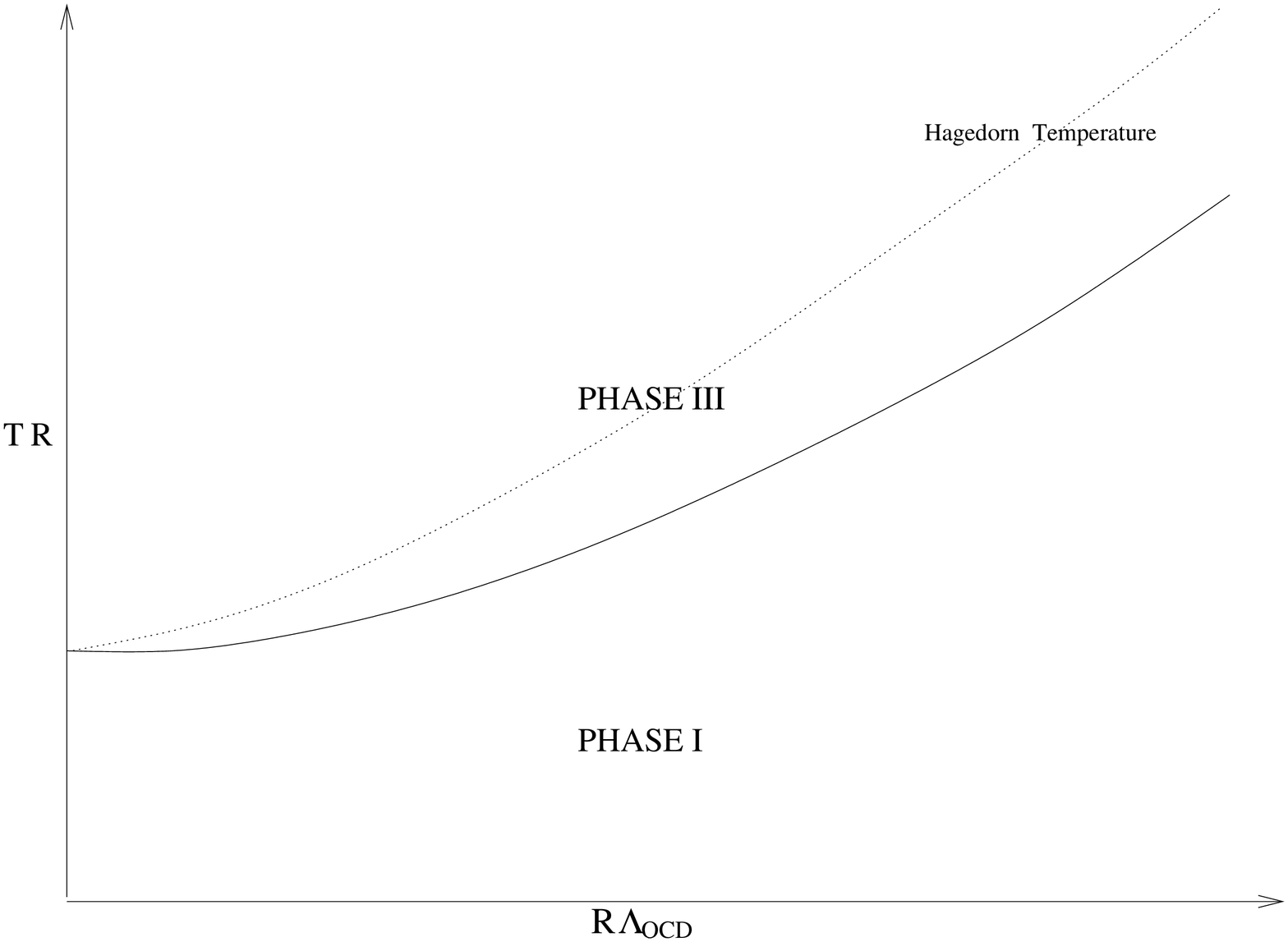}{4.0truein}
\figlabel{\confoneone}

\fig{The simplest phase diagram for a compactified
confining theory with positive
$b$ and a first order phase transition at $R=\infty$ .
Only solid lines represent phase transitions.}
{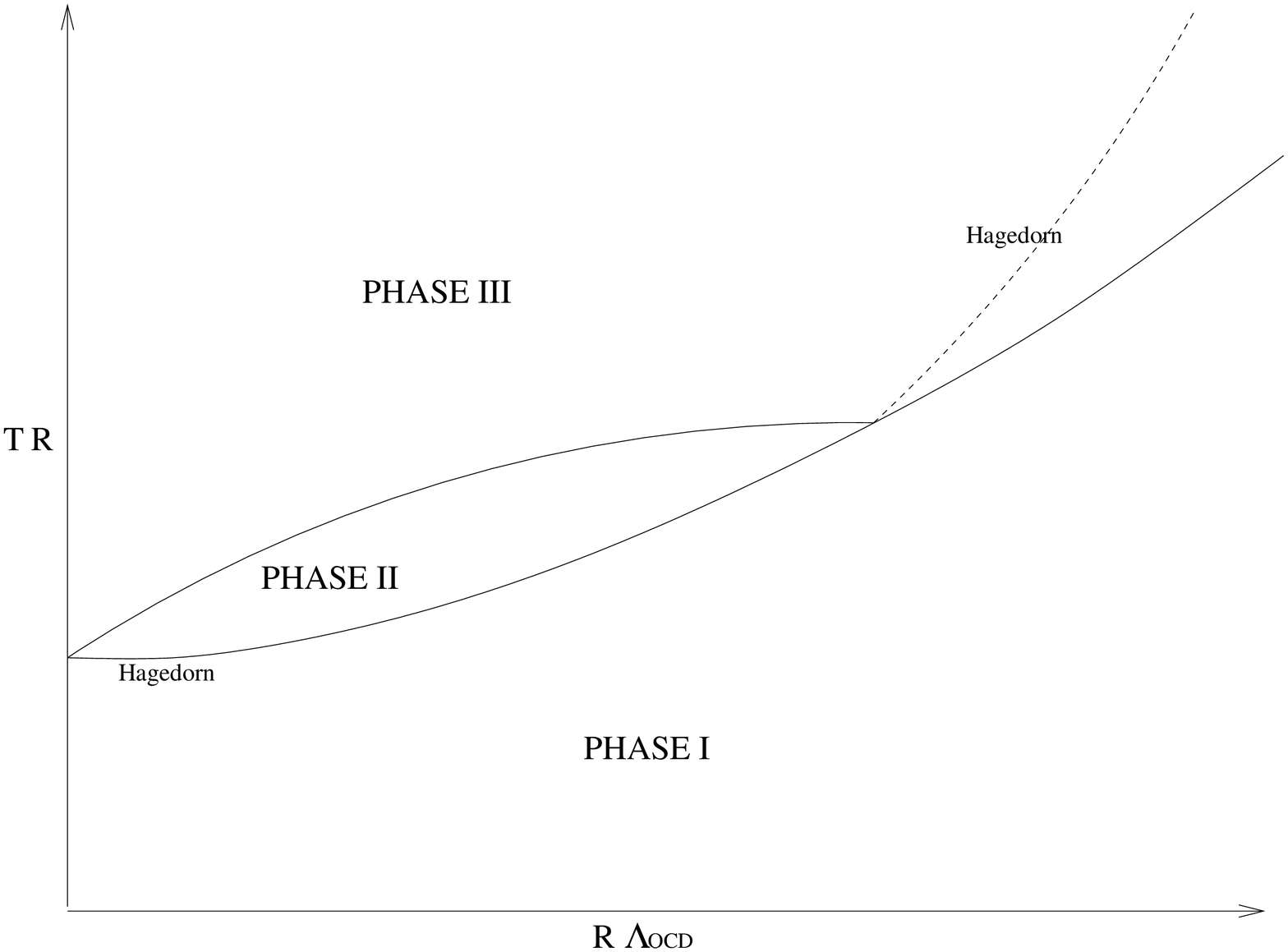}{4.0truein}
\figlabel{\conftwoone}

\fig{The simplest phase diagram for a compactified
confining theory with negative
$b$ and a second order phase transition at $R=\infty$ .
Only solid lines represent phase transitions.}
{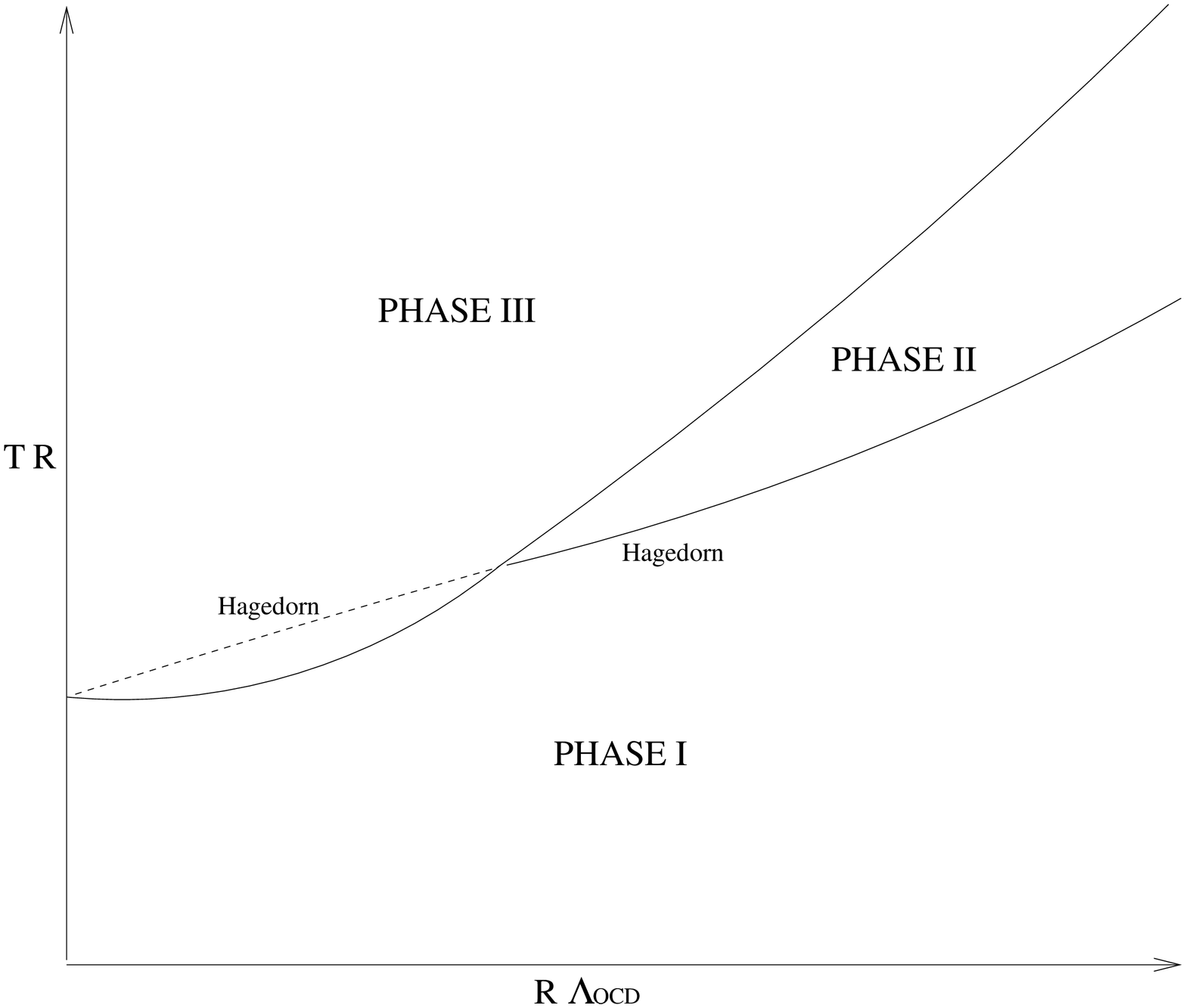}{4.0truein}
\figlabel{\confonetwo}

\fig{The simplest phase diagram for a compactified
confining theory with positive
$b$ and a second order phase transition at $R=\infty$ .}
{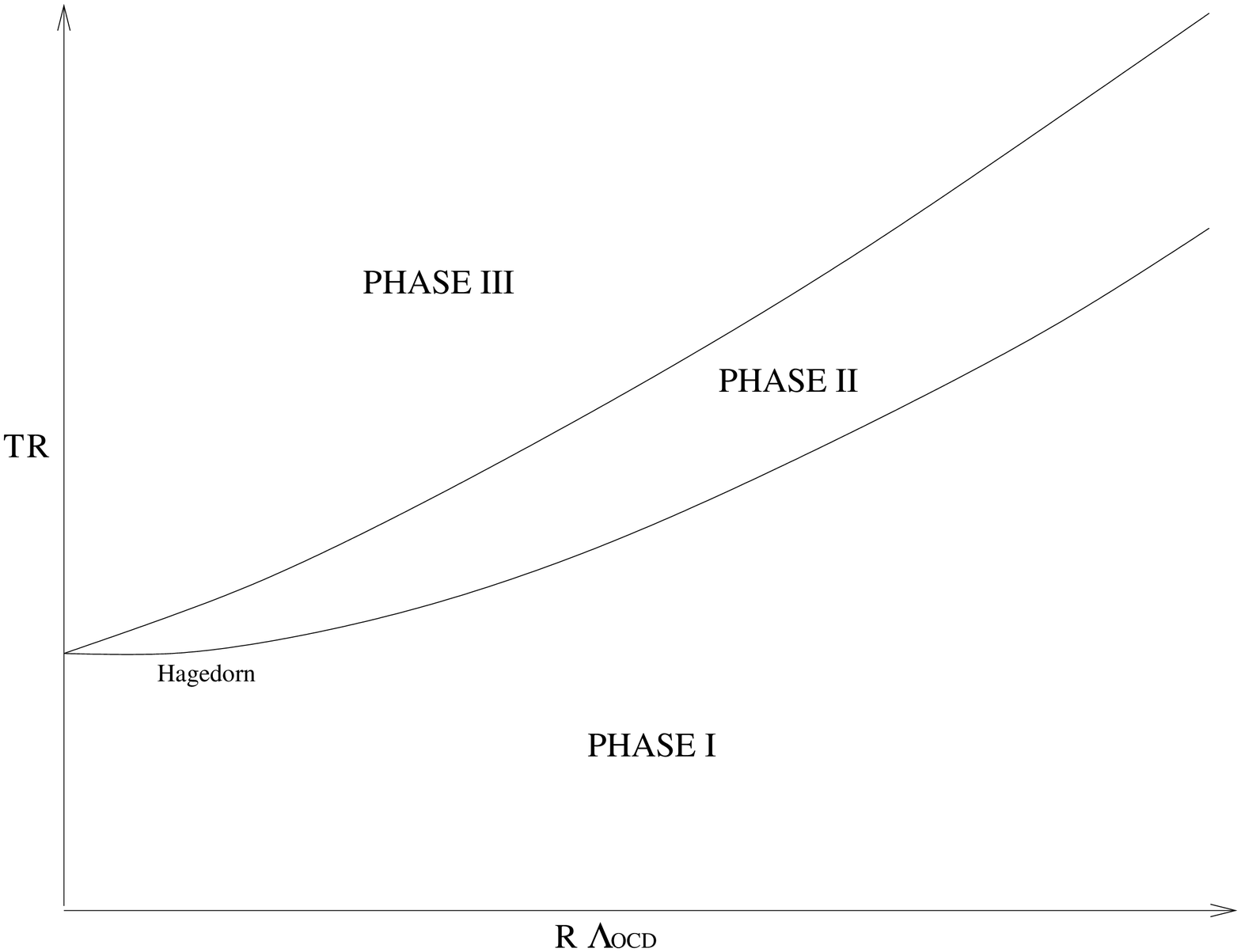}{4.0truein}
\figlabel{\conftwotwo}

On the other hand, as we saw in the previous section, at weak coupling
(small $R\Lambda_{QCD}$)
the transition may be of either first or second order, depending on
the (perturbatively computable) sign of $b$ for the theory in question.
For each of the four possibilities for the order of the phase transitions
at large and small $R\Lambda_{QCD}$ we have sketched
the simplest possible interpolation between the two limits,
and the corresponding phase
diagrams are displayed in figures \confoneone-\conftwotwo. 
Recall that
Phase I is a confined phase, characterized by a uniform distribution
of eigenvalues for the holonomy matrix $U$.\foot{In confining theories 
away from weak
coupling, when the size of the compact space becomes large compared to the
scale set by $\Lambda_{QCD}$, it is not as
useful to focus on the zero mode of the gauge field, since other modes
become light compared to $\Lambda_{QCD}$. 
However, while $S_{eff}(U)$ may be less relevant in
these more general cases, we can consider instead the effective action
$S_{eff}(U(y))$, where $U(y)$ is a spatially-dependent unitary matrix
given by the Wilson loop around the thermal circle at a point $y$ (see
\parteig\ for a related discussion). For large $N$ theories, we expect
that the saddle point configurations for every temperature (in all phases)
should be spatially
homogeneous, so that the corresponding $U(y)$ is constant. Thus, we
may still characterize the various phases by a single eigenvalue
distribution corresponding to this constant $U$.} On the other hand
Phases II and III are both deconfined phases, which are distinguished by the
fact that in Phase II the eigenvalue distribution is non-vanishing
on the whole unit circle, while it vanishes outside an interval in 
Phase III. 
\foot{
Alternatively, as mentioned for instance in \GattringerFI, they can be
distinguished by the behavior of the Polyakov loops for higher 
representations of the gauge group, such as the $k$'th product of 
fundamental representations for large $k$.}

Note that in each of the figures \confoneone-\conftwotwo\
all phase transition temperatures
at strong coupling are of $\CO(\Lambda_{QCD})$; consequently, in this
regime the dimensionless phase transition temperature $TR$ increases
linearly with the coupling parameter $R \Lambda_{QCD}$.

Since lattice data indicates that $3+1$-dimensional
pure Yang-Mills theory undergoes a single
first order deconfining phase transition at large $N$,
the simplest possibilities for the
phase diagram of the $3+1$-dimensional large $N$ pure Yang-Mills theory
on a compact space such as $S^3$
are depicted in either figure \confoneone\ or figure \conftwoone;
as above, the value of $b$ for this
theory (which we will report on in \future) will
distinguish between these two options.

Above we drew the simplest possible interpolations between the weak coupling
and strong coupling behaviors.
These simple interpolations are consistent with the values of all of the
order parameters we discussed at weak and strong coupling, but
several other rather natural phase diagrams may also be drawn. For instance, 
figures \confoneone-\conftwotwo\ do not apply to any theory that undergoes 
a phase transition as a function of the coupling at zero temperature. 
Such a phase transition is certainly possible, and could be inserted 
into any of the diagrams of figures \confoneone-\conftwotwo\ as a vertical 
line that divides Phase I into two regions, distinguished, presumably, 
by an order parameter unrelated to the Polyakov loop. A case where we
know that this happens is when the infinite volume theory exhibits global
symmetry breaking due to quantum effects (for example, the $d=4$ $\CN=1$ 
$SU(N)$ SYM theory, in which a chiral $\IZ_{2N}$ symmetry is spontaneously
broken to $\IZ_2$ at infinite volume). Since we see no sign of such a 
symmetry breaking in our analysis 
at weak coupling\foot{Of course, as in our discussion 
above of the breaking of the $\IZ_N$ symmetry associated with confinement, 
we do not
expect to see symmetry breaking at finite $N$ and finite volume, but
we expect to see a sum over different configurations in which the
symmetry is broken.}, it seems clear that in such cases
there are additional phase boundaries separating 
weak coupling and strong coupling regimes at low temperature. In general, we do not expect any such boundaries for asymptotically free gauge theories at high temperatures, since in this limit, the theories may be studied perturbatively for all values of $R \Lambda_{QCD}$.

We now turn to the large $N$ $SU(N)$ $\CN=4$ SYM theory on an $S^3$ of unit
radius and at strong 't Hooft coupling $\lambda$. Using the AdS/CFT
correspondence, this theory is equivalent to type IIB string theory at
weak string coupling on an $AdS_5\times S^5$ space (using global coordinates
for $AdS$) with a large radius
of curvature. The thermodynamical analysis of this theory shows that it
undergoes a single first order phase
transition as a function of temperature, which occurs at $T_{HP}={3\over 2
\pi} \simeq 0.477465$ \refs{\HawkingDH,\WittenZW}. In addition, since
the string coupling is small,
we have a large range of energies with a Hagedorn behavior of the
spectrum, with the Hagedorn temperature scaling as $T_H \propto 1 /
\sqrt{\alpha'} \propto \lambda^{1/4}$ and going
to infinity in the limit of strong 't Hooft coupling.

On the other hand, as we have described in detail above,
at $\lambda=0$ this theory also undergoes a single phase transition as a
function of temperature. This phase transition, at the temperature
$T_H = -1/\ln(7-4\sqrt{3}) \simeq 0.379663$, coincides with the Hagedorn
transition and it is weakly of first order. And, as discussed in \S6, the
behavior of the theory at small $\lambda$ depends crucially on the sign of
$b$ (defined in \compseffu). If $b$ turns out to be negative, the theory
undergoes a single first order phase transition below the Hagedorn
temperature. If $b>0$, the theory undergoes two continuous phase
transitions, the first of which is at the Hagedorn temperature.

\fig{Conjectured phase diagram for the $\CN=4$ SYM theory on a
sphere if $b$ is negative. Only solid lines represent phase transitions.}
{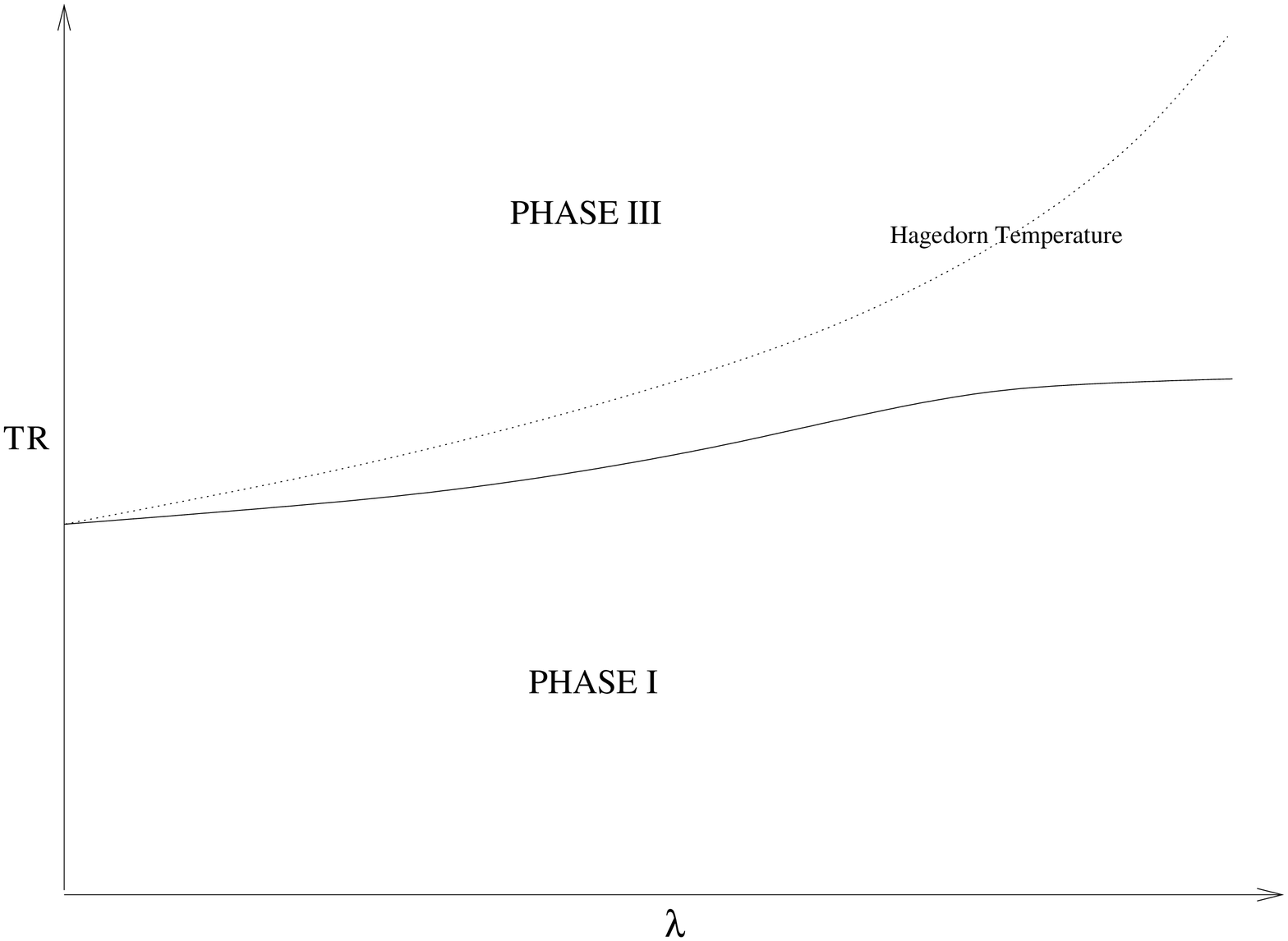}{4.0truein}
\figlabel{\nfourfirst}
As we have described above, the value of
$b$ for this theory is determined by a set of two-loop and
three-loop graphs.
We will present the result of this computation in \future.
If $b$ turns out to be negative, the behaviors at weak and strong
coupling are similar and the phase diagram of the theory has a natural
interpolation (shown in figure \nfourfirst) for all
$\lambda$ (though more complicated phase diagrams are also possible).

\fig{Conjectured phase diagram for the $\CN=4$ SYM theory on a
sphere if $b$ is positive. Only solid lines represent phase transitions.}
{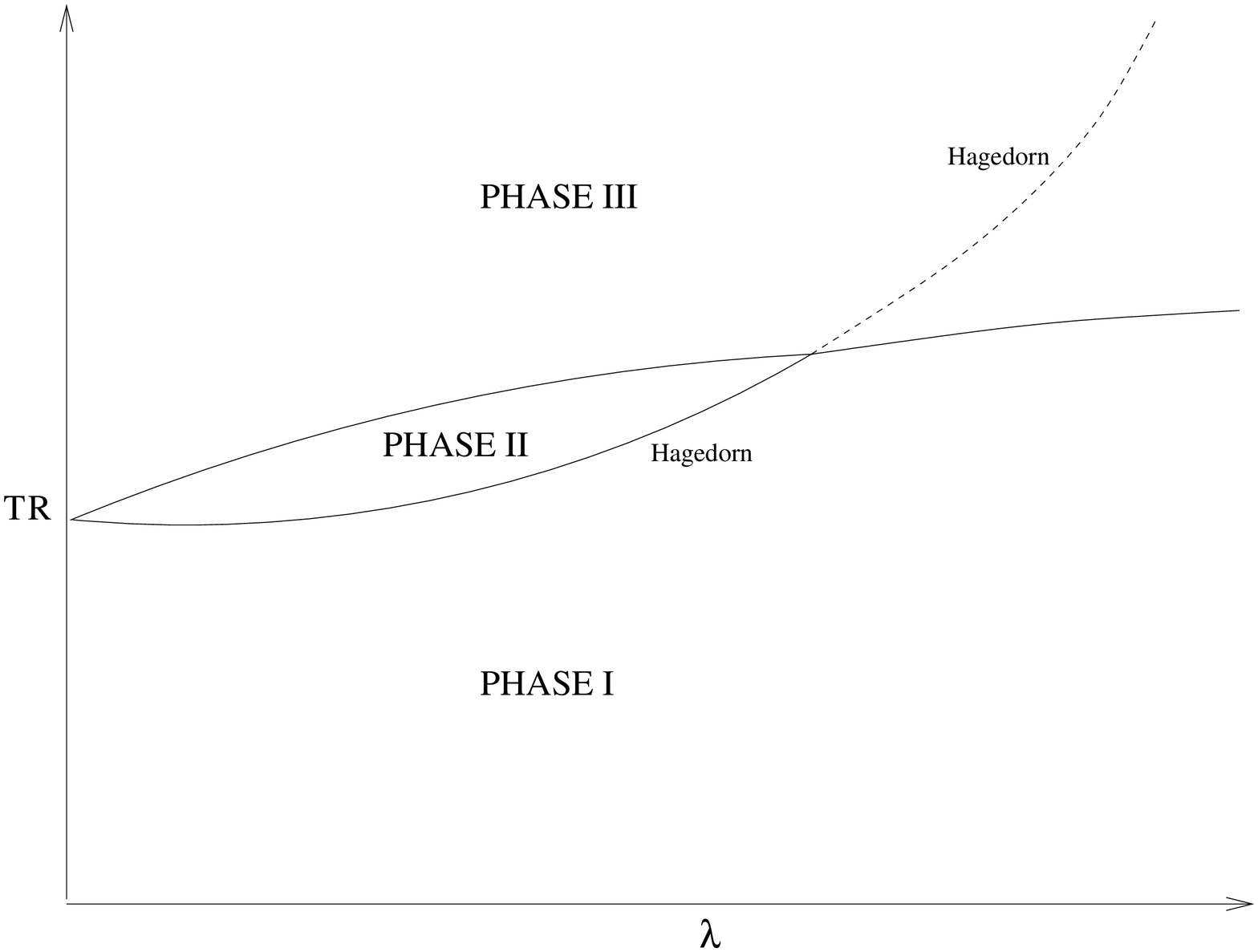}{4.0truein}
\figlabel{\nfoursecond}
On the other hand, if $b$ turns out to be positive, the simplest 
possible phase
diagram takes the form shown in figure \nfoursecond. Note the existence
of a tri-critical point at a special value of the 't Hooft coupling,
at which the deconfinement
phase transition changes from being second order to being
first order, and the
existence of a new phase at weak coupling and intermediate
temperatures.
\foot{Figures \nfourfirst\ and \nfoursecond\ have been drawn under the
assumption that the large $N$ $\CN=4$ theory in flat space does not undergo a
phase transition as a function of the coupling. Such a transition could
perhaps occur, and then it would be depicted by a vertical line that 
would divide Phase III into two different regions in figures 
\nfourfirst\ and \nfoursecond. 
We thank R. Gopakumar for emphasizing 
this possibility to us.
Similarly, it is possible in principle
that Phase I would be divided into different regions. }

Figures \nfourfirst\ and \nfoursecond\ are very similar to figures
\confoneone\ and \conftwoone, with one important
difference : the phase transition temperature, measured in units of $1/R$,
has a finite strong coupling limit for the $\CN=4$ theory,
but increases without
bound for confining theories.

Note that the interaction potential between a quark and an anti-quark,
at a distance $L \ll R$,
is expected to behave rather differently in the various phases
described above. In the case of confining gauge theories at low temperatures, 
this potential will be Coulomb-like at weak coupling ($R \Lambda \ll 1$)
but linear (string-like) for $R \Lambda \gg 1$, as long as $L \gg 1/\Lambda$.
However, in Phase III
(and presumably in Phase II as well) the quark-anti-quark force is
exponentially screened by the intervening plasma. In the case of the
$\CN=4$ SYM theory the quark-anti-quark force is Coulomb-like at every
coupling in the `confining' Phase I, but decays exponentially at
sufficiently large distances (compared to a scale set by the
temperature) in the deconfined phases.

\subsec{Dual interpretation of the $\CN=4$ SYM thermodynamics at strong
coupling}

We noted in \S2 that it should be possible to understand
the possible phase diagrams presented earlier in this section in terms of
a dual string theory description of the relevant gauge theories. Unfortunately,
among the $3+1$ dimensional gauge theories with adjoint fields, the only one
whose string dual is known is the $\CN=4$ supersymmetric Yang-Mills theory (and
theories related to it by renormalization group flows),
and this dual is mostly understood only in the limit of strong 't Hooft
coupling. In this subsection we will argue that, at least in this one case,
the thermodynamics of the string theory dual
fits rather nicely with the picture presented in this paper. Previous
discussions of the thermodynamics of $\CN=4$ SYM as a function of the
coupling, which are consistent with ours, appear in \eliezer, and a detailed
discussion of the transitions discussed in this subsection appears in
\refs{\BermanAB,\BarbonDI,\BarbonNW}.

The string dual to $\CN=4$ SYM on $S^3 \times \IR$ (where the $S^3$ is
taken to be of unit radius) is type IIB string theory on $AdS_5\times S^5$,
whose metric may be written as
\eqn\gm{ds^2=R_0^2\left(-\cosh^2\rho\,d\tau^2+d\rho^2+\sinh^2\rho\,
d\Omega_3^2 + d\Omega_5^2\right). }
Under this duality the Hamiltonian of the Yang-Mills theory is
identified with the generator of global time translations
$\partial_{\tau}$ in the geometry \gm. The energy $E$ 
of the gauge theory on a sphere of unit radius is related to the proper
energy in string
theory at $\rho=0$ by $E_{prop}= E/R_0$. The radius
$R_0$ of the $AdS_5\times S^5$ space is related to the 't Hooft coupling
$\lambda$ and the inverse string tension $\apm$ by $R_0
\simeq \lambda^{1\over 4} \sqrt{\apm}$ (we will ignore all numbers of
order unity in the qualitative discussion of this subsection).

For large $R_0/\sqrt{\apm}$ (large 't Hooft coupling)
the density of states $\rh(E)$ of type IIB string theory on \gm\
has four distinct regimes (see \adsreview\ and references therein).
The only states in the spectrum with proper
energy\foot{Recall that states of finite energy
on \gm\ are all localized about $\rho=0$. $AdS_5$ space behaves effectively
like a four dimensional box of physical radius $R_0$, as is apparent from the
fact that the spectrum of $\partial_{\tau}$ is discrete with
discretization step unity.} smaller than the string scale,
$E \ll \lambda^{1/4}$, are ten
dimensional supergravitons\foot{In terms of the gauge theory,
the only states in this regime are those created by the chiral primary
operators, their products and their descendants.} whose entropy scales as
$S(E)\propto E^{9/10}$.  For $E\gg\lambda^{1/4}$, excited string
states are added to the spectrum; the contribution of these states to
the entropy is $S\simeq E_{prop} \sqrt{\apm} \simeq \lambda^{-1/4}E$
(see \S2.1), and this contribution
dominates over the
graviton gas for $E\gg\lambda^{5/2}$. When the proper energy exceeds the
Planck mass (namely, $E\gg m_sR_0/g_s^{1/4}=N^{1/4}$) small ten dimensional
Schwarzschild
black holes are also added to the spectrum (note that these energies are
inaccessible in the strict $N \to \infty$ limit). The entropy of such black
holes is proportional to $S\sim\left(l_PE/R_0\right)^{8/7}=
\left(E/N^{1/4}\right)^{8/7}$; they have negative specific heat and
positive free energy. This entropy of black holes dominates over that of the
Hagedorn strings for $E \gg N^2/\lambda^{7/4}$. Finally, at an energy
$E_1\sim N^2$ (an energy at which the radius of these
Schwarzschild black holes becomes comparable to the $AdS$ radius $R_0$) the
black hole horizon covers the whole $S^5$, and the
specific heat of these black holes becomes positive. For $E>E_1$ these black
holes are referred to as  $AdS$-Schwarzschild or big black holes. At a
higher energy $E_2$ (also of $\CO(N^2))$ the free energy of these black holes
becomes negative. Finally, for $E\gg N^2$, the entropy of these black holes
is $S\propto N^{1/2}E^{3/4}$, so the thermodynamics of these big black holes
resembles that of a four dimensional conformal field theory.

\fig{$\log(T)$ as a function of $\log(E)$ (in the microcanonical ensemble)
for all the `phases' of type
IIB string theory on $AdS_5\times S^5$, when $R_0 \gg \sqrt{\apm}$ .}
{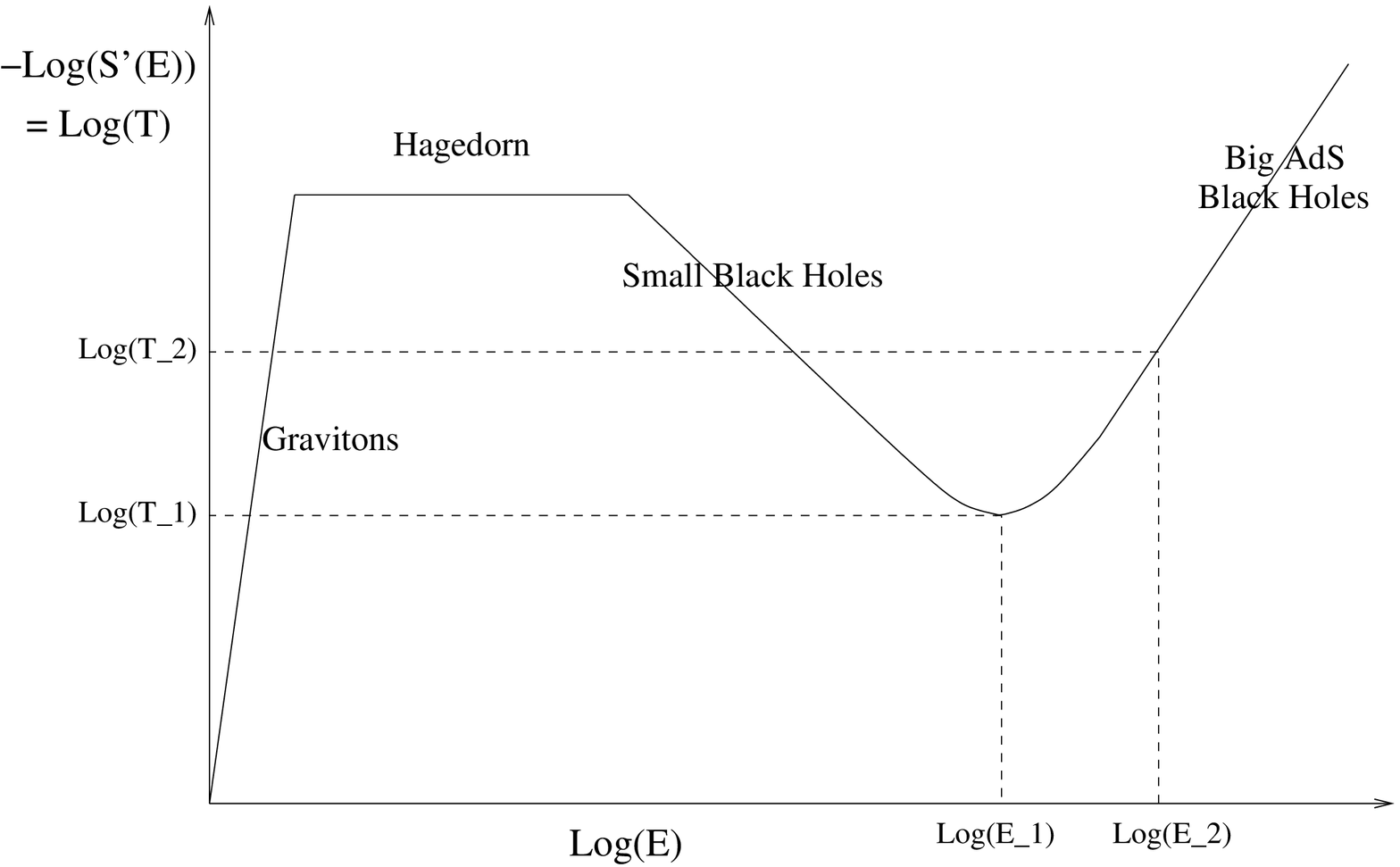}{4.0truein}
\figlabel{\adsphases}

For comparison with the canonical ensemble it is most convenient
(see \S6.5) to display the density of states described in the previous
paragraph as a plot (see figure \adsphases) of $\ln(T(E))=-\ln(S'(E))$
\BarbonDI. Quite
remarkably, figure \adsphases\ is identical in its general form to
figure \negbmicro, even though
figure \negbmicro\
was derived for a general gauge theory (with $b < 0$) at weak coupling while
figure \adsphases\ applies to the $\CN=4$ Yang Mills theory
at strong coupling, and
has been derived using its stringy dual.

To complete this subsection we now use figure \adsphases\ to discuss
the behavior of
the $\CN=4$ SYM theory at strong coupling in the canonical ensemble \BarbonDI\ 
(see \S6.5 
for a very similar discussion). For $T<T_1 (={\sqrt{2} \over \pi})$ the
only available saddle point is the thermal gas of gravitons. Over the
temperature range $T_1<T < T_H \simeq \lambda^{{1\over 4}}$ the theory
has three saddle points to  choose from. One of these saddle points
(small black holes) is unstable, so it cannot appear in the canonical
ensemble. The other two saddle points are stable and compete with each
other. The gas of gravitons has lower free energy for
$T_1<T<T_2 (={3 \over 2 \pi})$, while the big black hole dominates the
canonical ensemble for $T_2<T< T_H$. Consequently, the theory should undergo
a first order phase transition at $T=T_2$, which (as we mentioned above) is
indeed the case \refs{\HawkingDH,\WittenZW}.
$T=T_H$ is the Hagedorn
temperature at which the graviton gas saddle point stops existing. For
$T>T_H$ the generic state of the theory continues to be a big black hole,
which is now the only available phase.

\subsec{Deconfinement and black holes}

We do not, as yet, understand the string dual of weakly coupled gauge
theories, such as pure Yang-Mills or $\CN=1$ supersymmetric Yang-Mills on
a small $S^3$, or even the $\CN=4$ SYM theory at weak 't Hooft
coupling. Nonetheless, each of these theories is continuously related to
the $\CN=4$ theory at strong coupling, so the AdS/CFT
correspondence provides a demonstration (in the physicist's sense of the
term) of the existence of string duals for all these gauge theories.
Furthermore, these string duals may be expected to qualitatively resemble
type IIB string theory on $AdS^5\times S^5$, in the sense that they should
all include at least one additional ``holographic'' dimension beyond the
dimensions of the Yang-Mills theory. Consequently, even though we
understand little about the details of the string duals of arbitrary
gauge theories, we will use intuition from the previous subsection to
boldly speculate on the stringy dual interpretation of the phase diagrams in
figures \confoneone-\nfoursecond.

As we reviewed above, the deconfined phase of the
strongly coupled $\CN=4$ theory admits a dual description in terms of black
holes \foot{  The connection between
deconfinement transitions and black holes has also been verified in
examples of confining field theories in flat space,
such as the duality between a cascading $d=4$ $SU(N+M) \times SU(N)$ $\CN=1$
supersymmetric gauge theory
and type IIB string theory on a manifold which includes the resolved 
conifold \refs{\kleb,\conifold}. 
Similar results for the Maldacena-Nunez background
\MaldacenaYY\ were obtained in \mnbh, and for other theories in (for
example) \others.}. 
We would now like to argue that this is a general feature (see 
\SundborgUE\ for related comments):
large $N$ deconfined phases (Phases II or III in figures
\confoneone-\nfoursecond) should always
be associated with black holes \foot{By this we mean that if the background
has a
geometrical interpretation it would resemble a black hole; we generalize
this to other backgrounds by calling such phases ``black holes''.}.
Our simple argument follows directly from ideas presented
in \WittenZW.

In order to measure the Polyakov loop we need to put an external quark
on a trajectory that wraps around the circle of the time direction. Since
in the large $N$ gauge/string correspondence quarks appear at the boundaries
of the worldsheet, we expect the Polyakov loop in string theory to be realized
as the partition function (in Euclidean space) for a string whose boundary
wraps around the time-like loop, and this is indeed the case in the
AdS/CFT correspondence \refs{\maldawil,\reyyee}\foot{In this
case the boundary of the string is
actually not precisely the Polyakov loop, as it contains also couplings to
scalar fields. We expect that these additional couplings will not change the
qualitative features of the order parameter, and that these ``generalized
Polyakov loops'' will vanish if and only if the usual Polyakov loop vanishes.
Computations of such ``generalized Polyakov loops'' in the AdS/CFT 
correspondence first appeared in \refs{\WittenZW,\ReyBQ,\BrandhuberBS}.
}.
This partition function clearly vanishes if the
Euclidean time circle is non-contractible\foot{Again, this is true when we
have a good geometrical description of the background, and otherwise we take
the vanishing of the Polyakov loop to be the definition of a non-contractible
circle.}, but it is generically non-zero (when suitably defined,
as in our discussion of \S5.7 and the analogous discussions in 
\refs{\WittenZW,\AharonyQU})
if the time circle is contractible. In all the examples that we know
of, a static Euclidean time cycle is contractible in general 
relativity only when the 
corresponding Minkowskian solution has a horizon. Outside the range of
validity of general relativity it would seem reasonable to define black
holes by this requirement. Putting all this together, we conclude that suitably
defined Polyakov loops are non-zero only in a black hole phase. Thus, we
generally expect that the string theory (at finite temperature)
in Phases II and III will be in
a generalized black hole background, while the theory in Phase I (where the
time cycle is non-contractible) should be in a background similar to the naive
gauge theory background (in which the time direction is simply compactified).

A similar conclusion follows by analyzing the other order parameter for
confinement,
$\lim_{N\to \infty} F(T) / N^2$. In string theory this order parameter maps
to the sphere partition function, which naively always vanishes. However, this
is not true even in the low-energy general relativity approximation for
some types of unbounded spaces \refs{\HawkingDH,\WittenZW}, where the naive
classical action diverges and one needs a subtraction procedure to define it.
This can lead to a non-zero result for the relative classical partition
function between two spaces with the same asymptotics but a different interior.
This again suggests that deconfined phases must involve different space-times
than the confined phase (which are the same asymptotically but differ in the
interior), which have different classical actions. The discussion of the
previous paragraph suggests that these spaces should be generalized black
holes.

\subsec{Dual description at a general point in the phase diagram}

At any fixed value of the coupling in each of the diagrams of figures 
\confoneone-\nfoursecond, 
the compactified large $N$ gauge theory either
\item{(1)} Undergoes a single first order phase transition as a function of
the temperature away from a tri-critical point.
\item{(2)} Undergoes a single first order phase transition as a function of
the temperature at a tri-critical point.
\item{(3)} Undergoes two successive continuous phase transitions as a
function of the temperature.

\noindent
It is possible (see below for a caveat) that the density of states 
takes the form
shown in figures \negbmicro, \zerobmicro\ and 
\posbmicro, respectively, in every realization of the three 
cases listed above. We have already argued that this statement is true at weak 
coupling (see \S6.5), and for the $\CN=4$ SYM theory at strong coupling. 

If this picture turns out to be correct, it would be
tempting to further speculate that the dual description of Phase III
always involves a stable `big black hole'. Following this train of
thought, one would also be led to speculate (following \S7.2) 
that Phase II has a dual
description in terms of a different type of strange `black
holes' (see \S7.3 for our usage of the term black holes). It would
then follow that  the discovery of theories of type (3) 
(theories with a second order large $N$ deconfinement 
phase transition) would imply the
existence of a new class of {\it stable} `strange black holes' in a class of 
string theories. This would certainly be intriguing; it would be very
interesting to understand the string theory interpretation of the
distinction between the two classes of `black holes' (Phase II and Phase III).

Before concluding this section, we should note that the different
theories and phases which we grouped together in our discussion above do
seem to have some qualitative differences, despite sharing some
similar features. Let us first study the behavior of a confining
gauge theory (like $d=4$ pure Yang-Mills theory) as a function of coupling in
more detail. On a very large compact space 
(in units set by $\Lambda_{QCD}$) the pure
Yang-Mills theory behaves locally just like the infinite volume theory, 
and we expect that the deconfinement transition proceeds (as energy is 
added to the system at the transition temperature) by nucleation of 
bubbles of the deconfined phase, which grow and eventually coalesce to 
cover the whole space. In particular, in this strong coupling limit 
there should be stable configurations at the
transition temperature characterized by the coexistence of confined
and deconfined phases covering different regions of the compact 
space.\foot{Note that 
in our discussion above of microcanonical phase diagrams (and, indeed, in our
whole analysis which was based on the effective action $S_{eff}(U)$), 
we were implicitly discussing only homogeneous phases. It seems that
at least in some cases non-homogeneous configurations dominate at 
strong coupling (in the microcanonical ensemble) over the 
configurations we described above.} In such mixed configurations we
would expect the size of every phase bubble to be at least of order 
$\Lambda_{QCD}^{-1}$; in particular, mixed phases may be expected to
be absent on a compact space that is small in QCD units, namely at weak
coupling $\Lambda_{QCD} R \ll 1$.\foot{The fact that the 
qualitative behavior of Yang-Mills theories  changes in this respect  
in the transition from weak coupling to strong coupling may indicate 
that the naive interpolation we suggested between these two regimes 
is too optimistic.}
 This is consistent with the analysis in our
paper; recall that the order parameter we used for the phase
transition is (roughly) the constant mode of  $\vev{{1\over N}\tr(U(y))}$ 
on the compact space, and that at weak coupling
modes that describe inhomogeneous configurations of $\tr(U(y))$ are
massive with a mass of the same order as the phase transition 
temperature, and may safely be integrated out, as we have done.

The situation is completely different in the case of the $d=4$ ${\cal N}=4$ 
SYM theory. In this theory the existence of a phase
transition relies upon the compactness of the space at all values of
the coupling, and the behavior at weak coupling and at strong coupling
seem similar. Recall that, 
in the strongly coupled $\CN=4$ theory, the dual description of the
high temperature phase is a single large black hole. Since there
would seem to be no such thing as half of a black hole,
the high temperature phase cannot coexist with a low temperature 
phase at the transition temperature at strong coupling.

The contrasting behaviors of the systems described in the previous two
paragraphs suggest that the
``large black hole'' phase (Phase III) of strongly coupled pure 
Yang-Mills theory may be
significantly different from the large AdS black hole.
It would be interesting to understand what sort
of bulk description the coexistence of confined and deconfined phases in 
confining theories such as the pure Yang-Mills theory could have.
One is reminded of the appearance of non-homogeneous configurations in
the context of the Gregory-Laflamme transition \GregoryVY.

\newsec{Discussion and future directions}

In this paper we have analyzed the thermodynamics of weakly coupled
large $N$ gauge theories compactified on a sphere of radius $R$, or any other
compact manifold on which the theory has no zero modes. Our analysis
applies both to conformal gauge theories at small values of a tunable
coupling constant (such as the $\CN=4$ SYM theory) and to confining theories
(such as pure or $\CN=1$ Yang-Mills theory) with $R \Lambda_{QCD} \ll 1$.
We have shown that in the microcanonical ensemble
these theories exhibit an exponential (Hagedorn) density of states,
which is cut off at an energy of order $E \sim N^2$ (this was previously
shown for free gauge theories in \refs{\SundborgUE,\PolyakovAF}). The Hagedorn
temperature at zero coupling is easily determined from the field content of the
theory, and the corrections to it can be computed in perturbation theory.
We have demonstrated that these theories undergo deconfinement
phase transitions, with different possible phase diagrams discussed in \S6.

Our analysis has several points of interest. First, putting a confining
gauge theory on a compact space introduces a new dimensionless 
parameter into the
game; varying this parameter may continuously deform 
the flat space deconfinement
transition into a regime where the transition (which remains non-trivial)
may be reliably studied in perturbation theory\foot{This is somewhat
analogous to the continuous relation between confining phases and
perturbatively-accessible Higgs phases which can occur \BanksFI\ in theories
with fields in the fundamental representation upon changing coupling constants
or vacuum expectation values of fields.}. The analysis of this
weakly coupled problem leads to several insights. For instance, we have
argued that if the large $N$ deconfining phase transition is of
second order then it must be Hagedorn-like (as first shown from different
arguments in \PisarskiDB),
and it must be followed by a second phase transition at
a higher temperature (where the eigenvalue distribution corresponding to a 
holonomy around the thermal circle develops a gap). This suggests that a 
second order deconfining transition
in any large $N$ gauge theory
implies the existence of a previously unsuspected
intermediate temperature phase. Our analysis of the
weak coupling deconfinement transition
led to us suggest the four simplest possible phase diagrams (figures
\confoneone-\conftwotwo) for
confining large $N$ gauge theories; which diagram is actually implemented
depends on the details of the theory, including the sign of a perturbatively
computable number. More speculatively,
we have also suggested a stringy
interpretation (in terms of nucleation of black holes in the dual string
theory) of this deconfinement transition.

Turning now to the $\CN=4$ supersymmetric
Yang-Mills theory, our results imply that the
thermal phase transition in this theory at weak coupling is either of first
or second order, depending on the sign of a coefficient that we are now in
the process of computing. If the transition turns out to be of first order,
the simplest possible conclusion would be that the Hagedorn spectrum of
string theory on $AdS_5 \times S^5$ does not dominate the thermodynamics
of $\CN=4$ SYM at any temperature and at any non-zero value of
the coupling -- this was called ``Hagedorn censorship'' in \eliezer.
In such a case our computations would
enable us to study an unstable saddle point representing substringy
unstable (`Schwarzschild') black holes in this background.
On the other hand, if the weak coupling
transition turns out to be of second order, that would suggest the
existence of a tri-critical point at some finite 't Hooft coupling
(probably of order one)
in this supersymmetric Yang-Mills theory, 
as well as the existence of a previously unsuspected
intermediate temperature phase in this theory at weak coupling.
This new intermediate temperature phase should then have a dual description
in terms of a new set of bulk objects: mysterious new stable
black holes.

The picture we present in this paper supports the view that 't
Hooft's relation between large $N$ gauge theories and string
theories \thooft\ may extend also to weakly coupled gauge theories. The
naive picture of this relation is that at strong coupling the sum
over planar Feynman diagrams is dominated by graphs with
arbitrarily many interactions, which become dense and close up the
holes in the worldsheet of the Feynman diagram to form a
continuous $1+1$ dimensional field theory on a spherical Riemann
surface. This view has already been seriously challenged by the
AdS/CFT correspondence, which establishes a relationship between
string theories and gauge theories on a sphere at all values of
the gauge coupling\foot{Although the worldsheet theories dual to
perturbative gauge theories are strongly coupled, as a consequence
of large background curvatures, they presumably exist, as they may
be obtained from the well-understood duals to strongly coupled
gauge theories by tuning worldsheet parameters. See
\refs{\TseytlinGZ\KarchVN\joe\DharFI\ClarkWK\BianchiWX-\deMedeirosHR}
for recent attempts to understand this weakly coupled limit.}, and
by the consequent derivation of dualities between (topological)
open and closed string theories \refs{\rajesh, \OoguriGX}. Our
findings provide further evidence for the existence of a stringy
dual of weakly coupled gauge theories.  We have found that weakly
coupled gauge theories on compact manifolds share at least one
qualitative feature of string theory in flat space, namely a
string-like spectrum, in a regime in which the Feynman diagrams do
not seem to look like continuous Riemann surfaces. It would be
very interesting to understand more directly how and why weakly
coupled gauge theories manage to rearrange themselves as string
theories (see \refs{ \sundfree\GopakumarNS-\thorn} for some
attempts in this direction).

It is an interesting challenge to identify the string theories
which are dual to various weakly coupled (or free) gauge theories. For
the free gauge theories our analysis of \S3 gives us an explicit
formula \sumex\ for the spectrum of (free) single-string states, and perhaps
in some cases this may be enough to reconstruct the corresponding
worldsheet theory. By expanding our result for the partition function
in a power series in $1/N^2$, we can also extract information about
string coupling corrections in the putative dual string theory (though
it may turn out that real string interactions always involve the
Yang-Mills interaction, as in the study of string interactions on plane
waves \ppint).

The results of the computations in this paper (and our upcoming work
\future) may turn out to have independent interest. For conformal theories
such as the $d=4$ $\CN=4$ supersymmetric Yang-Mills theory, 
the partition function
computed in this paper and in \future\ encodes, in principle, the scaling
dimension of every operator in the theory, to the order
of computation (in $\lambda$)
\foot{This is true at least in the absence of first order phase 
transitions.}. 
The information about these anomalous
dimensions is packaged in an interesting fashion in $S_{eff}$ and may lead
to insights, perhaps in combination with recent speculations on
integrability and on the existence of a
large Yangian symmetry in the $\CN=4$ SYM theory \integrability.

The techniques of this paper may find application to several theories not
explicitly considered in this paper. It would be interesting to compare our
results to known results in strongly coupled two dimensional gauge theories
\refs{\td,\KutasovGQ,\tdstring}. On another note,
recall that large $N$ gauge theories in $d=1$ with a single scalar field
are believed to be holographically dual
to the $c=1$ and ${\hat c}=1$ string theories. Our analysis of free
one-matrix positive-sign harmonic oscillators at zero coupling shows
that they have neither phase transitions nor a Hagedorn-like spectrum;
however, the matrix models dual to the $c=1$ and ${\hat c}=1$ strings
are `negative mass'
harmonic oscillators, and it is possible that they somehow undergo a
`deconfining
transition' of the sort studied in this paper, described by some
effective action $S_{eff}(u_1)$, and that their high
energy behavior is dominated by two dimensional black holes (whose
spectrum is Hagedorn-like).\foot{Several people, including
A. Adams, P. Ho, G. Mandal, and S. Wadia,
have suggested a variant of this idea.} Another interesting example is
provided by the 't Hooft limit of the maximally supersymmetric
plane-wave deformation of Matrix theory, argued in \msv\ to be dual to
a little string theory compactified on $S^5$. For a small sphere, this theory
is weakly coupled, and one may study the little string theory
thermodynamics explicitly in this limit using methods similar to this
paper. This analysis is the subject of a paper \gordon\ to appear
shortly. Another interesting relation between the thermodynamics of
string theory, black holes and the thermodynamics of compactified 
gauge theories appears in the
context of Matrix theory (see, for instance, \martinec). It would be
interesting to try to relate our discussion to the results for toroidal
compactifications which are relevant there.
When we compactify a gauge theory on a space with a non-zero
fundamental group, we necessarily have additional zero modes coming
from the non-trivial Wilson loops. It would be interesting to
generalize our results to include such additional degrees of
freedom. Some recent results on the deconfinement transition of the
large $N$ gauge theory on a torus in the strong coupling limit appear
in \herbert.

We should also note that the connection between Hagedorn transitions,
black hole formation and some sort of `deconfinement' transition in a
dual field theory seems more general than the specific example of
gauge theories. For instance, string theory on $AdS_3$ is known to
undergo a phase transition as a function of temperature both
at `weak coupling' \justin\ (at the orbifold point) and at strong coupling (in
a geometric phase for the bulk). At weak coupling this transition (between a 
single ``long string'' phase and a multiple ``short string'' phase) is 
Hagedorn-like \malda, while at strong coupling it is
a first order Hawking Page transition \HawkingDH\ of the sort reviewed in this
paper. It would be interesting to analyze the extrapolation between
these two behaviors.

Finally, several technical and conceptual issues remain to be
addressed even within the direct line of attack of our paper. The
coefficient $b$ should certainly be computed for several gauge
theories \future. The analysis of this paper can be extended to search for
interesting features in generalized partition functions \foot{For
instance, the partition function generalized by the addition of a
chemical potential for an
$R$-symmetry charge in the strongly coupled $\CN=4$ SYM theory undergoes a
phase transition as a function of the chemical potential even at
zero temperature \refs{\ChamblinTK,\BuchelRE}. By generalizing our
analysis of this paper to this case, using the results of \S3.4,
we find that this zero temperature phase transition is absent in
the free Yang-Mills theory. Note that for finite temperature 
our analysis breaks down at a
critical value of the chemical potential when additional degrees of
freedom become light and charged scalars may condense \rotbh.
So (as noted for the compactified gauge theory 
in \HawkingDP) the strong coupling and weak coupling
regimes seem qualitatively different in this case.
Similar considerations may allow us
to make contact with string thermodynamics on
plane waves \GreeneCD.}.
And, last but not least, the dual interpretation of deconfining
phase transitions and their intriguing connection with black holes
certainly deserves further study.

\bigskip

\centerline{\bf Acknowledgements}

We would like to thank T. Banks, R. Dijkgraaf, A. Das, S. Elitzur,
D. Gross, M. Headrick, C. Johnson, S. Kachru, I. Klebanov,
J. Maldacena, G. Mandal, 
L. Motl, A. Neitzke, B. Pioline, J. Polchinski, E. Rabinovici, 
A. Schwimmer,
A. Sen, T. Senthil, S. Sethi, Y. Shamir, S. Shenker, A. Strominger,
T. Takayanagi, W. Taylor, A. Vishwanath, A. Yarom, X. Yin
and especially N. Arkani Hamed, R. Gopakumar, G. Semenoff, and S. Wadia 
for useful
discussions. We would like to thank Stanford University and the
Stanford-Weizmann workshop (partly financed by the
Israel-U.S. Binational Science Foundation) for hospitality during the
genesis of this project. OA would like to thank the University of
Chicago, the second Crete regional meeting on String Theory, the
Strings 2003 conference in Kyoto, the Aspen Center for Physics, and
the University of British Columbia for hospitality during the work on
this project.  JM would like to thank PIMS and UBC for hospitality during
the FMP school. SM would like to thank the Tata Institute of Fundamental
Research, ICTP Trieste, the organizers of the regional summer school at
Villa de Levya, Columbia, the second Crete regional meeting on String Theory,
the Benasque post-Strings workshop, and the Indian Institute for the
Cultivation of Sciences for hospitality while this work was in progress.
MVR would like to thank the Korean Institute for Advanced Study and
the Asia Pacific Center for Theoretical Physics for hospitality while
this work was in progress.
The work of OA was supported in part by the
Israel-U.S. Binational Science Foundation, by the ISF Centers of
Excellence program, by the European network HPRN-CT-2000-00122, and by
Minerva. OA is the incumbent of the Joseph and Celia Reskin career
development chair. The work of JM was supported in part by
an NSF Graduate Research Fellowship. The work of SM was supported  in part
by DOE grant DE-FG01-91ER40654 and a Harvard Junior Fellowship. The work
of KP was supported in part by DOE grant DE-FG01-91ER40654.
The work of MVR was supported in part by NSF grant
PHY-9870115, by funds from the Stanford Institute for
Theoretical Physics, by NSERC grant 22R81136 and by the Canada Research 
Chairs programme.

\appendix{A}{Properties of group characters}

In applications of group theory, it is often necessary to determine
the set of irreducible representations obtained in the tensor product
of some collection of other representations $R_i$. Characters provide
a powerful tool to achieve this. Given a group $G$, we may associate
with any representation $R_i$ a character $\chi_{R_i} : G \to
\IC$, defined such that $\chi_{R_i}(U)$ is equal to the trace of
the group element
$U$ in the representation $R_i$. From this definition, it follows that
the characters for sums and products of representations are given by
\eqn\sumprod{
\chi_{R_1 \oplus R_2} = \chi_{R_1} + \chi_{R_2},
\qquad \qquad \chi_{R_1 \otimes R_2} = \chi_{R_1} \cdot \chi_{R_2}.
}
The utility of the characters in decomposing tensor products follows
from the orthogonality of characters for irreducible representations
$R_i^I$ :
\eqn\orthorel{
\int [dU] \chi^*_{R^I_1}(U) \chi_{R^I_2}(U) = \delta_{R^I_1 R^I_2},
}
where $[dU]$ is the invariant (Haar) measure on the group manifold
normalized so that $\int [dU]=1$.  Thus, the number of irreducible
representations of type $R^I$ in the tensor product of the representations
$R_1, \dots , R_n$ is given by
\eqn\extractirred{
n_{R^I} = \int [dU] \chi^*_{R^I}(U) \prod_{i=1}^n \chi_{R_i}(U).
}
For the special case of the trivial (singlet) representation, we have
$\chi_{singlet}(U) = 1$, so that the number of singlets in the product
of representations $R_i$ is simply
\eqn\extractsing{
n_{singlet} = \int [dU] \prod_i \chi_{R_i}(U).  }
Finally, for the
applications of this paper, we require the character formulae for the
symmetrized and antisymmetrized products of $n$ identical
representations $R$. To obtain these, let $U_R$ be the matrix
representation of the group element $U$ in the representation
$R$. Then the trace of the matrix representation of $U$ in the
(anti)symmetrized tensor product of $n$ copies of $R$ is
\eqn\productrep{
(U_R)^{a_1}_{[a_1} \cdots (U_R)^{a_n}_{a_n]_{\pm}},
}
where the $[\cdots]_{\pm}$ indicates symmetrization or
anti-symmetrization of the indices with unit weight. This expression is
exactly the $t^n$ term in the expansion of the integral
\eqn\defofg{
G_{\pm}(U,t) = {1 \over \pi^{dim(R)}} \int [d \phi_\pm]
e^{-\bar{\phi}_\pm \phi_{\pm} \pm t \bar{\phi}_\pm U_R \phi_\pm}
}
in powers of $t$, where $\phi_+$ and $\phi_-$ are complex bosonic or
fermionic variables, respectively, in the representation $R$.
Thus, the result of this integral
\eqn\intres{
G_\pm (U,t) = (\det(1\mp t U_R))^{\mp 1}
}
serves as a generating function for the characters of
(anti)symmetrized products of arbitrary copies of the representation
R. More explicitly, by expanding the determinant we may express the
results directly in terms of the character for the representation $R$
as
\eqn\gensym{\eqalign{
G_+(U,t) &\equiv \sum_{n=0}^\infty t^n \chi_{sym^n(R)}(U) =
e^{\sum_{l=1}^\infty  t^l\; \chi_R(U^l) / l}, \cr
G_-(U,t) &\equiv \sum_{n=0}^\infty t^n \chi_{anti^n (R)}(U) =
e^{  \sum_{l=1}^\infty (-1)^{l+1} t^l\; \chi_R(U^l) / l }. \cr
}}

\appendix{B}{Counting states in $U(N)$ gauge theories}

In this appendix we derive the precise formula \sumex\ for the
counting of gauge-invariant states in a large $N$ theory with adjoint
fields, and we discuss the single-particle partition functions for
theories on a sphere with various field contents.

\subsec{Counting gauge-invariant states precisely}

In order to count the number of independent operators corresponding to
traces of products of fields in the large $N$ limit,
we wish to count the number of different arrangements of objects
subject to an identification of arrangements related by a cyclic
permutation. This can be done using Polya's theorem \refs{\SundborgUE,
\PolyakovAF}.

Consider a set of $m$ types of objects, and associate a weight $x_i$ with
each of these objects. The weight associated with a collection of
these objects is simply the product of the weights associated with
each of the individual objects. Polya solved the general problem
of summing over weights for all sets of $k$ of these objects, two
sets being treated as identical if they are related to each other
under the action of a specified subgroup of the permutation group.
The subgroup relevant to us is simply the cyclic subgroup of order
$k$; we will state Polya's result for this case. Define the
polynomial \eqn\poppol{p_k(y_1, y_2, \ldots , y_k)= { 1\over k}
\sum_{\pi} y_1^{n(\pi)_1} y_2^{n(\pi)_2} \ldots y_k^{n(\pi)_k},}
where the summation in \poppol\ runs over all elements $\pi$ of
the cyclic subgroup, and $n(\pi)_i$ is the number of cycles of
length $i$ in the permutation $\pi$. The answer to the question
addressed earlier in this paragraph is simply \eqn\polans{
p_k(\sum_{i=1}^m x_i,\sum_{i=1}^m x_i^2, \ldots, \sum_{i=1}^m
x_i^k).}

Applying this result to our problem, we find that the large $N$ partition
function
of single-trace states with $k$ oscillators is precisely given by
\eqn\sex{Z_k=p_k(z(x), z(x^2), z(x^3), \cdots , z(x^k)),}
where, as in \S3, $z(x)$ is the single-particle partition function.
This implies that
\eqn\sextwo{\eqalign{Z_{ST}=&\sum_{k=1}^{\infty}
p_k(z(x), z(x^2), z(x^3), \cdots ,
z(x^k))= \cr =&\sum_{k=1}^{\infty}{1\over k} \sum_{l=1}^{k}
z(x)^{n(k,l)_1} z(x^2)^{n(k,l)_2} \cdots z(x^k)^{n(k,l)_k},}}
where $n(k,l)_q$ refers to the number of cycles of length $q$ in the cyclic
permutation by $l$ shifts of $k$ objects.

It is easy to convince oneself that for specific values of $k$ and $l$,
$n(k,l)_q$ is non-zero for only one
value of $q$. At that value of $q$ it is given by $G(l,k)$, the greatest
common divisor of $l$ and $k$. It then follows that the $q$ for which
$n(k,l)_q$ is non-zero is given by $q =  k / G(l,k)$. Consequently, \sex\
may be rewritten as
\eqn\sext{Z_{ST} = \sum_{k=1}^{\infty}{1\over k} \sum_{l=1}^{k}
z(x^{k / G(l,k)})^{G(l,k)}.}
We now group together all terms with the same (fixed) $q=k /
G(l,k)$, so that $k=G(l,k) q$. Denoting $j=G(l,k)$,
we change the sum over $l$ and $k$ to a sum over $j$ and $q$, where
each term appears once for every $l \leq jq$ such that $G(l, jq) = j$. The
number of such $l$'s is precisely $\varphi(q)$, the number of positive
integers which are not larger than $q$ and are relatively prime to
$q$. Thus, we obtain
\eqn\sextn{Z_{ST} = \sum_{j=1}^{\infty} \sum_{q=1}^{\infty}
{\varphi(q) \over {jq}} z(x^q)^j = - \sum_{q=1}^{\infty} {\varphi(q)\over q}
\ln(1-z(x^q)),}
as in \sumex.

\subsec{Evaluating single-particle partition functions on spheres}

Next, we turn to a different topic which is the evaluation of the
single-particle partition functions for $d$-dimensional field
theories compactified on $S^{d-1}\times \IR$, with unit radius for
$S^{d-1}$. This may be carried out directly
by noting that the Laplacian on the sphere (or the spatial parts of
the other wave operators corresponding to particles with spin) may be
written directly in terms of angular momentum generators, which may be
diagonalized in the usual way.

Alternatively, since free field theories
are conformally invariant, and we are interested in conformally
coupled fields (though it is easy to generalize our results also to
other cases), we can use the conformal transformation that relates
$S^{d-1}\times \IR$ to $\IR^d$. This transformation takes states of the
field theory on $S^{d-1}\times \IR$ to local operators on $\IR^d$, with
the energy of the state becoming the scaling dimension of the
operator. Thus, an equivalent way to define the partition function in
such a case is by $z(x) = \sum_{local\ operators} x^{\Delta}$, where $\Delta$
is the scaling dimension of the operator.

We begin by considering a free scalar field $\ph$.
The local operators in the theory
are $\ph$, $\p_i \ph$, $\p_i \p_j \ph$, and so on, modulo the equation of
motion. Ignoring the equation of motion for a moment, these operators are all
generated by repeated application of the $d$ different derivative
operators $\p_1, \p_2
\ldots \p_d$, each of which is of unit dimension and so has the partition
function ${1 \over (1-x)}$, on the free field $\ph$ of dimension $(d/2-1)$.
Multiplying the various partition functions we find
\eqn\zprimes{z'_S(x)={x^{d/2-1} \over (1-x)^d}.}
In order to obtain $z_S(x)$ we must
subtract from this
the partition function for operators that vanish by the
equation of motion $\p^2 \ph = 0$. Such operators are generated by acting
with an arbitrary number of derivatives on $\p^2 \ph$, so their partition
function is $x^2 z'_S(x)$. Thus, we find
\eqn\pflets{z_S(x)=(1-x^2) z'_S(x) =
{x^{{d\over 2}}+x^{{d \over 2}-1} \over (1-x)^{d-1}.}}
As a check, we note that in $d=4$ the operators that we get by acting
with $k$ derivatives are in the $k^{th}$ traceless symmetric
representation of $SO(4)$ which has $j_1=j_2={k \over 2}$, and they
have dimension $\Delta = k+1$. These are simply the scalar spherical
harmonics on $S^3$. This implies that the number of
operators of dimension $\Delta$ is
$n_S(\Delta)=(2j_1+1)(2j_2+1)=\Delta^2$, consistent with the Taylor
expansion of \pflets\ for $d=4$.

Next, we turn to the free vector field. The number of gauge-invariant
operators is independent of the gauge, so we can fix an arbitrary
gauge for the counting. We will use
the gauge $A_0=0$ on $S^{d-1}\times \IR$, which becomes the gauge
$x^{\mu} A_\mu=0$ after the conformal transformation to $\IR^d$
(recall that, according to the state-operator map,
all operators are to be evaluated at $x=0$).
Differentiating the gauge condition at the point $x=0$ we find the relations
\eqn\gca{A_{\mu}=0, ~~~\p_{\mu} A_{\nu}+\p_{\nu} A_{\mu}=0,~ \cdots,~
\p_{\{ i_1}\p_{i_2}\ldots \p_{i_n} A_{i_{n+1}\} }=0,\cdots}
where the brackets $\{ \}$ denote symmetrization. To start with we ignore both
\gca\ and the equation of motion -- this leads to a single-particle partition
function $z'_V(x) = x^{2-{d\over 2}} d z'_S(x)$ (since the gauge field must
have scaling dimension one in any space-time dimension). 
Operators of dimension $\Delta$ that are
set to zero by \gca\ are given by symmetric tensors of rank 
$\Delta$; based on the previous paragraph the corresponding
partition function is $x^{1-{d\over 2}} z'_S(x)-1$, where the last
subtraction comes because there are no tensors of rank zero in \gca.
With the condition \gca, the Maxwell equation (at $x=0$) simply reduces to
$\p^2 A_\mu=0$. The number of independent operators set to zero by the
equation of motion is, therefore, counted by
$d x^{4-{d\over 2}} z'_S(x)$. Finally, the number
of operators set to zero by both the constraint \gca\ and the equation of
motion is encoded in the partition function $x^{5-{d\over 2}} z'_S(x)$. 
Putting it all together, using \zprimes, we find
\eqn\pfletv{\eqalign{z_V(x)&= {dx \over (1-x)^d}
-{1\over (1-x)^d} -{dx^3 \over (1-x)^d} +{x^4 \over (1-x)^d} +1 = \cr
&= 1-{(1+x)(1+x^2-dx) \over (1-x)^{d-1}}.}}
As a check, we note that in four dimensions, the set of operators
formed by acting with $k$ derivatives on $A_{\mu}$, obeying
\gca\ and the equation of motion, transform in the $SO(4)$ representation
$(j_1, j_2)=({{k+1}\over 2}, {{k-1}\over 2})\oplus ({{k-1}\over 2},{{k+1}\over
2})$. These are the vector spherical harmonics on $S^3$. It
follows that the number of operators at dimension $\Delta$ is $n_V(\Delta) =
2(\Delta^2-1)$,
consistent with \pfletv.

Finally, we turn to free fermions. For concreteness we work in even dimensions
with complex spinors of no chirality restrictions. Such a spinor has
$2^{{d\over 2}+1}$ real components. Ignoring the equation of motion, the
partition function for spinors is
$z'_F(x)=2^{{d\over 2}+1}\sqrt{x} z'_S(x)$.
The partition function that counts the operators which are set to zero
by the Dirac equation, is $2^{{d\over 2}+1} x^{{3\over 2}}
z'_S(x)$. Subtracting the second from the first we find
\eqn\pfletsp{z_F(x)={2^{{d\over 2}+1}x^{{d\over 2}- \half} \over (1-x)^{d-1}}.}
Of course, \pfletsp\ should be divided by two for chiral spinors or real
spinors, and by four for spinors that are both chiral and real.
As a check on \pfletsp, note that, in $d=4$, the operators made from
a complex chiral fermion field, at dimension $k+\half$, transform in
the $SO(4)$ representation with
$(j_1, j_2)=({k\over 2}, {{k-1}\over 2})$;
there are $2k(k+1)$ such operators
(the factor of 2 is because the spinors are complex), in agreement with
\pfletsp.

Note that each of \pflets, \pfletv, and \pfletsp\ tends
as $x \to 1$ (the high temperature limit)  to
\eqn\Fht{z(x) \to {2{\cal N}^{dof} \over {(1-x)^{d-1}}},}
where ${\cal N}^{dof}$ is the number of physical real degrees of freedom in the
corresponding field.

The formulas in this section, used for $d=4$, imply the formula
\sthreefields\ of \S3.

\appendix{C}{Hagedorn transitions at finite string coupling}

There have long been speculations that at finite values of the string
coupling $g_s$, one could have a phase transition at (or near) the
Hagedorn temperature, with a different description of the theory at
high temperatures. In this appendix we analyze, following \AtickSI,
the effect of turning on a small string coupling $g_s$ on the partition
function. Recall that, as reviewed in \S2.1, the Euclidean partition
function includes a winding mode $W$ that becomes tachyonic above the Hagedorn
temperature.
On general grounds, the perturbative effective action
that describes the interaction between $W$ and all other modes (let us call
them $\ph_n$) takes the form
\eqn\seffstr{\eqalign{ S_{eff} =\int d^dx & \left( |\p W|^2 +
{2 \over \apm} ({1 \over 8 \pi^2 \apm T^2} -1) |W|^2
+ \sum_n \half (\p \ph_n)^2 +{M_n^2 \over 2} \ph_n^2 + \cdots \right) + \cr
&+ \left(g_s  \sum_n I_n \ph_n |W|^2  + g_s^2 \theta |W|^4 + \cdots \right).}}
If all other fields $\ph_n$ are massive we can integrate them out, and
remain with an effective action for $W$ near the Hagedorn temperature.
After rescaling $W$ this takes the
form
\eqn\seffw{S_{eff}= {1\over g_s^2} \int d^d x \left(|\p W|^2 + {2 \over \apm}(
{1 \over 8 \pi^2  \apm T^2}- 1) |W|^2 + b|W|^4 + \cdots
\right) + \cdots,}
where
\eqn\fb{b=\theta -\sum_{n} {I_n^2 \over 2 M_n^2} .}

The dynamics of this theory depends crucially on the sign of $b$. For $b>0$,
$S_{eff}$ is the Landau Ginzburg free energy for a system that undergoes
 a second order phase transition at $T=T_H$.
In this case, for $T<T_H$,
 $S_{eff}$ is minimized at $W=0$. The `saddle point' contribution to the free
energy vanishes and the leading $\CO(g_s^0)$ contribution to $\ln(Z)$ is given
by the free string theory partition function.
However, at  temperatures just above $T_H$, $S_{eff}$ is minimized at
\eqn\wsaddle{|W|^2={ {T-T_H} \over {4\pi^2 b \apm^2 T_H^3}},}
and the $\CO(g_s^{-2}$) saddle point contribution to the free energy is
\eqn\zsaddle{{\ln(Z) \over V} = {1 \over g_s^2} \cdot
{ (T - T_H)^2 \over {32\pi^4 \apm^4 T_H^6 b}}.}

On the other hand, for $b<0$, $S_{eff}$ is the Landau Ginzburg free energy
for a system that potentially undergoes a first order phase transition at
a temperature lower than $T_H$. The $\CO(g_s^0)$ free energy below the phase
transition temperature is once again computed by free string theory. In this
case the high temperature behavior
is dominated by large values of $W$, so it cannot
be controlled in string perturbation theory; if there exists a high
temperature saddle point, ${\ln(Z) \over V}$ is again of $\CO(g_s^{-2})$,
but its precise value depends on the details of the terms we denoted by
``$\cdots$'' in \seffw.

For string theory in flat space the massless dilaton is always one of the
modes $\ph_n$, giving an infinite negative contribution to $b$ in \fb, so one
expects that if there is a phase transition in this case it would be of
first order. Unfortunately, presumably in this case (and for any string theory
with finite coupling in $\IR^{d-1,1}$) no sensible high-temperature
phase exists, both because there is a Jeans instability which cannot be
ignored once the free energy is of order $1/g_s^2$ \AtickSI,
and because we expect the
high-energy spectrum to include black holes with a density of states growing
faster than exponential (see, e.g., \AharonyTT).
However, while the formulas we wrote above strictly
apply only to the flat space
case, a similar Hagedorn behavior may be found in other spaces as well,
including spaces like anti-de Sitter space where the spectrum effectively
has a mass gap, and where these problems do not occur. We expect that a
similar effective action would arise also in these cases\foot{Though, in
theories that effectively have finite volume, we do not expect to find
strict phase transitions at finite coupling.}.
In such cases, for example in type IIB string theory on
$AdS_5\times S^5$, the arguments presented above may apply,
with the value of $b$
being either positive or negative, depending on the dynamics.
It turns out that the specific case
of string theory on anti-de Sitter space in global coordinates
corresponds to the case of $b < 0$, since it is known to exhibit
a first order phase transition which occurs (at small curvatures)
well below the Hagedorn temperature \refs{\HawkingDH,\WittenZW}. As
expected from the discussion above, this phase transition is not
visible in string perturbation theory around the $AdS_5$ background
but requires additional input.

It is interesting to note the similarity between the analysis
here and the analysis of \S6 of the Hagedorn transition in weakly
coupled large $N$ gauge theories. 

\listrefs
\end